\documentclass[
  reprint,
  superscriptaddress,
  amsmath,amssymb,
  aps,
]{revtex4-1}

\usepackage{lineno,hyperref}   
\usepackage[pdftex]{graphicx}  

\usepackage{dcolumn}
\usepackage{bm}
\usepackage{footmisc}
\usepackage{url}
\usepackage{color}   
\begin{document} 
%\preprint{APS/PhySH/XXX}

%%-----------------------------------------------------------------------------
%  Title, Author List, Abstract ... 
% 

\title{Measurement of neutrino and antineutrino neutral-current quasielastic-like 
       \\ interactions on oxygen by detecting nuclear de-excitation $\gamma$-rays}
    %\\ with data set from  ${\bf 3.16 \times 10^{21}}$~protons-on-target}
% 
%\author{Author list to be added later ...}
%%%%%%%%%%%%%%%%%%%%%%%%%%%%%%%%%%%%%%%%%%%%%%%%%%%%%%%%%%%%%%
% T2K author list generated on Sun Sep 15 19:06:46 2019
% setting: extra = False
%         author list from archive (starting July 11 2019 until September 30 2019)
%         exemption(s) granted to: svalder, abubak, munteanu, eeergo, prasad, kikawa, kuribayashi, odagawa,  tajima,  zoya, xiaoyueli
% Number of authors = 333
%%%%%%%%%%%%%%%%%%%%%%%%%%%%%%%%%%%%%%%%%%%%%%%%%%%%%%%%%%%%%%

\newcommand{\INSTHD}{\affiliation{University Autonoma Madrid, Department of Theoretical Physics, 28049 Madrid, Spain}}
\newcommand{\INSTEE}{\affiliation{University of Bern, Albert Einstein Center for Fundamental Physics, Laboratory for High Energy Physics (LHEP), Bern, Switzerland}}
\newcommand{\INSTFE}{\affiliation{Boston University, Department of Physics, Boston, Massachusetts, U.S.A.}}
\newcommand{\INSTD}{\affiliation{University of British Columbia, Department of Physics and Astronomy, Vancouver, British Columbia, Canada}}
\newcommand{\INSTGA}{\affiliation{University of California, Irvine, Department of Physics and Astronomy, Irvine, California, U.S.A.}}
\newcommand{\INSTI}{\affiliation{IRFU, CEA Saclay, Gif-sur-Yvette, France}}
\newcommand{\INSTGB}{\affiliation{University of Colorado at Boulder, Department of Physics, Boulder, Colorado, U.S.A.}}
\newcommand{\INSTFG}{\affiliation{Colorado State University, Department of Physics, Fort Collins, Colorado, U.S.A.}}
\newcommand{\INSTFH}{\affiliation{Duke University, Department of Physics, Durham, North Carolina, U.S.A.}}
\newcommand{\INSTBA}{\affiliation{Ecole Polytechnique, IN2P3-CNRS, Laboratoire Leprince-Ringuet, Palaiseau, France }}
\newcommand{\INSTEF}{\affiliation{ETH Zurich, Institute for Particle Physics and Astrophysics, Zurich, Switzerland}}
\newcommand{\INSTIE}{\affiliation{CERN European Organization for Nuclear Research, CH-1211 Genève 23, Switzerland}}
\newcommand{\INSTEG}{\affiliation{University of Geneva, Section de Physique, DPNC, Geneva, Switzerland}}
\newcommand{\INSTHJ}{\affiliation{University of Glasgow, School of Physics and Astronomy, Glasgow, United Kingdom}}
\newcommand{\INSTDG}{\affiliation{H. Niewodniczanski Institute of Nuclear Physics PAN, Cracow, Poland}}
\newcommand{\INSTCB}{\affiliation{High Energy Accelerator Research Organization (KEK), Tsukuba, Ibaraki, Japan}}
\newcommand{\INSTIB}{\affiliation{University of Houston, Department of Physics, Houston, Texas, U.S.A.}}
\newcommand{\INSTED}{\affiliation{Institut de Fisica d'Altes Energies (IFAE), The Barcelona Institute of Science and Technology, Campus UAB, Bellaterra (Barcelona) Spain}}
\newcommand{\INSTEC}{\affiliation{IFIC (CSIC \& University of Valencia), Valencia, Spain}}
\newcommand{\INSTHH}{\affiliation{Institute For Interdisciplinary Research in Science and Education (IFIRSE), ICISE, Quy Nhon, Vietnam}}
\newcommand{\INSTEI}{\affiliation{Imperial College London, Department of Physics, London, United Kingdom}}
\newcommand{\INSTGF}{\affiliation{INFN Sezione di Bari and Universit\`a e Politecnico di Bari, Dipartimento Interuniversitario di Fisica, Bari, Italy}}
\newcommand{\INSTBE}{\affiliation{INFN Sezione di Napoli and Universit\`a di Napoli, Dipartimento di Fisica, Napoli, Italy}}
\newcommand{\INSTBF}{\affiliation{INFN Sezione di Padova and Universit\`a di Padova, Dipartimento di Fisica, Padova, Italy}}
\newcommand{\INSTBD}{\affiliation{INFN Sezione di Roma and Universit\`a di Roma ``La Sapienza'', Roma, Italy}}
\newcommand{\INSTEB}{\affiliation{Institute for Nuclear Research of the Russian Academy of Sciences, Moscow, Russia}}
\newcommand{\INSTHI}{\affiliation{International Centre of Physics, Institute of Physics (IOP), Vietnam Academy of Science and Technology (VAST), 10 Dao Tan, Ba Dinh, Hanoi, Vietnam}}
\newcommand{\INSTHA}{\affiliation{Kavli Institute for the Physics and Mathematics of the Universe (WPI), The University of Tokyo Institutes for Advanced Study, University of Tokyo, Kashiwa, Chiba, Japan}}
\newcommand{\INSTID}{\affiliation{Keio University, Department of Physics, Kanagawa, Japan}}
\newcommand{\INSTIF}{\affiliation{King's College London, Department of Physics, Strand, London WC2R 2LS, United Kingdom}}
\newcommand{\INSTCC}{\affiliation{Kobe University, Kobe, Japan}}
\newcommand{\INSTCD}{\affiliation{Kyoto University, Department of Physics, Kyoto, Japan}}
\newcommand{\INSTEJ}{\affiliation{Lancaster University, Physics Department, Lancaster, United Kingdom}}
\newcommand{\INSTFC}{\affiliation{University of Liverpool, Department of Physics, Liverpool, United Kingdom}}
\newcommand{\INSTFI}{\affiliation{Louisiana State University, Department of Physics and Astronomy, Baton Rouge, Louisiana, U.S.A.}}
\newcommand{\INSTHB}{\affiliation{Michigan State University, Department of Physics and Astronomy,  East Lansing, Michigan, U.S.A.}}
\newcommand{\INSTCE}{\affiliation{Miyagi University of Education, Department of Physics, Sendai, Japan}}
\newcommand{\INSTDF}{\affiliation{National Centre for Nuclear Research, Warsaw, Poland}}
\newcommand{\INSTFJ}{\affiliation{State University of New York at Stony Brook, Department of Physics and Astronomy, Stony Brook, New York, U.S.A.}}
\newcommand{\INSTGJ}{\affiliation{Okayama University, Department of Physics, Okayama, Japan}}
\newcommand{\INSTCF}{\affiliation{Osaka City University, Department of Physics, Osaka, Japan}}
\newcommand{\INSTGG}{\affiliation{Oxford University, Department of Physics, Oxford, United Kingdom}}
\newcommand{\INSTGC}{\affiliation{University of Pittsburgh, Department of Physics and Astronomy, Pittsburgh, Pennsylvania, U.S.A.}}
\newcommand{\INSTFA}{\affiliation{Queen Mary University of London, School of Physics and Astronomy, London, United Kingdom}}
\newcommand{\INSTE}{\affiliation{University of Regina, Department of Physics, Regina, Saskatchewan, Canada}}
\newcommand{\INSTGD}{\affiliation{University of Rochester, Department of Physics and Astronomy, Rochester, New York, U.S.A.}}
\newcommand{\INSTHC}{\affiliation{Royal Holloway University of London, Department of Physics, Egham, Surrey, United Kingdom}}
\newcommand{\INSTBC}{\affiliation{RWTH Aachen University, III. Physikalisches Institut, Aachen, Germany}}
\newcommand{\INSTFB}{\affiliation{University of Sheffield, Department of Physics and Astronomy, Sheffield, United Kingdom}}
\newcommand{\INSTDI}{\affiliation{University of Silesia, Institute of Physics, Katowice, Poland}}
\newcommand{\INSTIA}{\affiliation{SLAC National Accelerator Laboratory, Stanford University, Menlo Park, California, USA}}
\newcommand{\INSTBB}{\affiliation{Sorbonne Universit\'e, Universit\'e Paris Diderot, CNRS/IN2P3, Laboratoire de Physique Nucl\'eaire et de Hautes Energies (LPNHE), Paris, France}}
\newcommand{\INSTEH}{\affiliation{STFC, Rutherford Appleton Laboratory, Harwell Oxford,  and  Daresbury Laboratory, Warrington, United Kingdom}}
\newcommand{\INSTCH}{\affiliation{University of Tokyo, Department of Physics, Tokyo, Japan}}
\newcommand{\INSTBJ}{\affiliation{University of Tokyo, Institute for Cosmic Ray Research, Kamioka Observatory, Kamioka, Japan}}
\newcommand{\INSTCG}{\affiliation{University of Tokyo, Institute for Cosmic Ray Research, Research Center for Cosmic Neutrinos, Kashiwa, Japan}}
\newcommand{\INSTHF}{\affiliation{Tokyo Institute of Technology, Department of Physics, Tokyo, Japan}}
\newcommand{\INSTGI}{\affiliation{Tokyo Metropolitan University, Department of Physics, Tokyo, Japan}}
\newcommand{\INSTHG}{\affiliation{Tokyo University of Science, Faculty of Science and Technology, Department of Physics, Noda, Chiba, Japan}}
\newcommand{\INSTF}{\affiliation{University of Toronto, Department of Physics, Toronto, Ontario, Canada}}
\newcommand{\INSTB}{\affiliation{TRIUMF, Vancouver, British Columbia, Canada}}
\newcommand{\INSTG}{\affiliation{University of Victoria, Department of Physics and Astronomy, Victoria, British Columbia, Canada}}
\newcommand{\INSTDJ}{\affiliation{University of Warsaw, Faculty of Physics, Warsaw, Poland}}
\newcommand{\INSTDH}{\affiliation{Warsaw University of Technology, Institute of Radioelectronics and Multimedia Technology, Warsaw, Poland}}
\newcommand{\INSTFD}{\affiliation{University of Warwick, Department of Physics, Coventry, United Kingdom}}
\newcommand{\INSTGH}{\affiliation{University of Winnipeg, Department of Physics, Winnipeg, Manitoba, Canada}}
\newcommand{\INSTEA}{\affiliation{Wroclaw University, Faculty of Physics and Astronomy, Wroclaw, Poland}}
\newcommand{\INSTHE}{\affiliation{Yokohama National University, Faculty of Engineering, Yokohama, Japan}}
\newcommand{\INSTH}{\affiliation{York University, Department of Physics and Astronomy, Toronto, Ontario, Canada}}

\INSTHD
\INSTEE
\INSTFE
\INSTD
\INSTGA
\INSTI
\INSTGB
\INSTFG
\INSTFH
\INSTBA
\INSTEF
\INSTIE
\INSTEG
\INSTHJ
\INSTDG
\INSTCB
\INSTIB
\INSTED
\INSTEC
\INSTHH
\INSTEI
\INSTGF
\INSTBE
\INSTBF
\INSTBD
\INSTEB
\INSTHI
\INSTHA
\INSTID
\INSTIF
\INSTCC
\INSTCD
\INSTEJ
\INSTFC
\INSTFI
\INSTHB
\INSTCE
\INSTDF
\INSTFJ
\INSTGJ
\INSTCF
\INSTGG
\INSTGC
\INSTFA
\INSTE
\INSTGD
\INSTHC
\INSTBC
\INSTFB
\INSTDI
\INSTIA
\INSTBB
\INSTEH
\INSTCH
\INSTBJ
\INSTCG
\INSTHF
\INSTGI
\INSTHG
\INSTF
\INSTB
\INSTG
\INSTDJ
\INSTDH
\INSTFD
\INSTGH
\INSTEA
\INSTHE
\INSTH

\author{K.\,Abe}\INSTBJ
\author{R.\,Akutsu}\INSTCG
\author{A.\,Ali}\INSTCD
\author{C.\,Alt}\INSTEF
\author{C.\,Andreopoulos}\INSTEH\INSTFC
\author{L.\,Anthony}\INSTFC
\author{M.\,Antonova}\INSTEC
\author{S.\,Aoki}\INSTCC
\author{A.\,Ariga}\INSTEE
\author{Y.\,Ashida}\INSTCD
% ...

\collaboration{The T2K Collaboration}\noaffiliation

\date{\today}

\begin{abstract}

Neutrino- and antineutrino-oxygen neutral-current quasielastic-like interactions 
are measured at Super-Kamiokande using 
nuclear de-excitation $\gamma$-rays to identify signal-like interactions
in data from a $14.94 \ (16.35)\times 10^{20}$ protons-on-target exposure of 
the T2K neutrino (antineutrino) beam. 
The measured flux-averaged cross sections on oxygen nuclei are 
$\langle \sigma_{\nu {\rm \mathchar`-NCQE}} \rangle = 
1.70 \pm 0.17 ({\rm stat.}) ^{+ {\rm 0.51}}_{- {\rm 0.38}} ({\rm syst.}) 
\times 10^{-38} \ {\rm cm^2/oxygen}$ with a flux-averaged energy of 0.82~GeV and 
$\langle \sigma_{\bar{\nu} {\rm \mathchar`-NCQE}} \rangle = 
0.98 \pm 0.16 ({\rm stat.}) ^{+ {\rm 0.26}}_{- {\rm 0.19}} ({\rm syst.})
\times 10^{-38} \ {\rm cm^2/oxygen}$ with a flux-averaged energy of 0.68~GeV, 
for neutrinos and antineutrinos, respectively. 
These results are the most precise to date, and the antineutrino result 
is the first cross section measurement of this channel. 
They are compared with various theoretical predictions.
The impact on evaluation of backgrounds to searches for supernova relic neutrinos 
at present and future water Cherenkov detectors is also discussed.

\end{abstract}
%

%\pacs{\textcolor{red}{Valid PACS appear here}} 
\maketitle
%\tableofcontents

%%-----------------------------------------------------------------------------
%  Main Part ... 
% 

%------------------------------------------------------------------------------
%  Introduction 
   \section{Introduction}
   \label{sec:intro}
%..............................................................................

Measurements of neutrino neutral-current (NC) processes give insight into 
neutrino-nucleus interactions and are important for understanding the nucleon itself 
as well as improving the sensitivity of searches for a variety of physics phenomena.
%% providing complementary information to 
%% charged-current (CC) interaction studies, .
%
The strange quark content of the nucleon ($\Delta s$), for instance, 
can be probed via NC interactions (see Ref.~\cite{bib:revnuint} and references therein), 
and its measurements have been demonstrated by the BNL E734 experiment~\cite{bib:xsecbnle734} 
and the MiniBooNE experiment~\cite{bib:mininuncqe,bib:mininubarncqe}. 
Precision measurements of the neutrino- and antineutrino-oxygen NC 
interactions in the sub-GeV region, where the quasielastic process is expected to be dominant,
also benefit a diverse array of searches with water Cherenkov detectors, 
such as Super-Kamiokande (SK)~\cite{bib:superk}, its future upgrade, SK-Gd~\cite{bib:skgd}, 
and its successor, Hyper-Kamiokande~\cite{bib:hyperk}. 
In supernova relic neutrino (SRN) searches~\cite{bib:sksrn12,bib:sksrn123,bib:sksrn4}, 
the present uncertainty on these interactions induces a large error on 
atmospheric neutrino backgrounds, limiting the sensitivity at low energies 
where the SRN flux is predicted to be large.
When searching for dark matter in accelerator neutrino experiments, as suggested 
in Refs.~\cite{bib:tofdm1,bib:tofdm2}, the rate of NC interactions must be
accurately estimated as they are indistinguishable from the signal.
Another motivation arises in the search for sterile neutrinos in accelerator neutrino experiments
\cite{bib:minossterile,bib:novasterile,bib:t2ksterile}. 
The fact that the NC interaction cross section does not depend on the neutrino flavor makes it 
possible to search for a deficit of NC events, which would be interpreted
as transitions from active to sterile neutrinos. 
%The situtation is similar for sterile neutrino searches with accelerators~\cite{bib:minossterile,bib:novasterile,bib:t2ksterile}. 
%The fact that the NC cross section does not depend on the neutrino flavor makes it 
%possible to search for a deficit of NC events, which would be interpreted
%as transitions from active to sterile neutrinos. 

NC interactions at the neutrino energies of interest here ($E_\nu \lesssim 1$~GeV) 
are difficult to observe in water Cherenkov detectors because 
%%% PRD 
their final state particles are either neutral or charged but often below the Cherenkov threshold.
Instead, the present work seeks to identify these interactions using Cherenkov light 
arising from the electromagnetic cascade produced by $\gamma$-rays emitted from the de-excitation 
of the recoil nucleus~\cite{bib:folomeshkin,bib:gershtejn,bib:nussinov,bib:ankowski}.  
%In order to overcome this, nuclear de-excitation $\gamma$-rays and 
%% Cherenkov light arising from subsequent electromagnetic cascades 
% 
At $E_\nu \gtrsim 200$~MeV, the NC quasielastic nucleon knock-out (NCQE) processes, 
 
  %%% Equation : NCQE
  \begin{eqnarray}
  \label{eq:ncqeformula}
   \nu (\bar{\nu}) + {\rm ^{16}O} &\rightarrow& \nu (\bar{\nu}) + n + {\rm ^{15}O^{*}}, \\ [+10truept] 
   \nu (\bar{\nu}) + {\rm ^{16}O} &\rightarrow& \nu (\bar{\nu}) + p + {\rm ^{15}N^{*}}, 
  \end{eqnarray}
  \vspace{2truept}
  %%%

\noindent
%$\nu (\bar{\nu}) + {\rm ^{16}O} \rightarrow \nu (\bar{\nu}) + p + {\rm ^{15}N^{*}}$ or 
%$\nu (\bar{\nu}) + {\rm ^{16}O} \rightarrow \nu (\bar{\nu}) + n + {\rm ^{15}O^{*}}$, 
become dominant over NC inelastic processes without nucleon knock-out,  
%
%  %%% Equation : other NC inelastic 
%  \begin{eqnarray}
%  \label{eq:ncqeformula}
%   \nu (\bar{\nu}) + {\rm ^{16}O} \rightarrow \nu (\bar{\nu}) + {\rm ^{16}O^{*}}, 
%  \end{eqnarray}
%  \vspace{2truept}
%  %%%
%
%\noindent
$\nu (\bar{\nu}) + {\rm ^{16}O} \rightarrow \nu (\bar{\nu}) + {\rm ^{16}O^{*}}$ \cite{bib:ankowski}.
%
%This is referred to as the NCQE interaction. 
%%% PRD
The resulting excited nuclei relax to the ground state with the emission 
of $\gamma$-rays promptly.
These $\gamma$-rays are available as a probe to study the NCQE interaction 
as has been demonstrated at T2K~\cite{bib:t2kncqe1to3} and SK~\cite{bib:skncqe}.
%\textcolor{red}{in a short time-scale, usually less than O(1)~ns.
%These de-excitation $\gamma$-rays are reconstructed at the primary neutrino interaction 
%timing and thus available as a probe to study the NCQE interaction
%as has been demonstrated at T2K~\cite{bib:t2kncqe1to3} and SK~\cite{bib:skncqe}.}
%
Previous studies at T2K measured the neutrino-oxygen NCQE interaction cross section 
with a data set of $3.01\times10^{20}$~protons-on-target (POT) and  
SK measured this process with its atmospheric neutrino data, which is a mixture of 
neutrino and antineutrino interactions. 
Both measurements suffer from large statistical and systematic uncertainties. 

This paper reports the updated result from T2K using neutrinos and
the first measurement using antineutrinos.
% 
%We note that here the signal definition has been changed relative to 
%previous studies~\cite{bib:t2kncqe1to3,bib:skncqe} from ``NCQE'' to ``NCQE-like",
%because the present selection may contain contributions from 
%NC two-particle-two-hole (2p2h) interactions.
%
In this work the signal is termed ``NCQE-like", to highlight the fact that the event selection 
may contain contributions from NC two-particle-two-hole (2p2h) interactions 
%%% PRD 
where two nucleons are involved in the interaction via meson-exchange currents.
Previous studies \cite{bib:t2kncqe1to3,bib:skncqe} may have also included such events, 
though they were not addressed specifically. 
Further descriptions will be given in Section~\ref{sec:discuss}.
In the analysis, data taken with exposures of $14.94\times10^{20}$~POT 
in neutrino mode and $16.35\times10^{20}$~POT in antineutrino mode are used. 
Both the statistical and systematic errors have been reduced with 
the present analysis.

The paper is structured as follows.  
First, Section~\ref{sec:t2kexp} details the experimental setup of T2K. 
Section~\ref{sec:montecarlo} explains the Monte Carlo (MC) simulation
and is followed by descriptions of the event reconstruction and selection in Section~\ref{sec:reconsel}.
Estimates of uncertainties in the analysis are described in Section~\ref{sec:uncertainty} 
before cross section results are given in Section~\ref{sec:xsec}. 
After discussion of the results in Section~\ref{sec:discuss} 
concluding remarks are given in Section~\ref{sec:conclude}.

%\clearpage
%------------------------------------------------------------------------------
%  The T2K Experiment 
   \section{The T2K Experiment}
   \label{sec:t2kexp}
%..............................................................................

The T2K experiment \cite{bib:t2kexp} has been designed for precise measurement of 
neutrino oscillation parameters \cite{bib:t2k2017oa} 
and has a broad program of additional physics measurements.
It consists of the J-PARC neutrino beamline, near detectors, and 
SK as its far detector. 
T2K has taken data in nine separate run periods, termed Runs 1$-$9, and 
its beam intensity has increased throughout. 
Protons are bundled into eight bunches (six in Run~1), referred to as a spill, and 
accelerated to 30~GeV/c by the J-PARC Main Ring synchrotron. 
Bunches are approximately 100~ns wide and separated by about 580~ns and spills 
are delivered to the neutrino production target with a repetition rate of 2.48~s.
% 
% The proton beam is then directed onto a graphite target to produce 
% hadrons such as pions and kaons. 
% 
Hadrons produced in proton-target (graphite) interactions are efficiently focused and sign-selected 
by magnetic fields produced by three electromagnetic horns 
\cite{bib:horndesign,bib:hornmeasure}, before entering a decay volume.
The polarity of the magnetic horns can be changed, allowing 
selection and focusing of either positively or negatively charged hadrons
to produce beams composed of predominantly neutrinos or antineutrinos following the decay of the hadrons.
The former is referred to as forward horn current (FHC) mode while the latter 
is referred to as reverse horn current (RHC) mode.
Located 280~m away from the graphite target the two near detectors, INGRID~\cite{bib:ingrid} 
and ND280 \cite{bib:t2kfgd,bib:t2ktpc}, are placed on-axis and $2.5^{\circ}$ 
off-axis with respect to the proton beam direction, respectively. 
ND280 is used to measure the (anti)neutrino spectrum before the onset of neutrino 
oscillations and INGRID monitors the (anti)neutrino beam direction and intensity 
to ensure beam quality during data taking. 
In addition to the INGRID measurements 
a muon monitor placed just after the decay volume 
measures the beam direction and intensity on a bunch-by-bunch 
basis by detecting muons from pion and kaon decays~\cite{bib:mumondesign,bib:mumonmeasure,bib:mumonemt}.

Super-Kamiokande is located 295~km away from the target and $2.5^{\circ}$ off-axis. 
Beam timing information is shared between J-PARC and SK via a GPS system. 
It is a cylindrical water Cherenkov detector located 1,000~m under
Mt.~Ikeno in Kamioka, Japan. 
The detector is divided into two parts, an inner detector (ID) and an outer detector (OD).
The ID measures 33.8~m in diameter and 36.2~m in height and is instrumented 
with 11,129 20-inch inward-facing photomultiplier tubes (PMTs) on its wall, 
while the entire detector volume, which includes the $\sim$2~m thick OD region, 
extends 2.75~m radially and 2.6~m above and below the ID. 
Serving primarily as a veto, the 
OD is equipped with 1,885 8-inch outward-facing PMTs attached on the back side of the ID wall.
The entire volume is filled with 50~kton of ultra-pure water.
%while for the present analysis only the region 
%more than 200~cm inside from the ID wall is used. 
%This region is referred to as the fiducial volume (FV) and contains 22.5~kton water. 
% 
%The SK operation periods are separated into five as of June of 2019 
%since its beginning of data taking in 1996.  
In the present work, data from the fourth stage of the detector, known as SK-IV, are used. 
Further descriptions of SK can be found in Ref.~\cite{bib:superk}.

%------------------------------------------------------------------------------
%  Event Simulation 
   \section{Event Simulation}
   \label{sec:montecarlo}
%..............................................................................

Simulation of the signal and background processes are essential to 
the optimization of the event selection and determination of systematic uncertainties 
in this analysis.
Monte Carlo (MC) events generated according to models of 
neutrino beam, neutrino interactions, and the detector response including
the $\gamma$-ray emission are considered.

%%%%%
\subsection{Neutrino flux}

\begin{figure}[htbp]
  % 1st figure
   \begin{center}
    \includegraphics[clip,width=7.2cm]{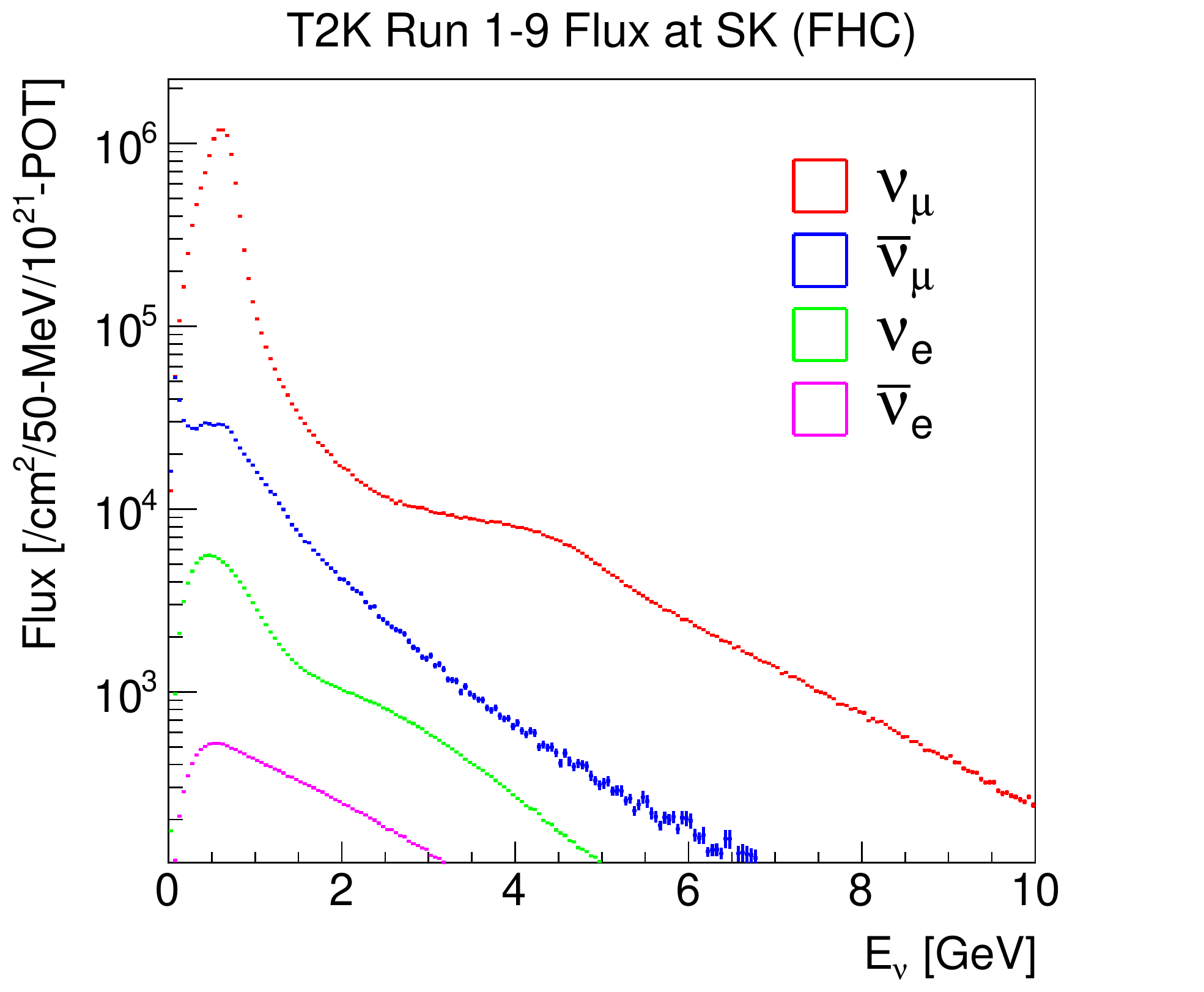}
   \end{center}
  % 2nd figure
   \begin{center}
    \includegraphics[clip,width=7.2cm]{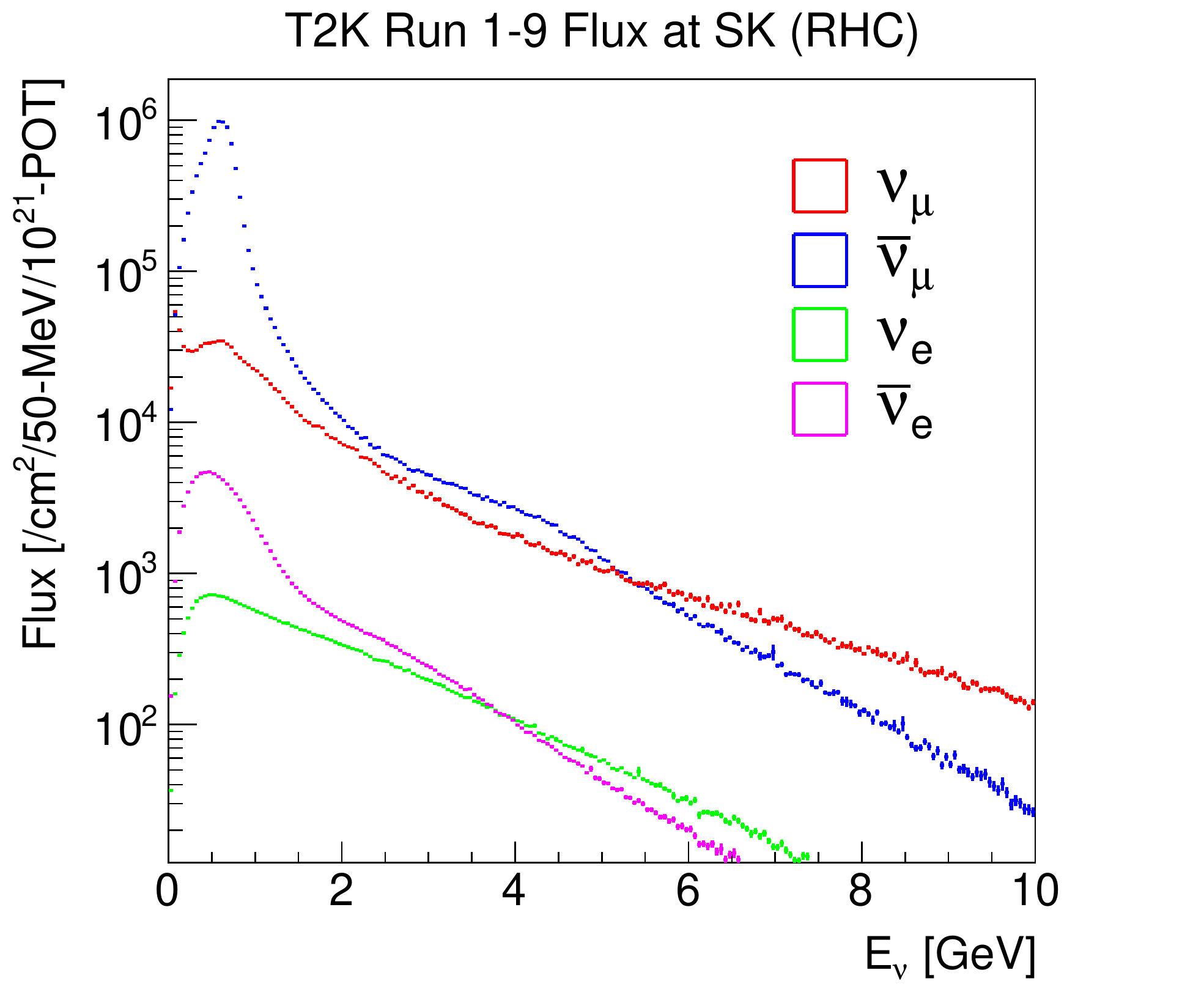}
   \end{center}
  \vspace{-5truept}
  \caption{T2K neutrino flux predictions at SK for the FHC (top) and RHC (bottom) 
           operation modes without neutrino oscillations.} 
  \label{fig:t2kflux}
  \end{figure}

The neutrino flux is estimated by simulation based on 
FLUKA2011~\cite{bib:fluka2011} and GEANT3~\cite{bib:geant3} 
for modeling hadronic interactions and particle transport and decays in the beamline. 
Pion and kaon production cross sections are renormalized using data from 
the NA61/SHINE experiment taken using both thin and T2K replica targets 
\cite{bib:t2kflux,bib:na61shinethin1,bib:na61shinethin2,bib:na61shinethin3,bib:na61shinereplica}.  
Oscillations are taken into account for neutrinos that produce charged-current (CC) interactions at SK, 
%as a weight factor multiplied to the final CC sample for each neutrino energy 
%and flavor type.
using parameters from the recent T2K measurements~\cite{bib:t2k2017oa}.
Figure~\ref{fig:t2kflux} shows the predicted T2K fluxes in the FHC and RHC modes
without neutrino oscillations.  
% 
%It is found that the wrong sign component, antineutrinos in FHC and 
%neutrinos in RHC, is small in both modes.

  %%% Figure : T2K Flux

%%%%%
\subsection{Neutrino interaction}

NEUT (version~5.3.3)~\cite{bib:neut} 
is used to simulate neutrino-nucleon interactions and subsequent 
final state interactions inside the target nucleus. 
For NCQE interactions the nominal nucleon momentum distribution 
is based on the Benhar spectral function \cite{bib:benhar,bib:ankowski},
while for CC quasielastic (CCQE) interactions 
the relativistic Fermi gas model~\cite{bib:smithmoniz} is used. 
The axial-vector mass is  $M_{\rm A}^{\rm QE} = 1.21~{\rm GeV/c^2}$ and  
the Fermi momentum for oxygen is 225~MeV/c. 
CC 2p2h interactions are modeled with the calculation in Ref.~\cite{bib:nieves},
but their neutral counterpart is not implemented in NEUT 
%%% PRD 
since no model is available in the literature.
The simulation uses BBBA05 vector form factors \cite{bib:bbba05} and a dipole axial-vector form factor.
Single pion production is based on the model of Rein and Sehgal~\cite{bib:reinsehgal}.
The axial-vector mass in the resonance interaction is $M_{\rm A}^{\rm RES} = 0.95~{\rm GeV/c^2}$. 
Deep inelastic scattering is simulated using the GRV98 parton 
distribution~\cite{bib:grv98} with corrections by Bodek and Yang~\cite{bib:bodekyang}. 
The final state interactions of hadrons inside the nucleus are simulated 
with a cascade model as described in Refs.~\cite{bib:t2k2016oa,bib:neut}.
Further simulation details are given in Ref.~\cite{bib:t2k2016oa}.

%%%%%
\subsection{$\gamma$-ray emission and detector response}

\begin{figure}[htbp]
   \begin{center}
    \includegraphics[clip,width=9.0cm]{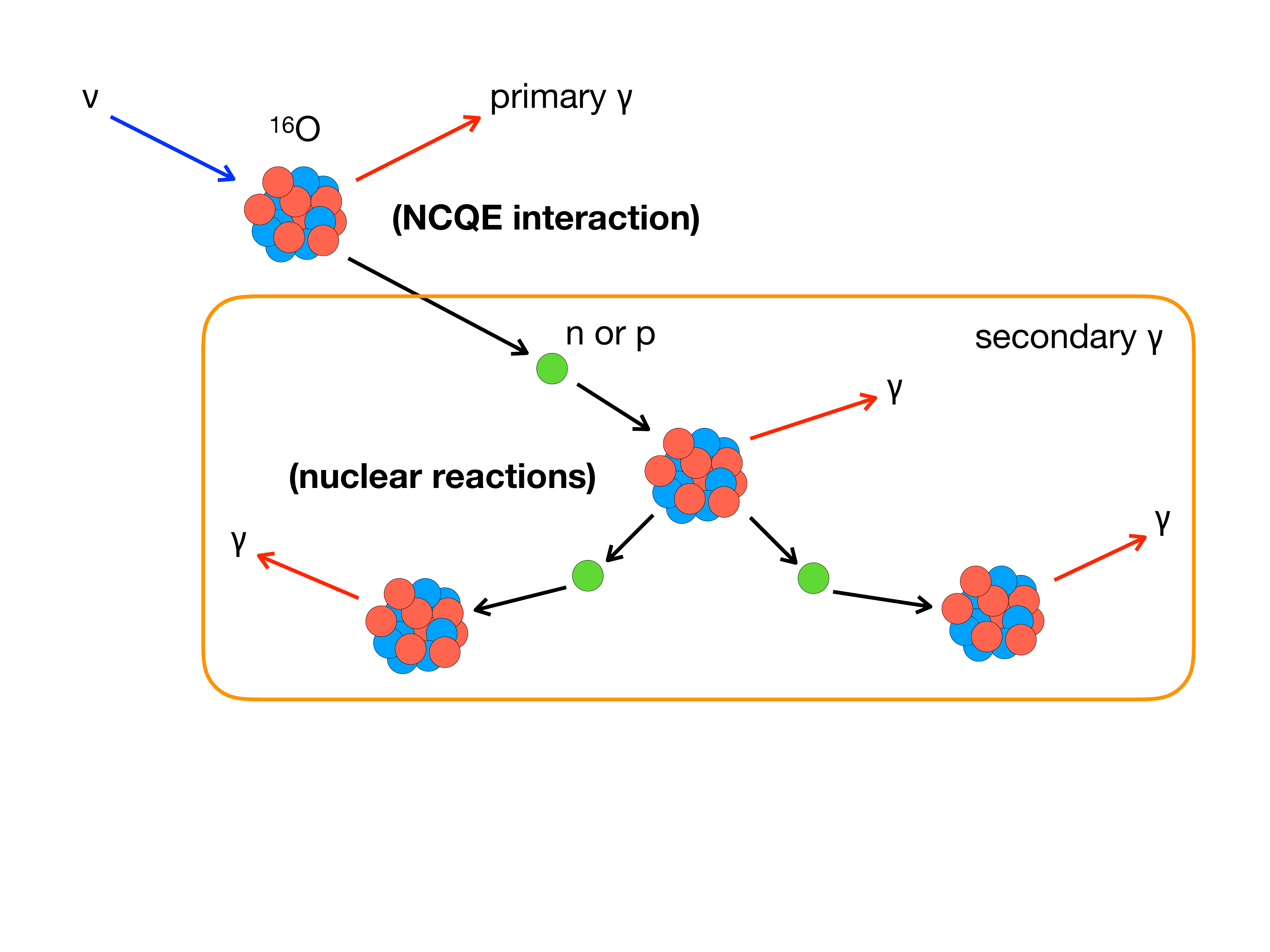}
   \end{center}
  \vspace{-55truept}
  \caption{Schematic of primary and secondary $\gamma$-rays in the NCQE interaction.}
  \label{fig:ncgammaschematic}
  \end{figure}

The emission of $\gamma$-rays from nuclear de-excitation is a key part of this analysis and is simulated 
separately for those produced by the neutrino-nucleus interactions (primary-$\gamma$) 
and those from nucleon-nucleus interactions (secondary-$\gamma$). 
These processes are schematically illustrated in Figure~\ref{fig:ncgammaschematic}.

  %%% Figure : NCgamma Schematic

%%% Primary-gamma
After the initial neutrino interaction an excited state of 
the remaining nucleus is selected based on the probabilities 
calculated in Ref.~\cite{bib:ankowski}. 
There are four possible states,
${\rm (p_{1/2})^{-1}}$, ${\rm (p_{3/2})^{-1}}$, ${\rm (s_{1/2})^{-1}}$, and {\it others}.
Here ${\rm (state)^{-1}}$ represents the state of the nucleus after 
a nucleon that initially occupied ${\rm states} = $ ${\rm p_{1/2}}$, ${\rm p_{3/2}}$, ${\rm s_{1/2}}$ is 
removed from the nucleus. 
The probability for each of four states to be produced is 0.158, 0.3515, 0.1055, and 0.385, 
respectively~\cite{bib:ankowski}. 
The ${\rm (p_{1/2})^{-1}}$ state is the ground state of ${\rm ^{15}O}$ or ${\rm ^{15}N}$
and therefore leads to no $\gamma$-ray emission.
Conversely, ${\rm (p_{3/2})^{-1}}$ almost always emits one $\gamma$-ray,
with 6.18~MeV from ${\rm ^{15}O}$ and 6.32~MeV from ${\rm ^{15}N}$ being the most likely.
Since ${\rm (s_{1/2})^{-1}}$ is a higher excited state,
the branching fraction to decays including nucleons or alpha particles may be large.
After such decays, the resulting nuclei may decay with $\gamma$-ray 
emission if it is still in an excited state thereafter.
The {\it others} state includes all other possibilities 
and mainly includes contributions from short-range correlations among nucleons. 
At present there is no data nor theoretical predictions of $\gamma$-ray emission for the states 
covered by {\it others} so in the nominal simulation 
they are integrated into ${\rm (s_{1/2})^{-1}}$. 
A systematic uncertainty stemming from this choice is described in Section~\ref{sec:uncertainty}. 
Further detailed descriptions on the treatment of these states are 
given in Ref.~\cite{bib:t2kncqe1to3}.

%%% Secondary-gamma
The interactions of secondary particles inside SK and the response of its PMTs are 
simulated with a GEANT3-based package~\cite{bib:geant3}. 
Hadronic interactions are of particular importance to the present analysis, 
especially models of neutron-nucleus reactions and the resulting $\gamma$-ray emission.
These are handled by GCALOR~\cite{bib:gcalor1,bib:gcalor2}, which 
implements the MICAP model for neutrons below 20 MeV and NMTC above 20 MeV.
%% (see Refs.~\cite{bib:gcalor1,bib:gcalor2} and references therein for details).
The MICAP model uses experimental cross sections from the ENDF/B-V library~\cite{bib:endf5},
while NMTC is based on an intra-nuclear cascade model.

%------------------------------------------------------------------------------
%  Analysis
   \section{Reconstruction and Selection}
   \label{sec:reconsel}
%..............................................................................

%%% Reconstruction 
Each event in SK is reconstructed with tools used for 
solar neutrino analysis~\cite{bib:bonsai,bib:sksolar3,bib:sksolar4}.
The visible energy ($E_{\rm rec}$) is reconstructed using the number of hit PMTs.
At these energies PMTs usually have registered only one photoelectron 
and there are typically between 10 and 200 hit PMTs in the current analysis window.
Note that the definition of energy in the present work differs from the previous 
T2K work \cite{bib:t2kncqe1to3}, where the electron mass (0.511~MeV) was added to 
the visible energy.
The current definition is consistent with recent low energy analyses in SK 
\cite{bib:sksolar4,bib:skncqe}.
The interaction vertex and direction are inferred from the PMT hit pattern and timing.
%% rvw 
A Cherenkov angle ($\theta_{\rm C}$) for each event is calculated 
as the most frequently occuring value in the distribution 
of opening angles to all three-hit combinations of PMTs. 
Various calibrations are used to evaluate the performance of the reconstruction 
as detailed in Refs.~\cite{bib:sklinac,bib:skdt}.

%%% Selection 
This analysis considers five event categories, 
neutrino NCQE interactions (``$\nu\mathchar`-$NCQE''), 
antineutrino NCQE interactions (``$\bar{\nu}\mathchar`-$NCQE''), 
all other NC interactions (``NC-other''), CC interactions, and accidental (beam-unrelated) backgrounds. 
Both the NC-other and CC categories include contributions from neutrinos and antineutrinos. 
Note that these event categories reflect the neutrino interaction prior to 
additional particle interactions within the nucleus.
This means that, for example, the NC-other sample contains pion production events 
where a pion was produced but was later absorbed in the nucleus.
The first four interactions are simulated using NEUT and beam-unrelated backgrounds are estimated using data 
outside of the T2K spill timing window. 
Event selection criteria are tuned to effectively select signal events, 
$\nu\mathchar`-$NCQE and $\bar{\nu}\mathchar`-$NCQE interactions, while removing 
other events as follows.

(1) 
Events are required to be in the energy range 3.49 to 29.49~MeV, 
above which CC interactions become dominant. 
Only data judged to be of good quality, based on the beam and 
detector conditions during each spill, are used~\cite{bib:t2k2016oa}.  
To select beam-induced events with high purity, the reconstructed event timing 
is required to be within $\pm 100$~ns of the expected timing of each bunch
%%% PRD
(``on-timing"). 
A sample of beam-unrelated events is selected by applying the same energy 
and quality cuts in a time window $[-500, -5]~\mu$s before the beam spill
%%% PRD
(``off-timing"). 
%An off-timing selection of $[-500, -5]~\mu$s
%with respect to the expected beam timing is applied instead while the other cuts are 
%identical when selecting the beam-unrelated events. 
% 
Events with hit clusters in a window spanning 20 to 0.2~$\mu$s before the event trigger
which are consistent with activity from electrons produced in the muon or pion decay chain (decay-$e$'s)
are removed.
The effect on the signal efficiency by this cut is negligible.
%In case the parent muon or pion, which originates from the neutrino interaction product 
%or the cosmic-ray muon, does not have large enough momentum to emit Cherenkov light, 
%while the electron or positron from their decay (called `decay-e') does, 
%such event may leave sign a few $\mu s$ later than the primary hits. 
% 
%If the triggered event is the decay-e, there should be some hits before. 
%Then such pre-activity is searched for 0.2 to 20~$\mu s$ before the event timing 
%and the event having more hits than background level is rejected.

\begin{figure}[htbp]
  % 1st figure
  \begin{center}
   \includegraphics[clip,width=6.0cm]{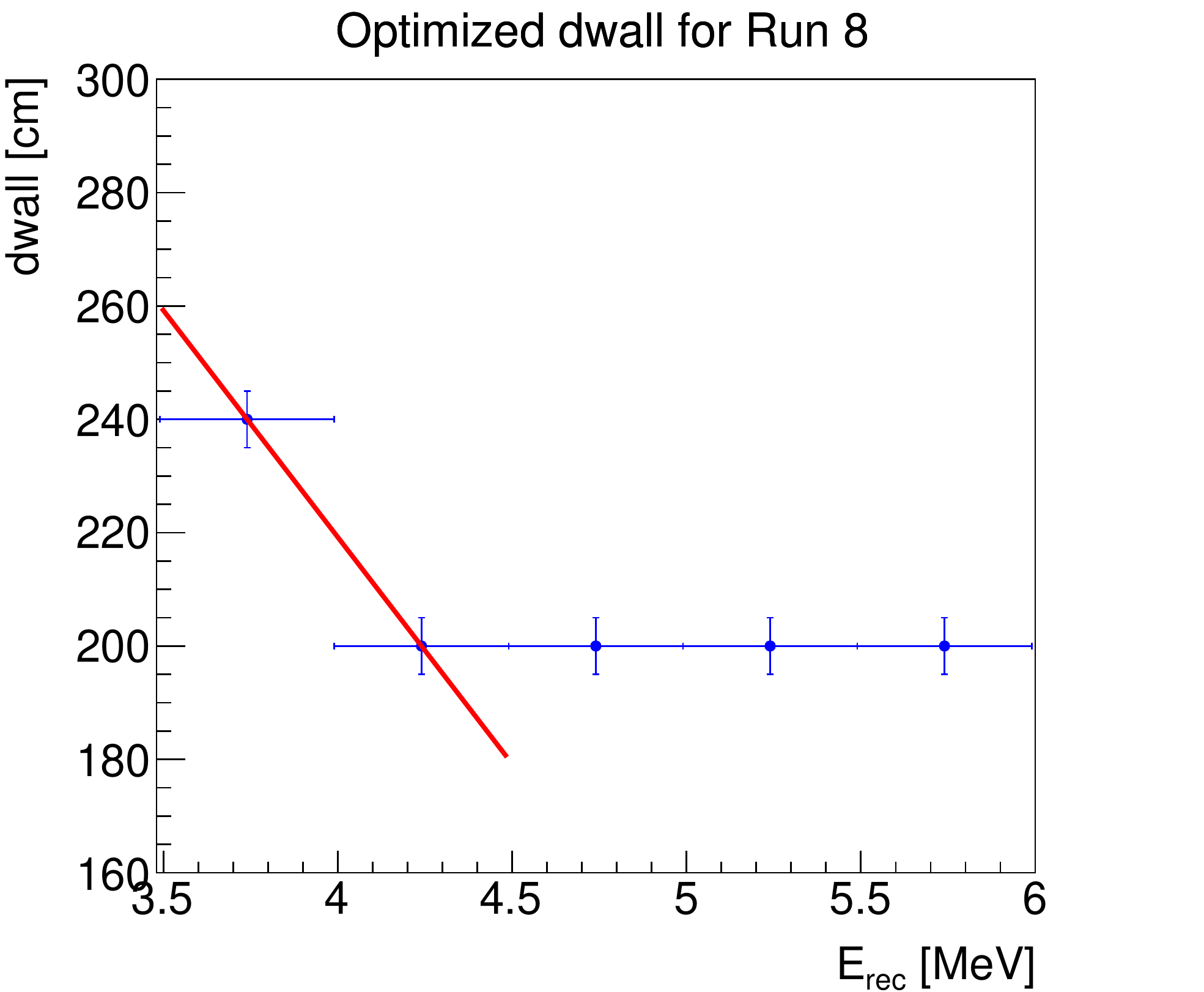}
  \end{center}
  % 2nd figure
  \begin{center}
   \includegraphics[clip,width=6.0cm]{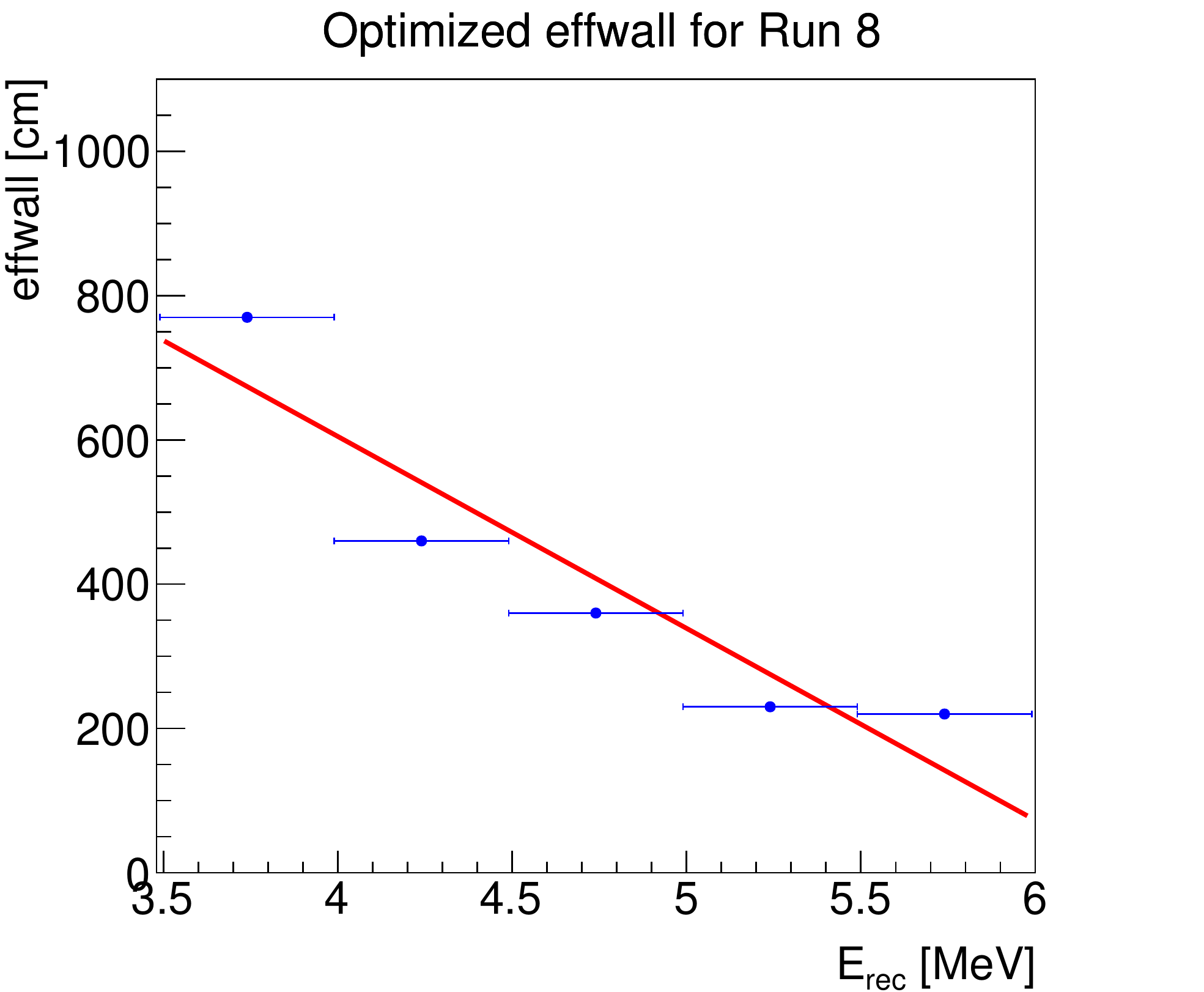}
  \end{center}
  % 3rd figure
  \begin{center}
   \includegraphics[clip,width=6.0cm]{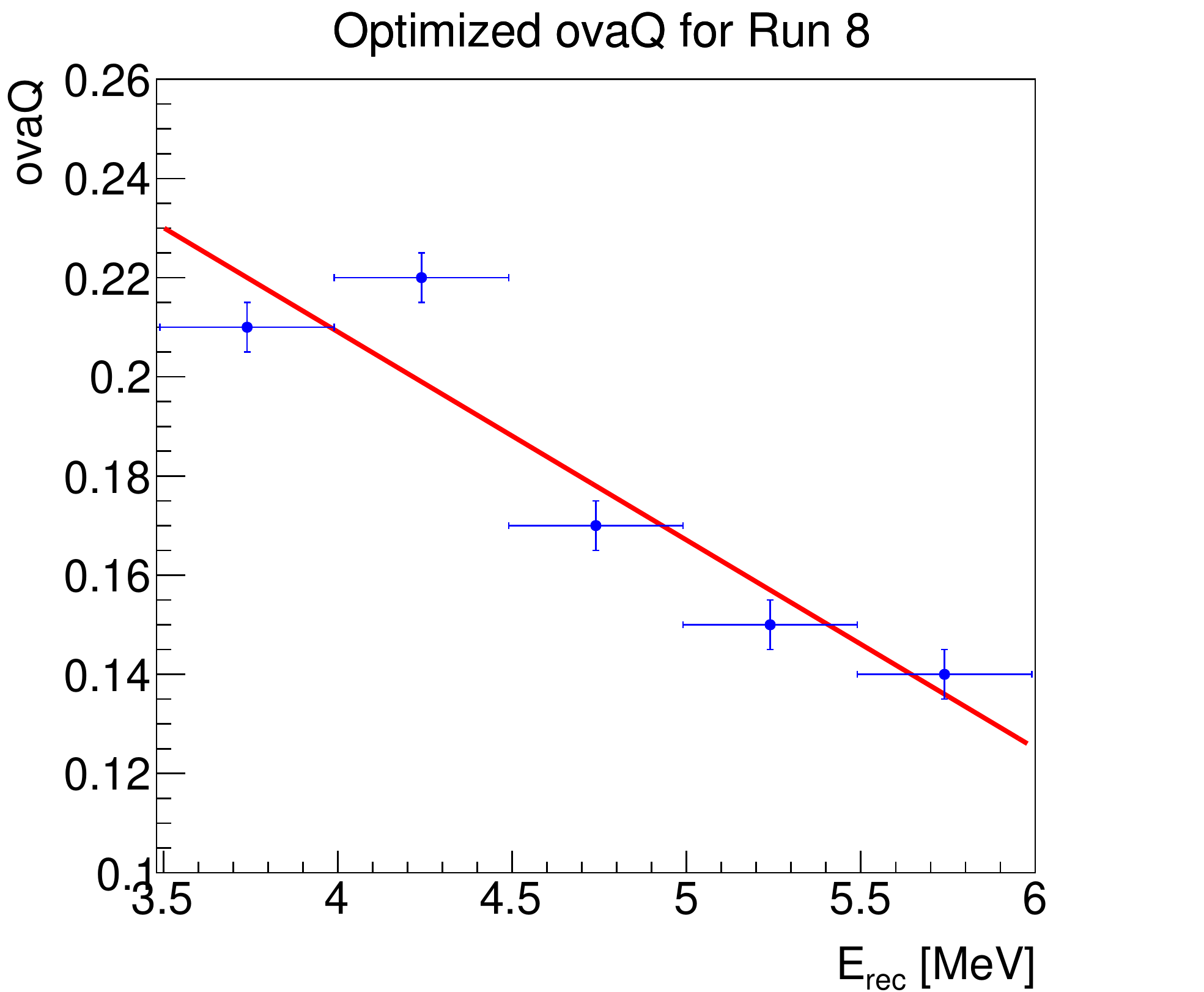}
  \end{center}
  \vspace{-5truept}
  \caption{Optimized cut values for $dwall$ (top), {\it effwall} (middle),
           and $ovaQ$ (bottom), in each low energy bin for one of the FHC mode runs (Run~8).
	   Vertical bars on each point represent the bin width used in parameter scans.
	   Red lines represent linear fits to the distributions and are used for the cut values.
           Events with parameter values above the lines are used in the analysis.
	   The fit regions for {\it dwall} and {\it effwall} are explained in the text.}
  \label{fig:cutoptexample}
  \end{figure}

\begin{figure}[htbp]
  % 1st figure
   \begin{center}
    \includegraphics[clip,width=7.2cm]{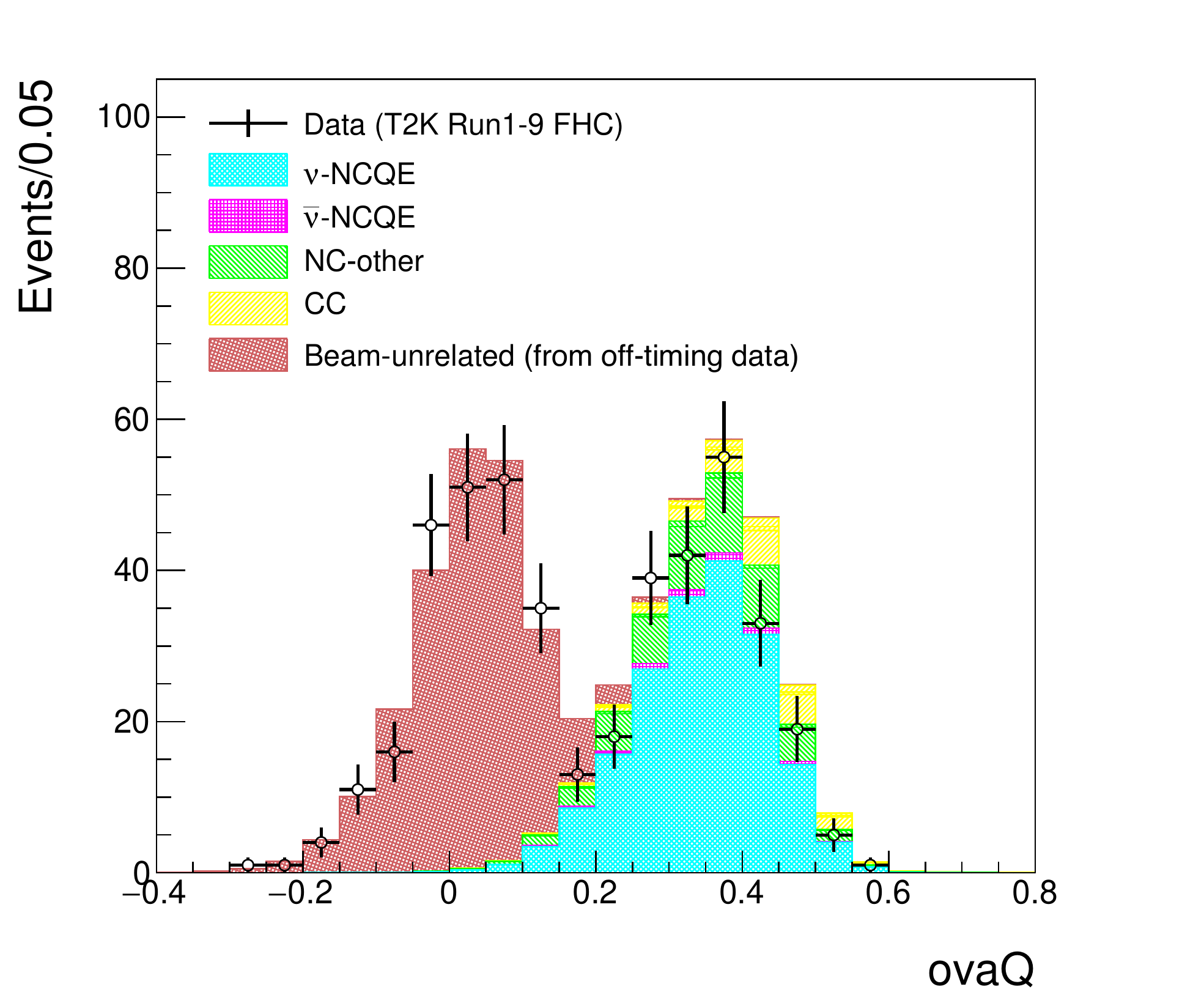}
   \end{center}
  \vspace{-15truept}
  % 2nd figure
   \begin{center}
    \includegraphics[clip,width=7.2cm]{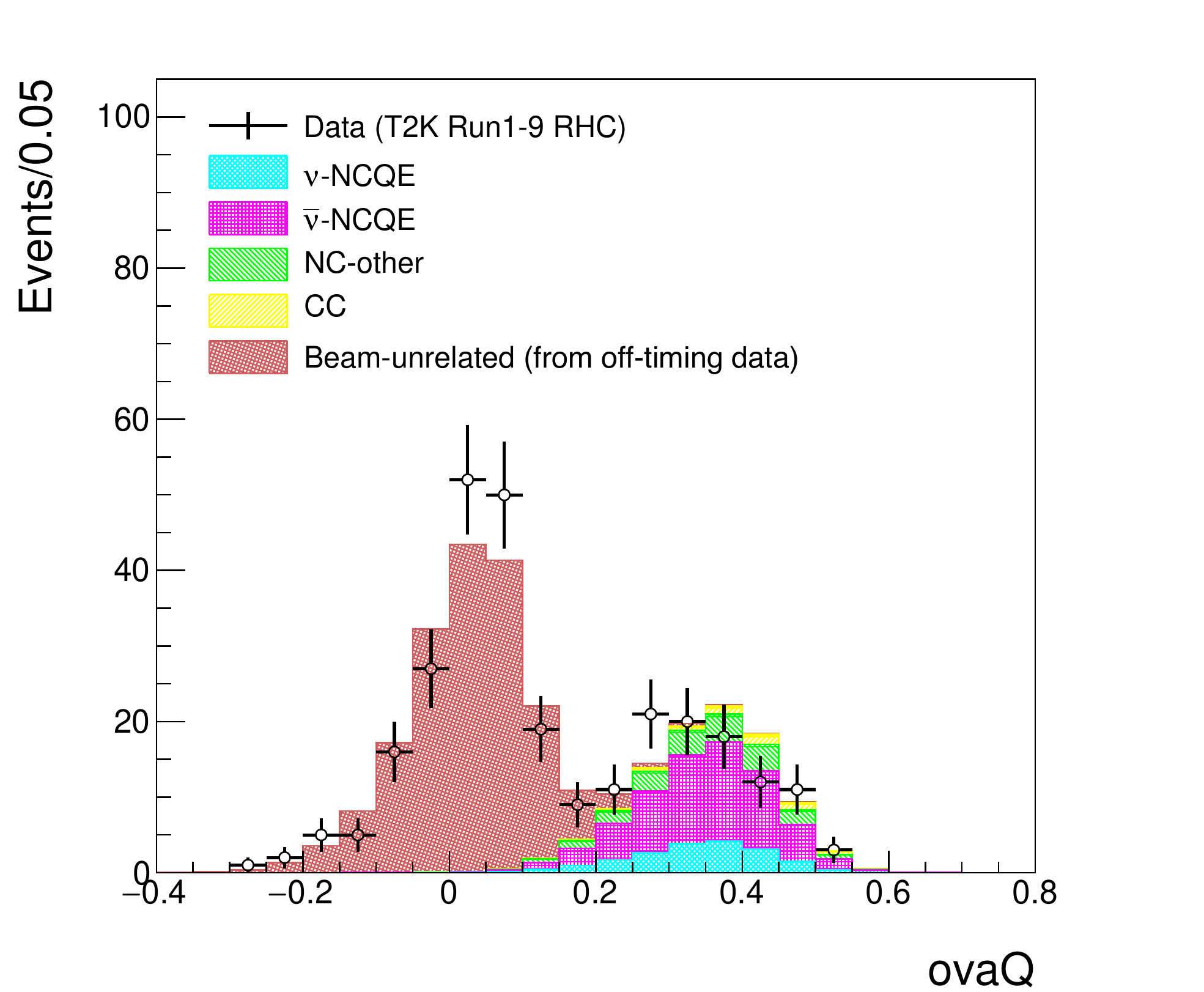}
   \end{center}
  \vspace{-5truept}
  \caption{Distributions of {\it ovaQ} for FHC (top) and RHC (bottom) 
           after the cuts in (1), the FV cut, the {\it dwall} cut, and the {\it effwall} cut. 
	   The MC prediction is broken down into four interactions: neutrino and 
	   antineutrino NCQE, NC-other, and CC. 
           Beam-unrelated events are obtained from the off-timing data 
	   as explained in the text.} 
  \label{fig:ovaqdist}
  \end{figure}

(2) 
Several additional event selection cuts are applied to remove backgrounds 
from radioactive impurities from the detector walls.
First, a fiducial volume (FV) cut is applied to all events, 
which requires the distance between the reconstructed vertex position and 
the ID wall ({\it dwall}) to be more than 200~cm.
% 
%%In order to further reduce backgrounds, several energy dependent cuts 
%%are applied subsequently, as explained below. 
% 
Below 6~MeV radioactive backgrounds increase considerably, 
requiring tighter {\it dwall} and reconstructed event quality cuts.
Cuts in this energy region are tuned (discussed below) using 
three variables, {\it dwall}, {\it effwall}, and {\it ovaQ}. 
Here {\it effwall} is the distance from the event vertex 
to the ID wall as measured backward along the reconstructed track direction.
The {\it ovaQ} parameter is a measure of the reconstruction quality 
and is defined as the difference of two parameters, 
$ovaQ = g_{\rm vtx}^2 - g_{\rm dir}^2$, where $g_{\rm vtx}$ and $g_{\rm dir}$ 
are the vertex and direction fit quality parameters, respectively~\cite{bib:sksolar2}. 
Cuts on these parameters are optimized for five regions between 3.49 and 5.99~MeV 
with each 0.5~MeV bin width.
The optimization is performed separately for each T2K run period 
because the detector condition and the beam power differ from run to run. 
A figure-of-merit (FOM) designed to maximize sensitivity to the NCQE 
signal is defined as:

  %%% Equation : Figure-of-Merit
  \begin{eqnarray}
  \label{eq:fomdefine}
   {\rm FOM} = \frac{N_{\rm sig}}{\sqrt{N_{\rm sig} + N_{\rm bkg}}}, 
   %(N_{\rm bkg} = N_{\rm bkg}^{\rm MC} + N_{\rm bkg}^{\rm beam\mathchar`-unrelated}), %\nonumber
  \end{eqnarray}
  \vspace{2truept}
  %%%

\noindent 
where $N_{\rm sig}$ is the number of signal events predicted by the MC ($\nu$-NCQE for FHC and 
$\bar{\nu}$-NCQE for RHC)
and $N_{\rm bkg}$ is the total number of background events.
The latter is composed of two components, $N_{\rm bkg}^{\rm MC}$ and $N_{\rm bkg}^{\rm beam\mathchar`-unrelated}$,
which represent non-signal neutrino events such as NC-other and CC interactions, 
and beam-unrelated events from the off-timing data sample, respectively.
Cuts on the three parameters above are chosen to maximize the FOM in each energy region.
As an illustration the optimized values of {\it dwall}, {\it effwall}, and {\it ovaQ} 
for one of the FHC mode runs (T2K Run~8) are shown in Figure~\ref{fig:cutoptexample}.
A linear function is fit to each distribution 
to obtain the final cut criteria and is denoted by the red line in the figure.
For the {\it dwall} and {\it effwall} distributions, if the optimized value is
200~cm (the FV cut criterion) in two successive energy bins, 
the second and later bins are removed and the fit is repeated. 
In the end, each of these three parameters is required to be larger than the obtained line. 
That is, events with values in the upper right portion of the plots in the figure are kept. 
Note that at higher energies the optimum {\it dwall} and {\it effwall}
values fall below 200~cm, but such events are already removed by the initial FV cut.
Figure~\ref{fig:ovaqdist} shows the {\it ovaQ} distributions after 
the cuts described in (1), the FV cut, the optimized {\it dwall} cut, and the optmized {\it effwall} cut.
There is clear separation between signal and background.
Further description of the variables used in this selection are given in Refs.~\cite{bib:sksolar2,bib:t2kncqe1to3}.

  %%% Figure : Cut parameter for Run8
  %%%

  %%% Figure : ovaQ

\begin{figure*}[htbp]
  % 1st figure
  \begin{minipage}{0.23\hsize}
   \begin{center}
    \includegraphics[clip,width=4.2cm]{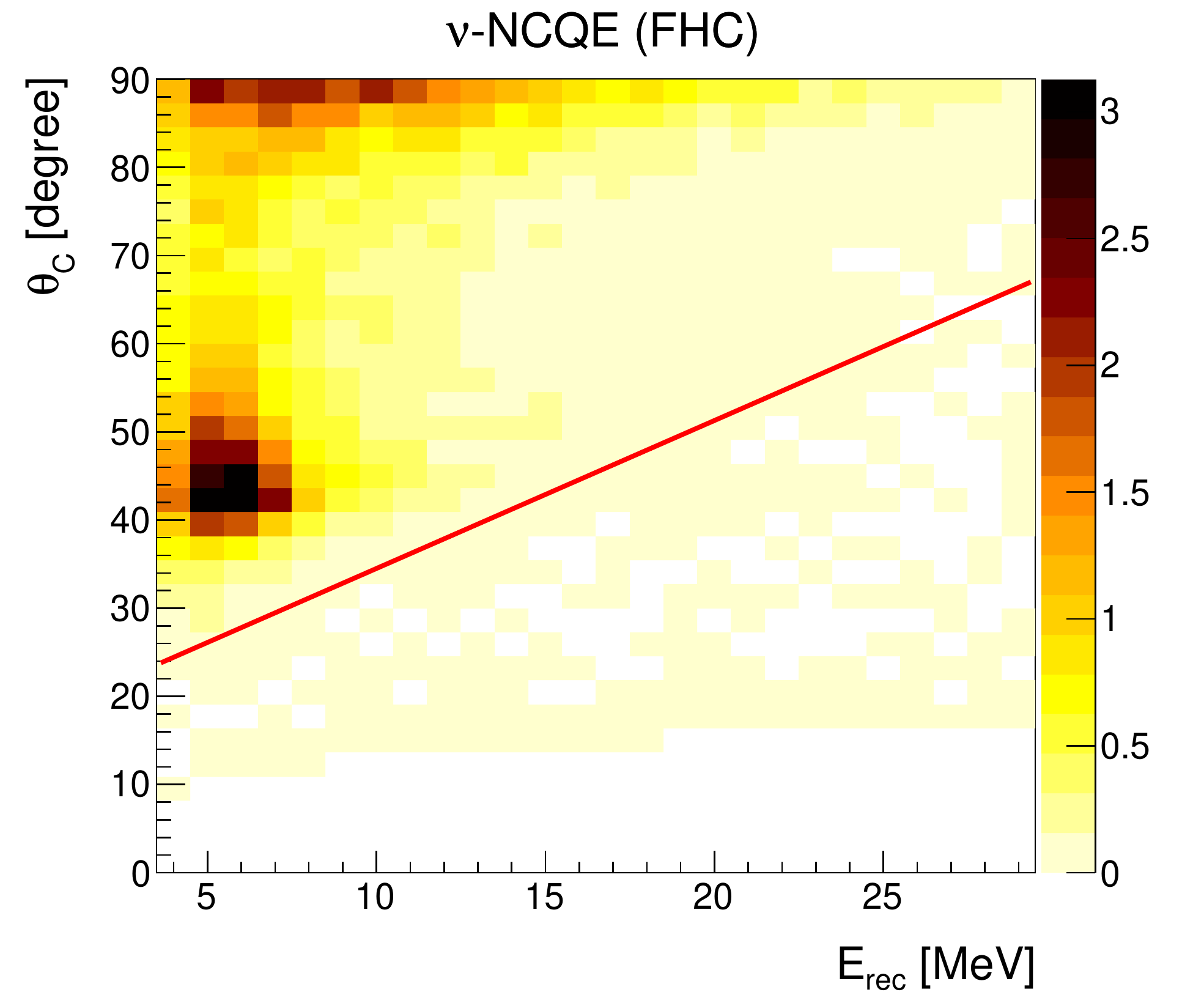}
   \end{center}
  \end{minipage}
  % 2nd figure
  \begin{minipage}{0.23\hsize}
   \begin{center}
    \includegraphics[clip,width=4.2cm]{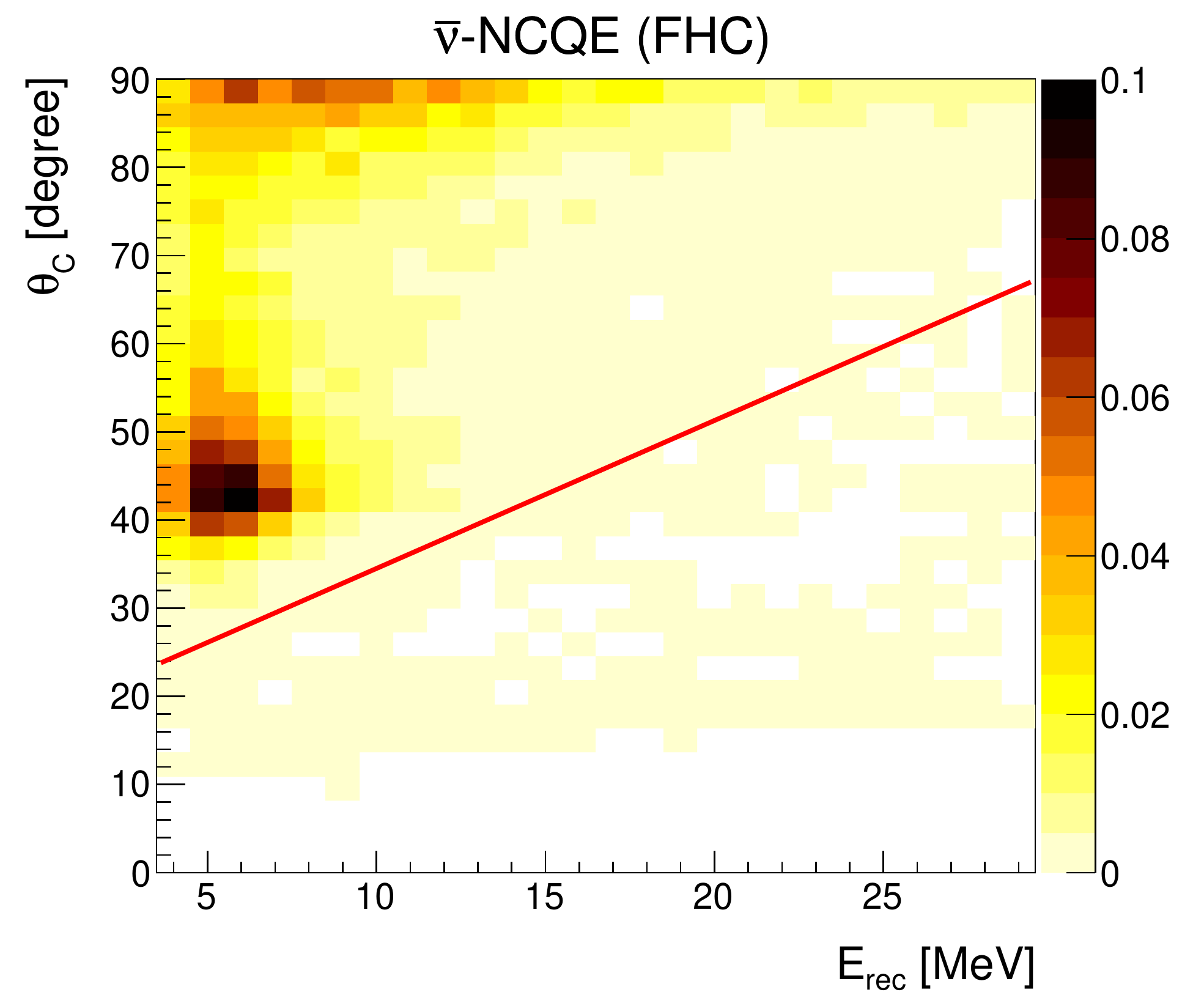}
   \end{center}
  \end{minipage}
  % 3rd figure
  \begin{minipage}{0.23\hsize}
   \begin{center}
    \includegraphics[clip,width=4.2cm]{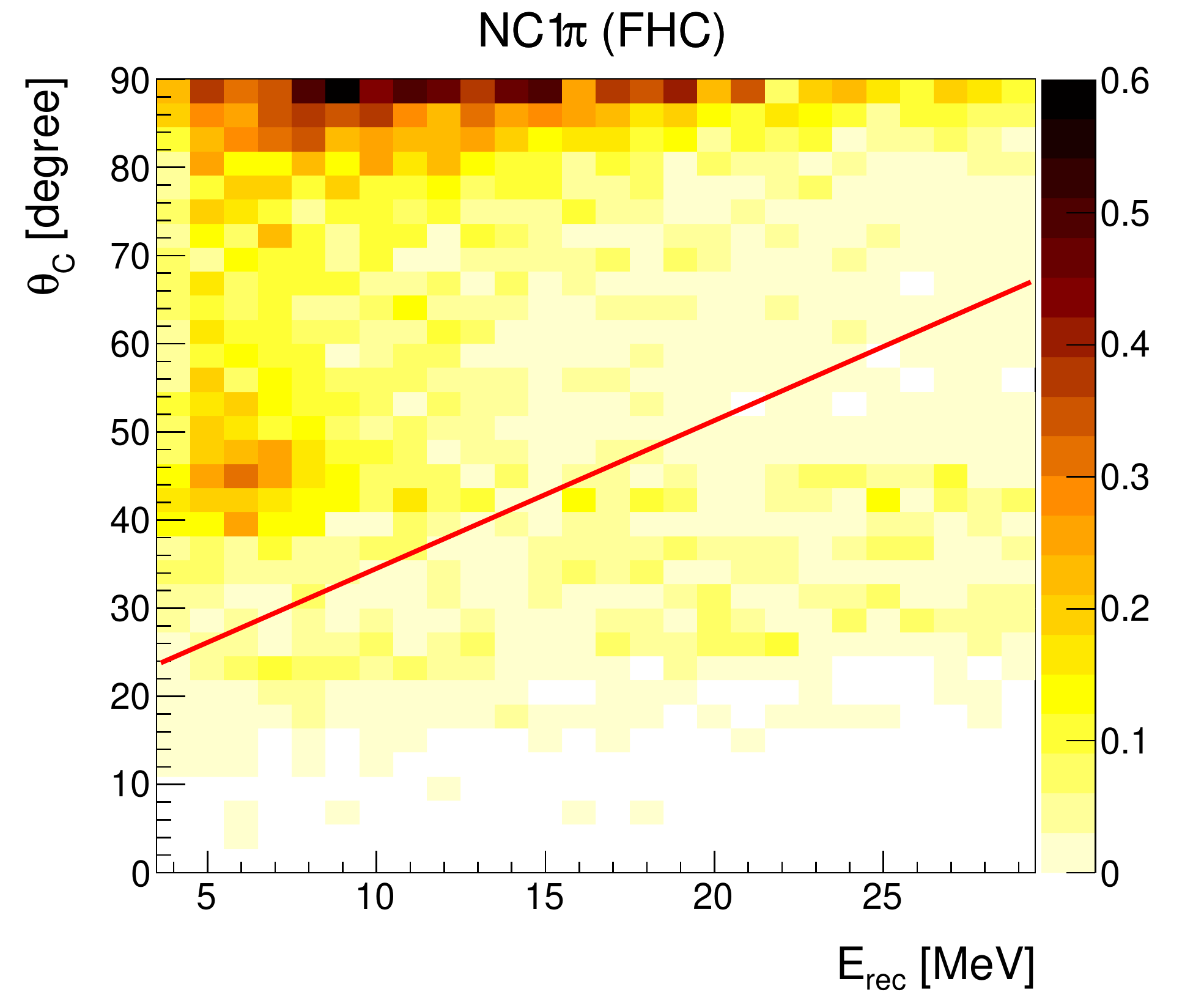}
   \end{center}
  \end{minipage}
  % 4th figure
  \begin{minipage}{0.23\hsize}
   \begin{center}
    \includegraphics[clip,width=4.2cm]{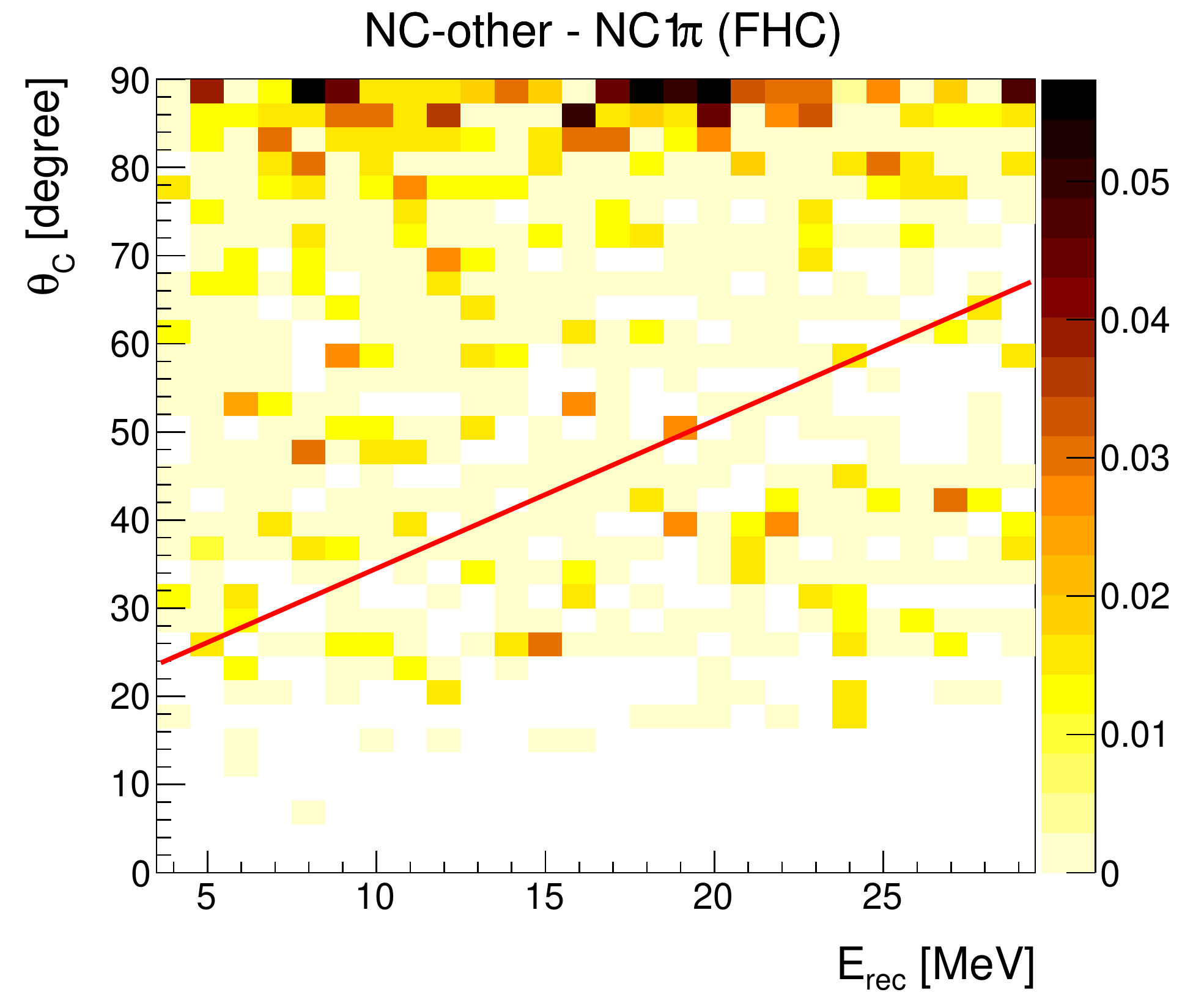}
   \end{center}
  \end{minipage}
  % 5th figure
  \begin{minipage}{0.23\hsize}
   \begin{center}
    \includegraphics[clip,width=4.2cm]{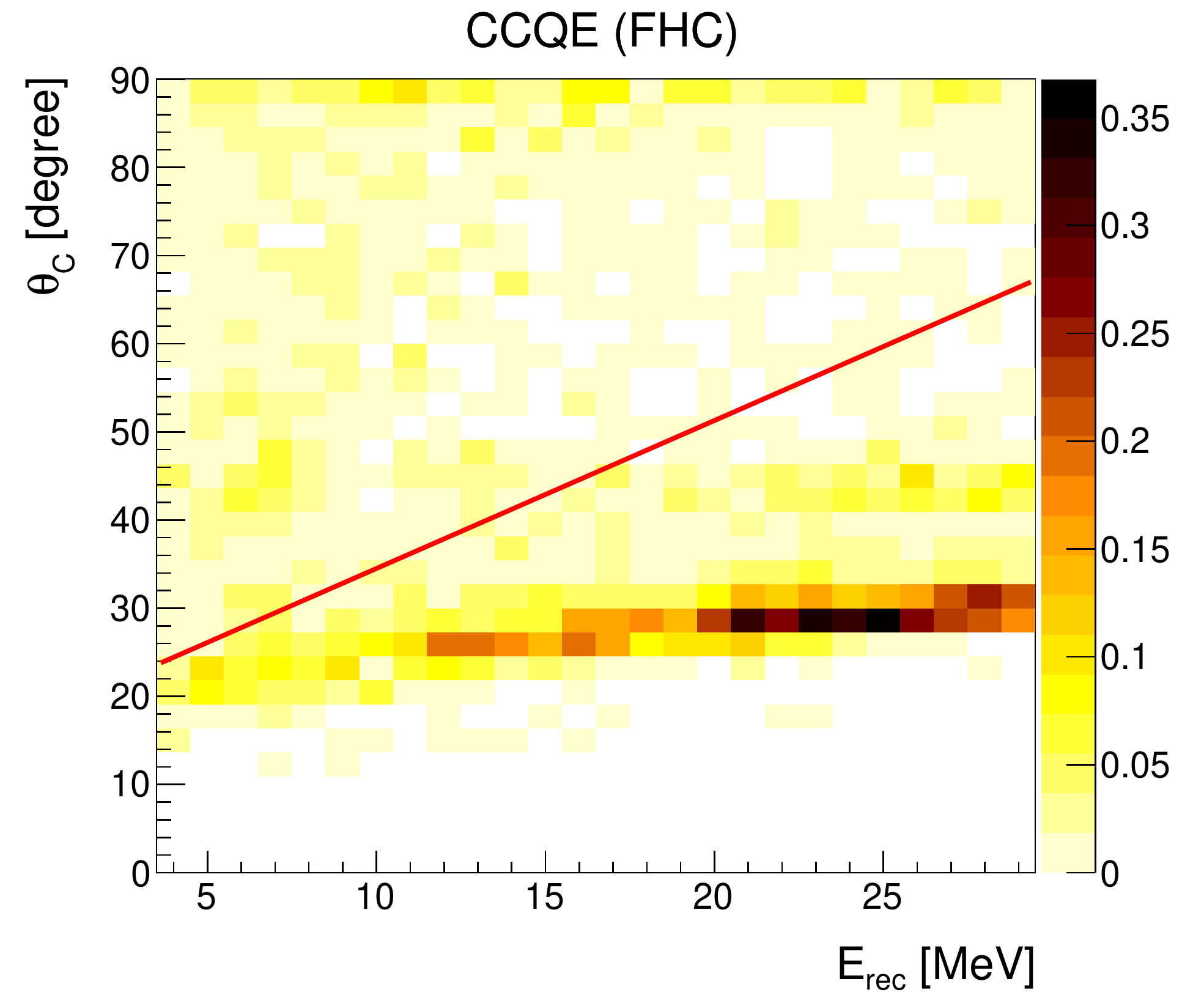}
   \end{center}
  \end{minipage}
  % 6th figure
  \begin{minipage}{0.23\hsize}
   \begin{center}
    \includegraphics[clip,width=4.2cm]{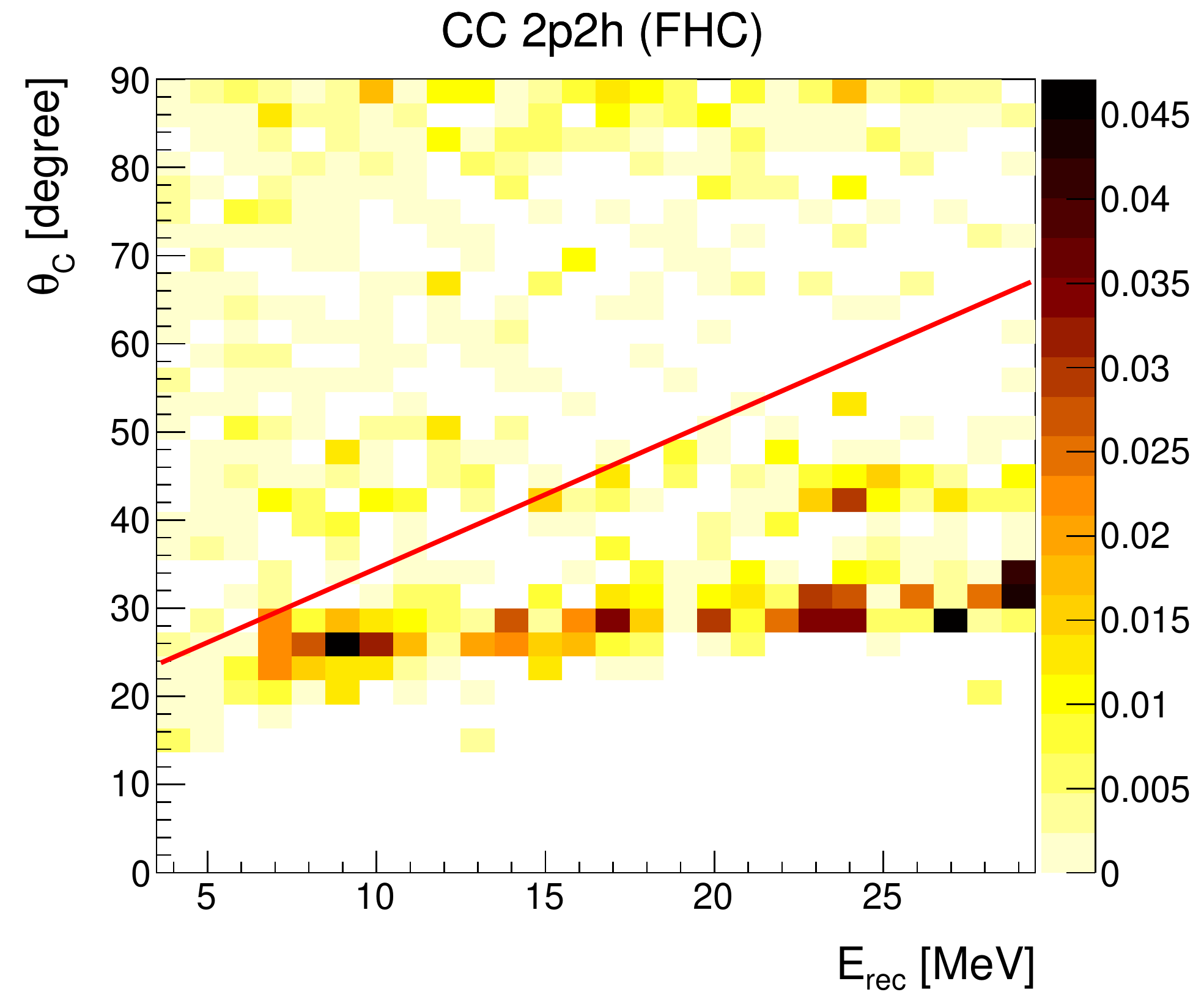}
   \end{center}
  \end{minipage}
  % 7th figure
  \begin{minipage}{0.23\hsize}
   \begin{center}
    \includegraphics[clip,width=4.2cm]{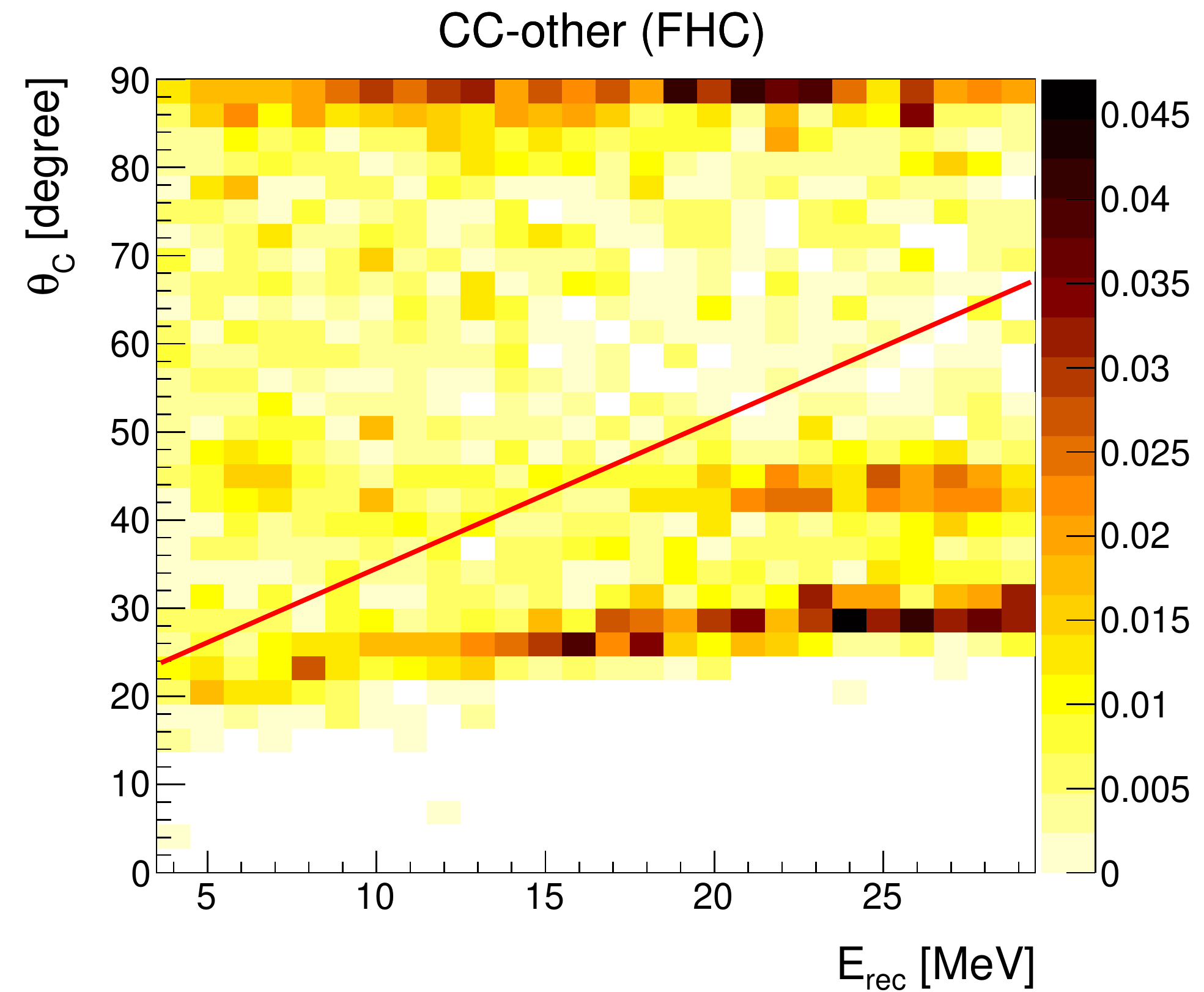}
   \end{center}
  \end{minipage}
  % 8th figure
  \begin{minipage}{0.23\hsize}
   \begin{center}
    \includegraphics[clip,width=4.2cm]{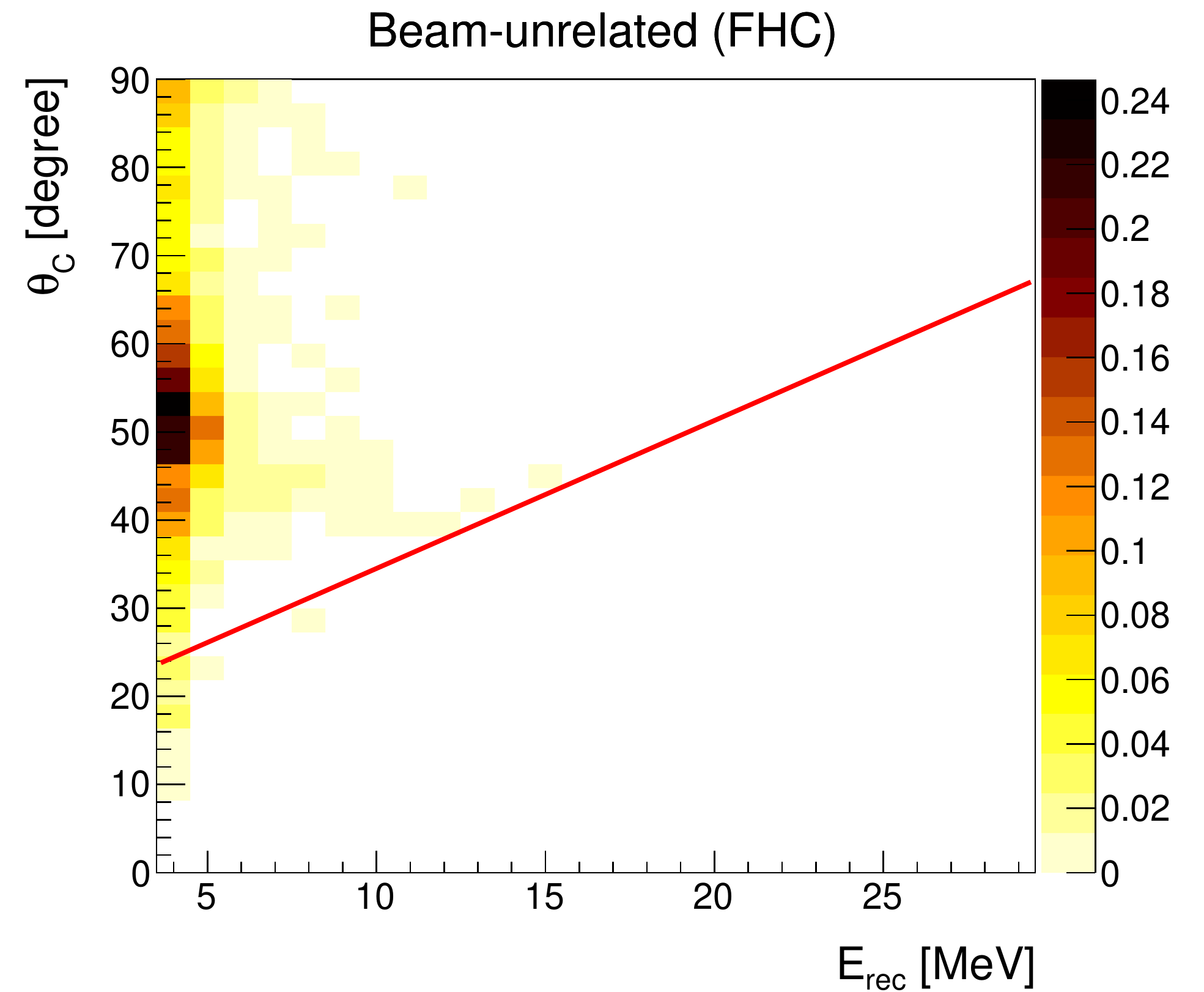}
   \end{center}
  \end{minipage}
  \vspace{+10truept}
  \caption{Two-dimensional $E_{\rm rec}$$-$$\theta_{\rm C}$ distributions 
           of each neutrino interaction channel by MC and beam-unrelated events by the off-timing data 
	   in FHC mode; 
	   the optimized linear function for the CC interaction cut is shown in red. 
	   Events above the line are used in the analysis.
	   The z-axis represents the predicted number of events [/MeV/2.7-degree] in the T2K Run~1$-$9 FHC mode.
	   NC1$\pi$ represents neutrino and antineutrino neutral-current interactions with a pion production,  
	   and CC-other represents all other CC interactions than CCQE and CC 2p2h.}
	   %The NC-other represents NC interactions other than NCQE 
	   %and ${\rm NC1\pi}$, whereas elsewhere in this paper NC-other includes NC$1\pi$.}
  \label{fig:ccintcut}
  \end{figure*}

(3) 
The final phase of the event selection is focused on the removal 
of CC interaction events.
A single charged particle whose momentum is large compared to its mass 
is likely to have a Cherenkov angle of $\sim$$42^\circ$ in water. 
On the other hand if the particle momentum is lower, the reconstructed 
Cherenkov angle decreases. 
In this analysis low energy muons from CC interactions and still above 
Cherenkov threshold distribute around $\theta_{\rm C} = 20^\circ$$-$$35^\circ$,
whereas decay-$e$'s have $\theta_{\rm C} \sim 42 ^\circ$.
The contribution of each can be seen in Figure~\ref{fig:ccintcut}. 
To reduce these CC events, a linear cut in 
the reconstructed energy and Cherenkov angle plane is chosen 
by maximizing the FOM defined in Eq.~(\ref{eq:fomdefine}). 
In the figure the resulting cut is shown with a red line. 
This is performed separately for the FHC and RHC samples. 
Using the optimized cut the signal efficiency is 99\% (99\%) 
while 63\% (58\%) of CC events are removed in FHC (RHC) mode.
%It is found that NCQE events are not reduced much while CC interactions are 
%effectively removed. 
%%% PRD
Some CC-other events still remain after this cut, 
which could be due to, for example, multiple-$\gamma$ emission via neutron production (as explained later), 
but this fraction is small with respect to the total number of selected events.
Similar population is seen also in the NC-other distribution. 

  %%% Figure : CC interaction cut
\begin{figure}[htbp]
  % 1st figure
   \begin{center}
    \includegraphics[clip,width=7.44cm]{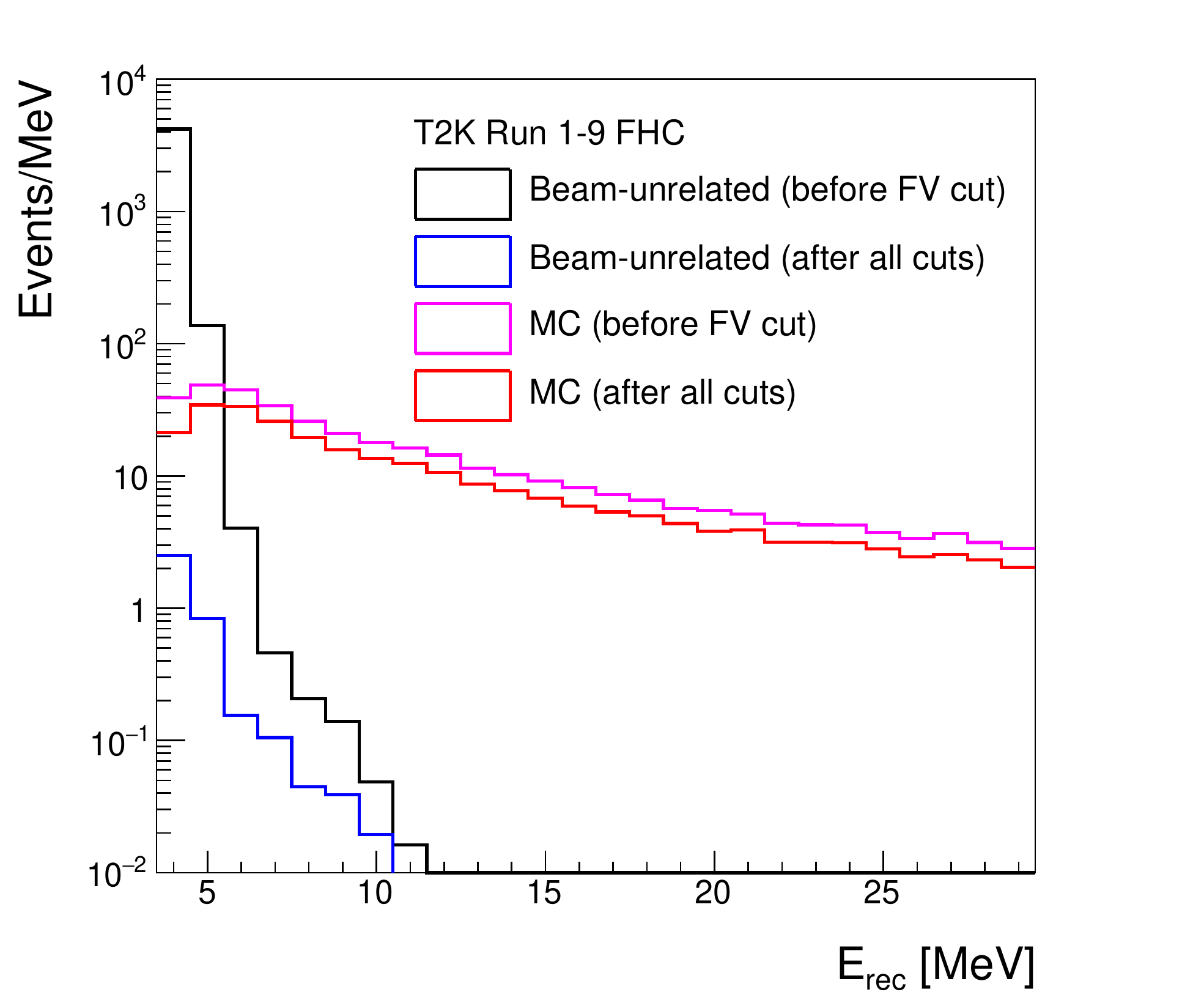}
   \end{center}
  \vspace{-15truept}
  % 2nd figure
   \begin{center}
    \includegraphics[clip,width=7.44cm]{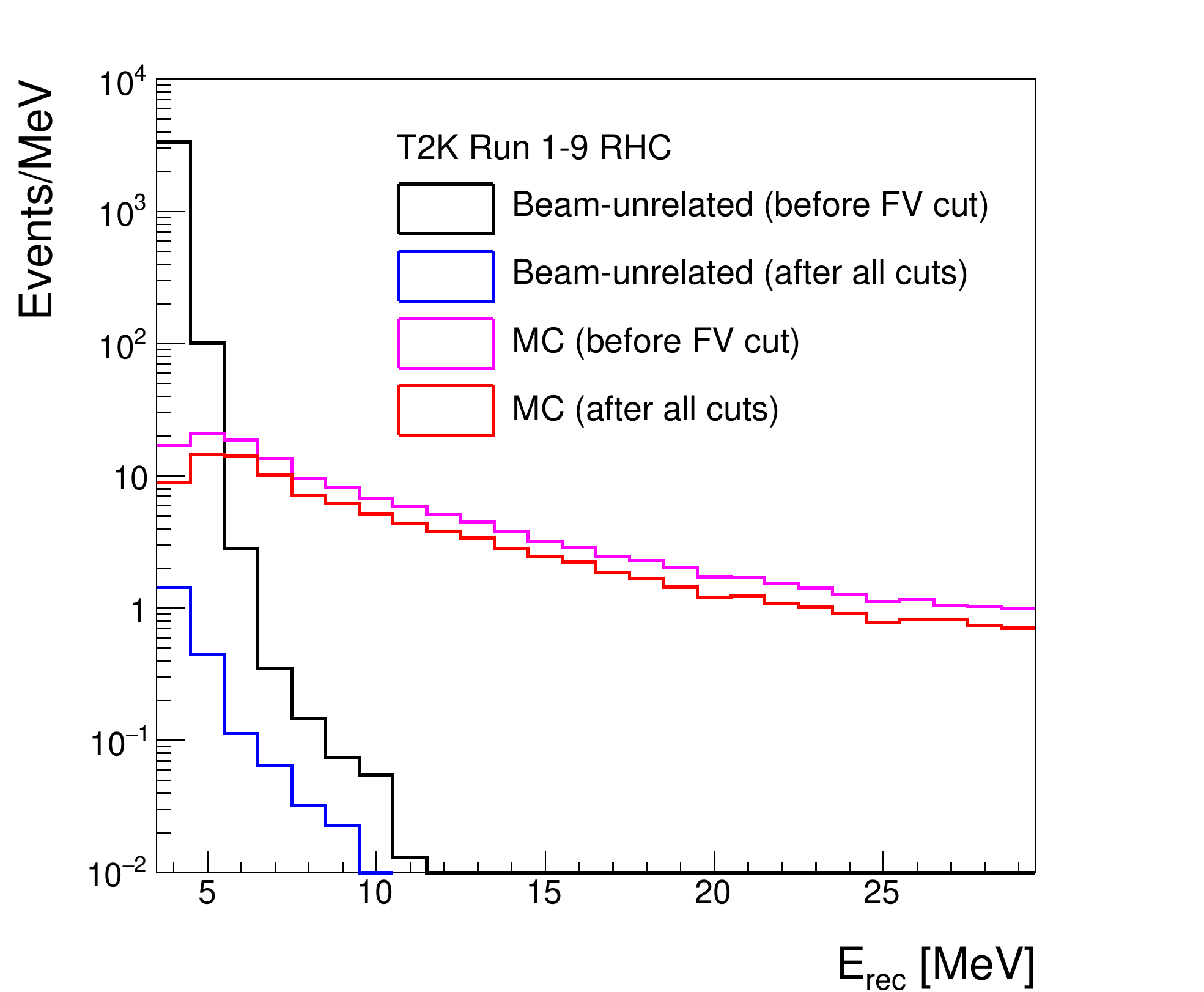}
   \end{center}
  \vspace{-5truept}
  \caption{Reconstructed energy distributions of MC and beam-unrelated events 
           before the FV cut and after all cuts for FHC (top) and RHC (bottom).}
  \label{fig:fhc_nmcandoffbeam}
  \end{figure}

\begin{figure*}[htbp]
  % 1st figure
  \begin{minipage}{0.325\hsize}
   \begin{center}
    \includegraphics[clip,width=6.6cm]{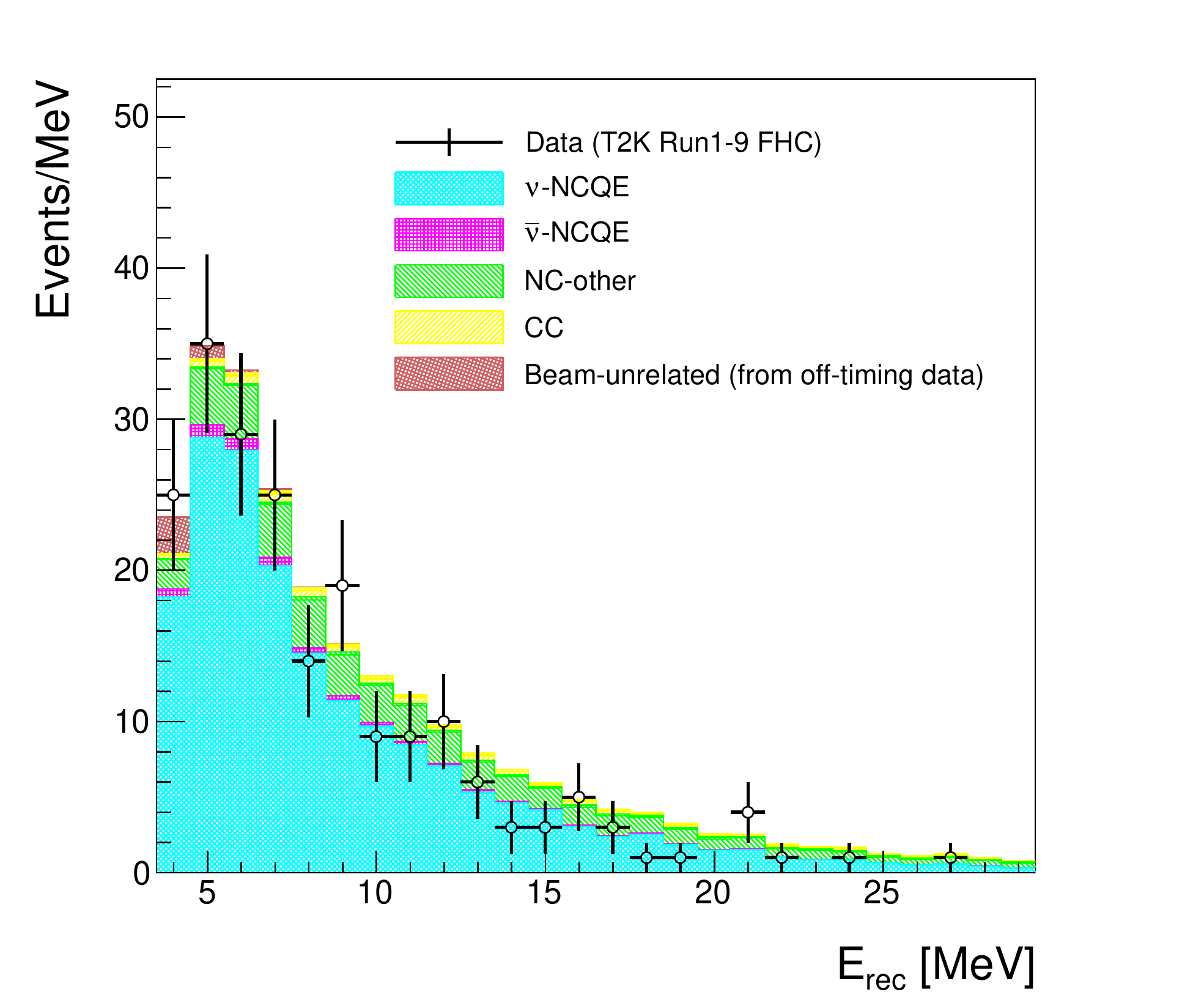}
   \end{center}
  \end{minipage}
  % 2nd figure
  \begin{minipage}{0.325\hsize}
   \begin{center}
    \includegraphics[clip,width=6.6cm]{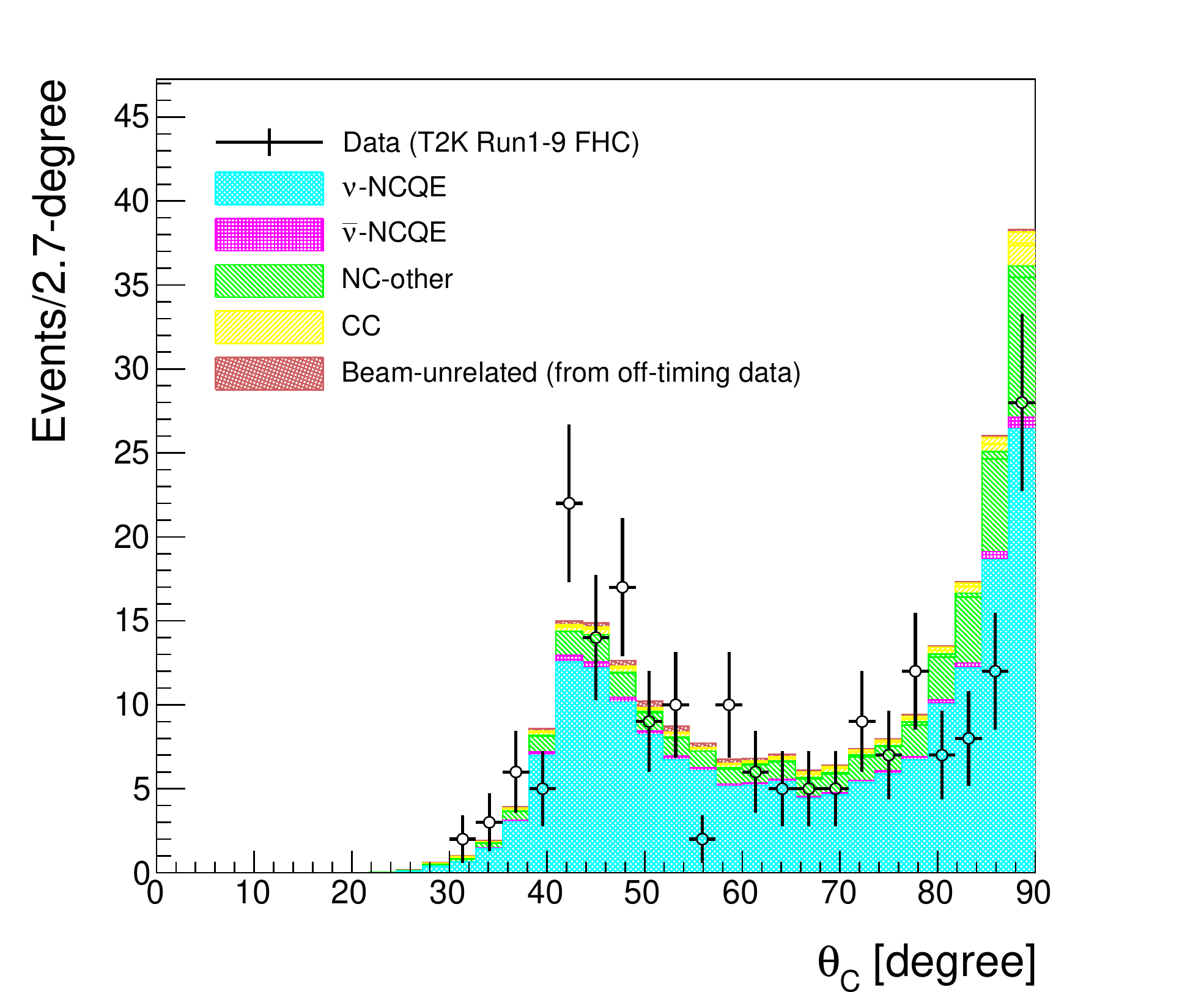}
   \end{center}
  \end{minipage}
  % 3rd figure
  \begin{minipage}{0.325\hsize}
   \begin{center}
    \includegraphics[clip,width=6.6cm]{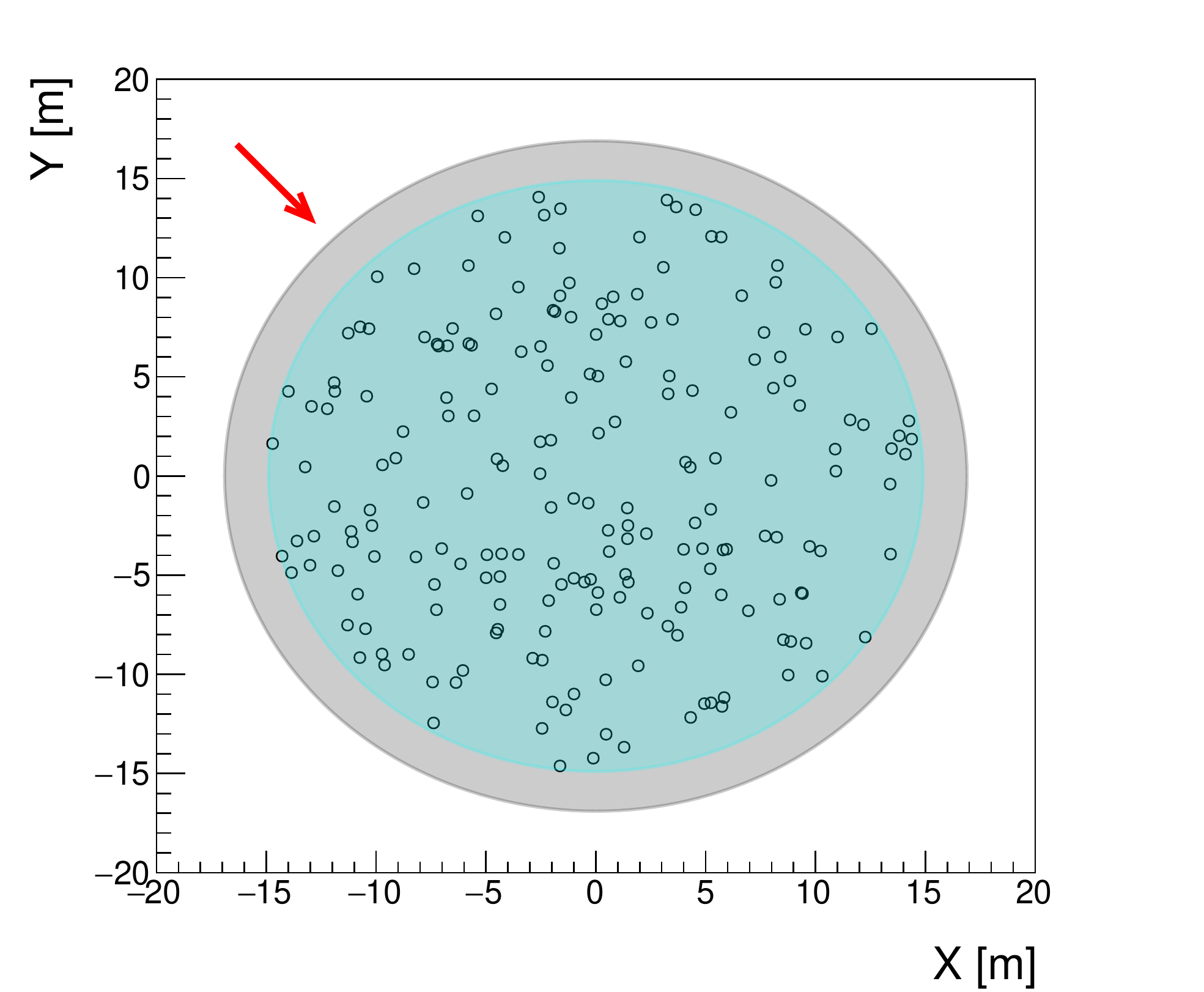}
   \end{center}
  \end{minipage}
  \caption{Distributions of $E_{\rm rec}$ (left), $\theta_{\rm C}$ (middle), and 
           vertex (right) from the FHC sample. 
	   In the right panel, the red arrow indicates the beam direction and 
	   the gray and sky blue regions correspond to the ID and FV, respectively.}
  \label{fig:fhcselected}
  \end{figure*}

\begin{figure*}[htbp]
  % 1st figure
  \begin{minipage}{0.325\hsize}
   \begin{center}
    \includegraphics[clip,width=6.6cm]{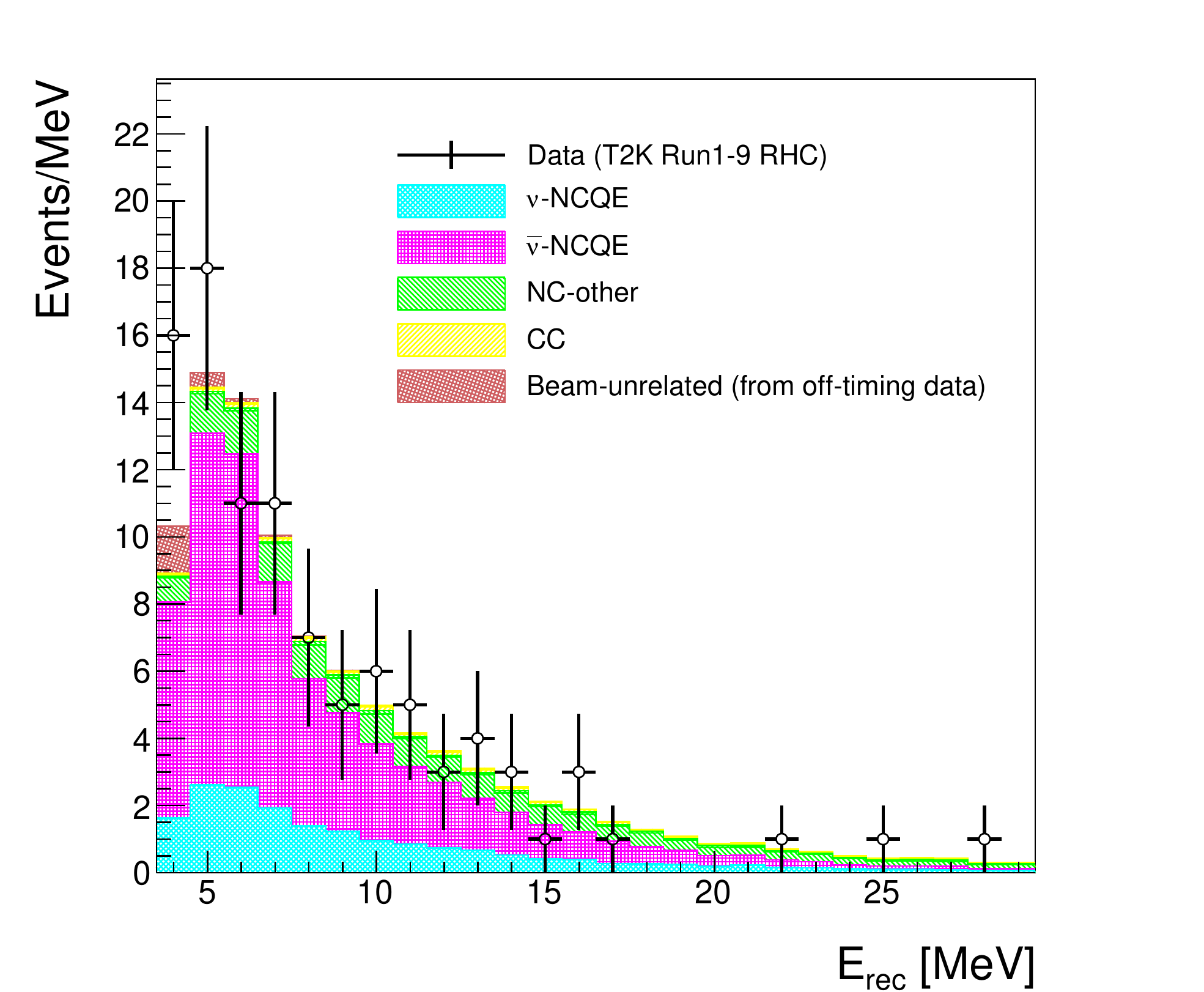}
   \end{center}
  \end{minipage}
  % 2nd figure
  \begin{minipage}{0.325\hsize}
   \begin{center}
    \includegraphics[clip,width=6.6cm]{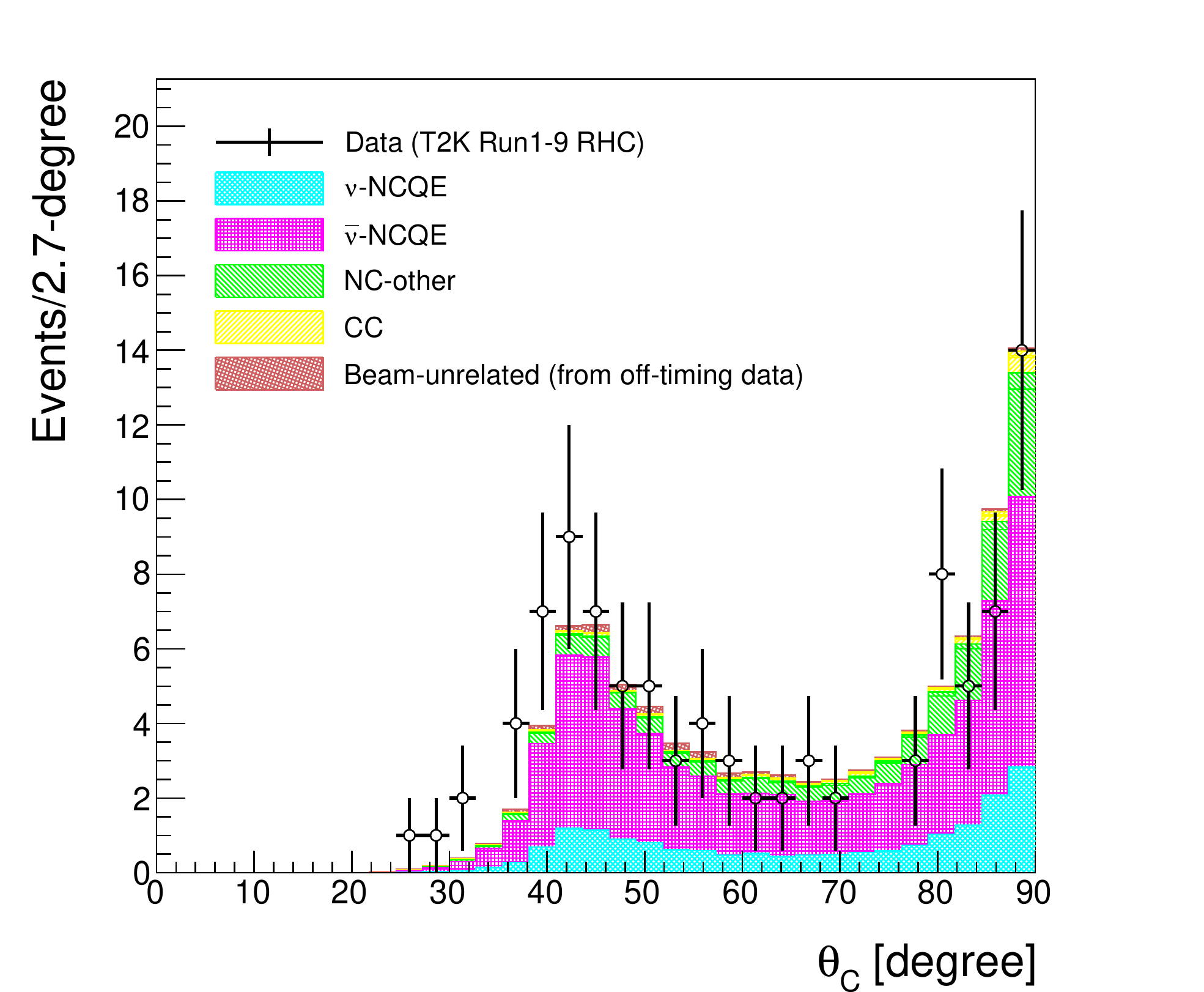}
   \end{center}
  \end{minipage}
  % 3rd figure
  \begin{minipage}{0.325\hsize}
   \begin{center}
    \includegraphics[clip,width=6.6cm]{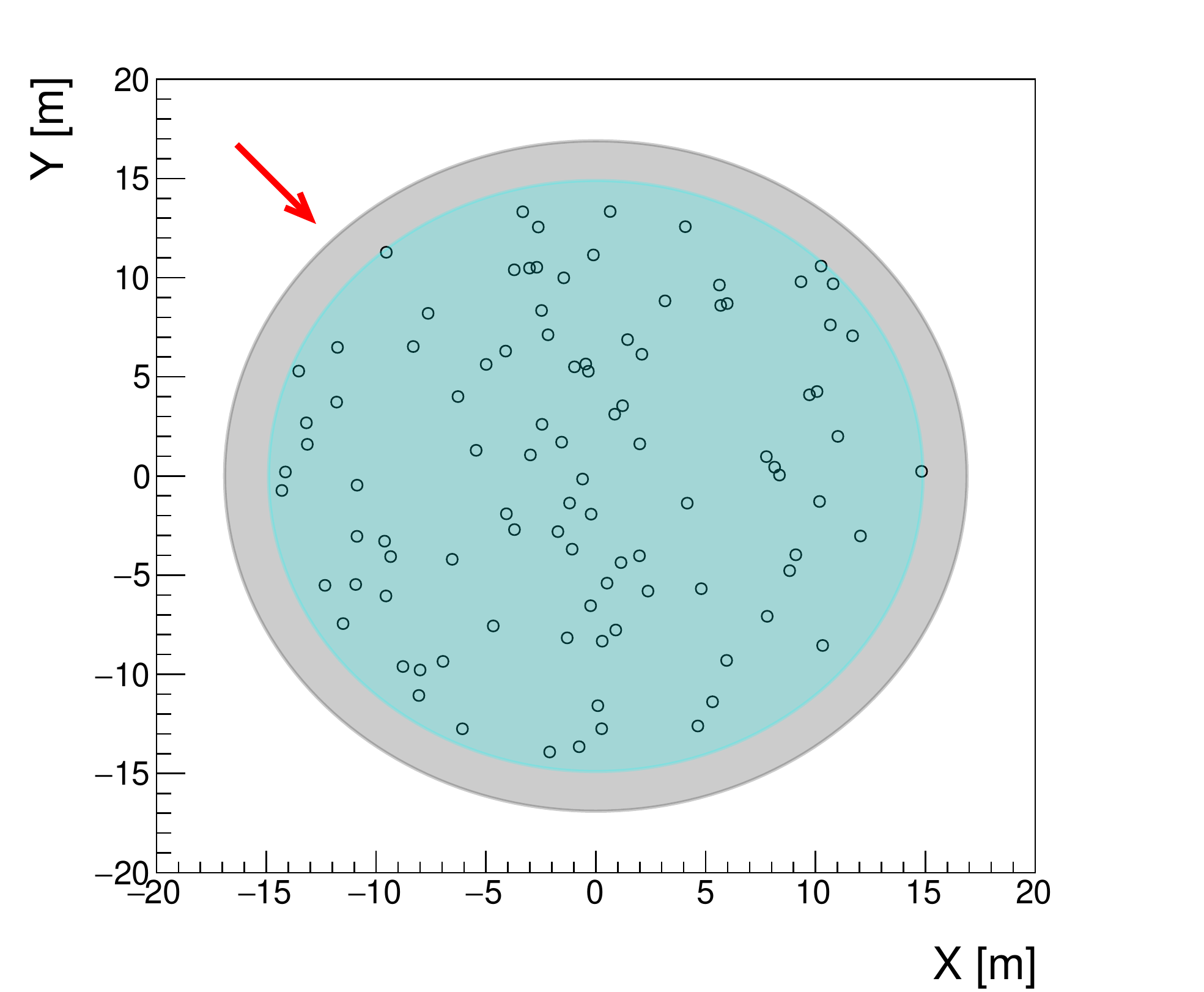}
   \end{center}
  \end{minipage}
  \caption{Distributions of $E_{\rm rec}$ (left), $\theta_{\rm C}$ (middle), and 
           vertex (right) from the RHC sample. 
	   In the right panel, the red arrow indicates the beam direction and 
	   the gray and sky blue regions correspond to the ID and FV, respectively.}
  \label{fig:rhcselected}
  \end{figure*}

After all cuts, the event selection is more than 80\% efficient for signal events,
while reducing background events by more than two orders of magnitude. 
Figure~\ref{fig:fhc_nmcandoffbeam} shows a comparison of the number of MC beam neutrino events  
against beam-unrelated events both before and after these cuts.
% 
%%% PRD
The event selection summary for the beam data and MC is shown in Table~\ref{tab:reduchist}.
%
%%% Final sample 
After the event selection, 204 events are observed in the FHC data and 
97 events are observed in the RHC data. 
These are compared with the number of predicted events in Table~\ref{tab:reduchist}. 
While the FHC sample has a high signal purity, the neutrino component forms nearly 20\% of the RHC sample
because of the difference between the neutrino and antineutrino cross sections. 
Figures~\ref{fig:fhcselected} and \ref{fig:rhcselected} show distributions of 
the reconstructed energy, Cherenkov angle, and vertex position for the FHC and RHC samples, respectively. 
The observed $E_{\rm rec}$ distributions agree well with the predictions
in both FHC and RHC modes.
A clear contribution from $\sim$6~MeV $\gamma$-rays is observed in both operation modes. 
In the FHC $\theta_{\rm C}$ distribution, the data at high angles 
is below the MC expectation, while no such MC excess is seen in the RHC data. 
This excess was also observed in the previous T2K measurement \cite{bib:t2kncqe1to3}
although the statistical error was larger. 
At high angles this distribution is dominated by events with multiple  $\gamma$-rays.
Such events are caused mainly by fast neutron interactions with nuclei in the water.
The excess in FHC may then be attributed to inaccurate modeling of secondary neutron reactions 
and their subsequent $\gamma$-ray emissions. 
The fact that the disagreement between observation and prediction is visible 
in FHC and not in RHC, may be understood by the difference in the out-going 
nucleon kinematics between neutrino and antineutrino interactions. 
Helicity conservation in antineutrino interactions produces more forward-going 
leptons in the final state and consequently lower momentum nucleons. 
The latter therefore goes on to produce fewer secondary $\gamma$-rays than that 
from its neutrino interaction counterpart. 
Comparing the ratio of the single-$\gamma$ peak ($\sim$$42^\circ$) 
to the multiple-$\gamma$ peak ($\sim$$90^\circ$) of the MC 
in each figure, there are relatively fewer events in the 
high-angle region of the RHC sample.
The vertex positions of selected events in the data 
are found to be uniform and no bias relative to the beam direction is observed.  

  %%% Figure : Noff vs. Nmc

  %%% PRD
  %%% Table : Reduction history 
  \begin{table*}[htbp]
  \begin{center}
  \caption{Number of events after each cut in data and MC.
	   Before the timing cut, only the beam quality and detector condition cuts are applied.}
  \label{tab:reduchist}
  \vspace{2truept}
   \begin{tabular}{l | c | c c c c c c} \hline \hline
    & Observation & \multicolumn{6}{c}{Prediction} \\ \hline \hline
    FHC \ & \ \ On-timing data \ \       & \ \ Total & \ $\nu$-NCQE & \ $\bar{\nu}$-NCQE & \ NC-other & \ CC   & \ Beam-unrelated \ \\ \hline
    Timing cut          \ & \ \ 4595 \ \ & \ \ -     & \ -          & \ -                & \ -        & \ -    & \ 4357.5         \ \\
    Decay-e cut         \ & \ \ 4553 \ \ & \ \ -     & \ -          & \ -                & \ -        & \ -    & \ 4350.8         \ \\
    FV cut              \ & \ \ 831  \ \ & \ \ 896.8 & \ 190.7      & \ 5.2              & \ 52.1     & \ 24.9 & \ 623.9          \ \\
    {\it dwall} cut     \ & \ \ 735  \ \ & \ \ 791.4 & \ 190.0      & \ 5.2              & \ 51.9     & \ 24.8 & \ 519.5          \ \\
    {\it effwall} cut   \ & \ \ 442  \ \ & \ \ 492.7 & \ 185.6      & \ 5.0              & \ 51.4     & \ 24.6 & \ 226.1          \ \\
    {\it ovaQ} cut      \ & \ \ 220  \ \ & \ \ 263.9 & \ 181.0      & \ 4.9              & \ 50.2     & \ 24.1 & \ 3.7            \ \\
    CC cut              \ & \ \ 204  \ \ & \ \ 238.4 & \ 178.6      & \ 4.8              & \ 42.5     & \ 8.9  & \ 3.6            \ \\ \hline \hline
    RHC \ & \ On-timing data \           & \ \ Total & \ $\nu$-NCQE & \ $\bar{\nu}$-NCQE & \ NC-other & \ CC   & \ Beam-unrelated \ \\ \hline
    Timing cut          \ & \ \ 3626 \ \ & \ \ -     & \ -          & \ -                & \ -        & \ -    & \ 3746.9         \ \\
    Decay-e cut         \ & \ \ 3597 \ \ & \ \ -     & \ -          & \ -                & \ -        & \ -    & \ 3470.0         \ \\
    FV cut              \ & \ \ 613  \ \ & \ \ 606.0 & \ 19.6       & \ 60.7             & \ 19.6     & \ 5.7  & \ 500.4          \ \\
    {\it dwall} cut     \ & \ \ 535  \ \ & \ \ 524.1 & \ 19.5       & \ 60.5             & \ 19.5     & \ 5.7  & \ 418.9          \ \\
    {\it effwall} cut   \ & \ \ 282  \ \ & \ \ 279.4 & \ 19.1       & \ 58.7             & \ 19.3     & \ 5.6  & \ 176.7          \ \\
    {\it ovaQ} cut      \ & \ \ 101  \ \ & \ \ 101.8 & \ 18.5       & \ 57.0             & \ 18.7     & \ 5.5  & \ 2.1            \ \\
    CC cut              \ & \ \ 97   \ \ & \ \ 94.3  & \ 17.9       & \ 56.5             & \ 15.5     & \ 2.3  & \ 2.1            \ \\ \hline \hline
   \end{tabular}
  \end{center}
  \end{table*}
  %%%

%  %%% Table : Selected event numbers
%  \begin{table*}[htbp]
%  \begin{center}
%  \caption{Number of predicted events (MC $+$ Beam-unrelated) and 
%           fractional composition of each interaction mode in parentheses for both operation modes. 
%	    The number of observed events in the on-timing window is also listed.}
%  \label{tab:selectedevt}
%  %
%  \vspace{2truept}
%   \begin{tabular}{l c c c c c c} \hline \hline
%    FHC \ \ & Total \ \ & $\nu$-NCQE \ \ & $\bar{\nu}$-NCQE \ \ & NC-other \ \ 
%            & CC \ \ & Beam-unrelated \\ \hline
%    Prediction \ \ & 238.4 (100\%) \ \ & 178.6 (75.0\%) \ \ & 4.8 (2.0\%) \ \ & 42.5 (17.8\%) \ \ 
%            & 8.9 (3.7\%) \ \ & 3.6 (1.5\%) \\ 
%    Observation \ & 204 &  &  &  &  &  \\ \hline \hline
%    %
%    RHC \ \ & Total \ \ & $\nu$-NCQE \ \ & $\bar{\nu}$-NCQE \ \ & NC-other 
%            & CC \ \ & Beam-unrelated \\ \hline
%    Prediction \ \ & 94.3 (100\%) \ \ & 17.9 (19.0\%) \ \ & 56.5 (59.9\%) \ \ & 15.5 (16.5\%) \ \ 
%            & 2.3 (2.5\%) \ \ & 2.1 (2.1\%) \\ 
%    Observation \ \ & 97 &  &  &  &  &  \\ \hline \hline
%   \end{tabular}
%  \end{center}
%  \end{table*}
%  %%% 

  %%% Figure : Distributions of selected events in FHC
  %%%

  %%% Figure : Distributions of selected events in RHC
  %%%

%------------------------------------------------------------------------------
%  Uncertainty Estimates
   \section{Uncertainty Estimates}
   \label{sec:uncertainty}
%..............................................................................

%%% Statistical error
Based on the observed number of events in Table~\ref{tab:reduchist}, 
the associated statistical error is 7.0\% for the FHC sample and 
10.2\% for the RHC sample.
%$\sqrt{204}/204 = 7.0\%$ in FHC and $\sqrt{97}/97 = 10.2\%$ in RHC.

%%% Systematic error
Systematic errors from six main sources are considered in the analysis, 
namely the neutrino flux prediction, the neutrino interaction model, 
the primary-$\gamma$ and secondary-$\gamma$ emission models,   
neutrino oscillation parameters, and the detector response. 
In this analysis, CC measurement results from the T2K near detectors are not used 
so as to ensure that flux and interaction systematics are treated independently. 
Only statistical uncertainties are considered for beam-unrelated events, 
3.0\% in the FHC sample and 3.9\% in the RHC sample, since  they are also 
part of the observed data and respond to detector uncertainties in the same way.
The effect of possible rate fluctuations between the on- and off-timing windows 
is negligible.
Table~\ref{tab:systerr} summarizes the impact of each of these error categories on the 
different interaction modes populating the samples.
Among them, systematic errors from the secondary-$\gamma$ production model are 
the leading uncertainties. 
The error sources are described in detail below. 

  %%% Table : Systematic error summary
  \begin{table*}[htbp]
  \begin{center}
  \caption{Summary of systematic uncertainties on the observed event rate in percent for each sample component. 
           The fraction of each component, listed as ``Event fraction", is also shown in percent.
           For beam-unrelated events the total error entry represents the statistical uncertainty.}
  \label{tab:systerr}
  \vspace{2truept}
   \begin{tabular}{l c c c c c} \hline \hline
    FHC                           & \ $\nu$-NCQE & \ $\bar{\nu}$-NCQE & \ NC-other & \ CC   & \ Beam-unrelated \\ \hline
    Event fraction                & \ 75.0       & \ 2.0              & \ 17.8     & \ 3.7  & \ 1.5 \\ \hline 
    Neutrino flux                 & \ 6.7        & \ 8.6              & \ 7.3      & \ 6.4  & \ - \\  
    Neutrino interaction          & \ 3.0        & \ 3.0              & \ 8.2      & \ 16.5 & \ - \\  
    Primary-$\gamma$ production   & \ 11.0       & \ 10.6             & \ 6.0      & \ 6.6  & \ - \\  
    Secondary-$\gamma$ production & \ 13.5       & \ 13.4             & \ 19.5     & \ 17.6 & \ - \\  
    Oscillation parameter         & \ -          & \ -                & \ -        & \ 4.1  & \ - \\  
    Detector response             & \ 3.4        & \ 3.4              & \ 2.0      & \ 5.2  & \ - \\ \hline
    Total error                   & \ 19.2       & \ 19.7             & \ 23.3     & \ 26.7 & \ 3.0 \\ \hline \hline
    RHC                           & \ $\nu$-NCQE & \ $\bar{\nu}$-NCQE & \ NC-other & \ CC   & \ Beam-unrelated \\ \hline
    Event fraction                & \ 19.0       & \ 59.9             & \ 16.5     & \ 2.5  & \ 2.1  \\ \hline
    Neutrino flux                 & \ 7.0        & \ 6.4              & \ 7.0      & \ 6.5  & \ - \\  
    Neutrino interaction          & \ 3.0        & \ 3.0              & \ 10.8     & \ 38.2 & \ - \\  
    Primary-$\gamma$ production   & \ 12.2       & \ 11.4             & \ 3.5      & \ 0.5  & \ - \\  
    Secondary-$\gamma$ production & \ 13.6       & \ 13.1             & \ 19.3     & \ 21.4 & \ - \\  
    Oscillation parameter         & \ -          & \ -                & \ -        & \ 3.1  & \ - \\  
    Detector response             & \ 3.4        & \ 3.4              & \ 2.0      & \ 5.2  & \ - \\ \hline
    Total error                   & \ 20.1       & \ 19.0             & \ 23.4     & \ 44.7 & \ 3.9 \\ \hline \hline
   \end{tabular}
  \end{center}
  \end{table*}
  %%% 

\subsection{Neutrino flux and interaction model uncertainties}

The impact of neutrino flux and interaction systematic uncertainties 
in this analysis is estimated by the change in the number of selected events 
relative to the nominal model under a $1\sigma$ shift in each error source.
The procedure follows previous T2K analyses~\cite{bib:t2kncqe1to3,bib:t2kccincpod,bib:t2kccinckoga}.

%%% Neutrino flux 
Flux uncertainties are evaluated for each neutrino flavor, horn polarity, and neutrino energy bin.
Uncertainties in the hadronic interaction cross section are the dominant 
contribution to the assigned 6$-$8\% flux uncertainties.
This represents a large improvement over previous T2K analyses,  
due to improved hadron production and interaction constraints 
from NA61/SHINE measurements using a replica of the T2K target~\cite{bib:na61shinereplica}.

%%% Neutrino interaction 
The value of the axial-vector mass used to generate quasielastic interactions 
with its $1\sigma$ error is $M_{\rm A}^{\rm QE} = 1.21 \pm 0.18~{\rm GeV/c^2}$. 
Similarly the Fermi momentum in oxygen is taken to be $225 \pm 31~{\rm MeV/c}$. 
Parameters describing contributions from 2p2h interactions, resonant pion production, 
and deep inelastic scattering follow the assignments 
in previous analyses~\cite{bib:t2kncqe1to3,bib:t2kccincpod,bib:t2kccinckoga}. 
These result in uncertainties of 8.2\% (10.8\%) for the NC-other and 
16.5\% (38.2\%) for CC interaction backgrounds in the FHC (RHC) measurement.
The larger uncertainty in the RHC CC component, as seen in Table~\ref{tab:systerr}, 
is attributed to the different effect of $M_{\rm A}^{\rm QE}$.
% 
%Note that in this treatment the NCQE cross sections are not changed by these errors.
%%% PRD 
Since $\gamma$-rays are emitted isotropically and SK has $4\pi$ acceptance, 
the signal efficiencies are unaffected by neutrino interaction model uncertainties.
%Signal efficiencies are not affected by these uncertainties because 
%the $\gamma$-rays are emitted isotropically and SK has $4\pi$ acceptance. 
%This assumption breaks down near the detector wall, but the effect is neglible here.
% 
%Systematic uncertainties relevant for the signal efficiency instead coming from 
%the production mechanism of those $\gamma$-rays are discussed below.
%% Errors for the signal interaction cross sections are not considered 
%% in the present work. 

It should be noted that while NC inelastic scattering without nucleon emission, 
$\nu (\bar{\nu}) + {\rm ^{16}O} \rightarrow \nu (\bar{\nu}) + {\rm ^{16}O^{*}}$, 
should be present in the selected sample, it is not simulated in this analysis.
According to Ref.~\cite{bib:kolbe}, the sum of cross sections leading to
${\rm ^{15}O^{*}}$ and ${\rm ^{15}N^{*}}$ after the ${\rm ^{16}O}(\nu,\nu')$ interaction
increases from $6.7\times10^{-42}$~${\rm cm^2}$
at $E_\nu = 50$~MeV to $481\times10^{-42}$~${\rm cm^2}$ at $E_\nu = 500$~MeV,
while it is almost constant above $\sim$200~MeV.
By comparing this to the expected NCQE cross section in Ref.~\cite{bib:ankowski}, 
it is found that the NCQE process dominates over the NC inelastic process without nucleon knock-out 
above $E_\nu \sim 200$~MeV. 
In addition, the former cross section is $\sim$40 times larger at 500~MeV 
and is expected to be even larger at higher energies. 
In the present measurement the signal is predominantly 
from neutrinos above $E_\nu \sim 500$~MeV.
Assuming that the detection efficiency of $\gamma$-rays produced from the de-excitation of 
nuclei recoiling from the NC inelastic interaction without nucleon emission 
is comparable to that of NCQE scattering, 
a 3\% error on the signal channel is assigned conservatively in consideration of the expected 
interaction cross section differences.
Another possible contribution is from NC interactions on hydrogen, 
$\nu (\bar{\nu}) + {\rm ^{1}H} \rightarrow \nu (\bar{\nu}) + {\rm ^{1}H}$, 
where the final state protons may produce $\gamma$-rays via reactions with water. 
However, the contribution from such interactions is expected to be less 
than 1\% of that from NCQE interactions on oxygen and therefore does not 
significantly affect the results of the present measurement.

\subsection{Primary- and secondary-$\gamma$ production uncertainties}

%%% Primary-gamma
Errors on the primary $\gamma$-ray emission come from the uncertainties 
on the spectroscopic factors. 
Calculation of the spectroscopic strength for the ${\rm p_{3/2}}$ state has 
been found to be consistent with electron scattering data
within 5.4\% \cite{bib:ankowski}, which leads to an error on 
the observed event rate at T2K of less than 3\%.
The uncertainty due to the {\it others} state (all other states than 
${\rm (p_{1/2})^{-1}}$, ${\rm (p_{3/2})^{-1}}$, and ${\rm (s_{1/2})^{-1}}$) 
being included into the ${\rm (s_{1/2})^{-1}}$ state in the nominal model 
is estimated by comparison with an extreme case. 
Since no significant deviation in the predicted ${\rm p_{3/2}}$ strength has been 
observed in ($e,e'p$) and ($p,2p$) experiments~\cite{bib:leuschner,bib:rcnpe148}, 
{\it others} cannot behave like the ${\rm (p_{3/2})^{-1}}$ state. 
%% like ${\rm (p_{3/2})^{-1}}$. 
% 
In contrast, the possibility that the {\it others} state behaves like the 
ground state, ${\rm (p_{1/2})^{-1}}$, emitting no $\gamma$-rays, is considered,  
since this would not contradict any existing data. 
To model this, the {\it others} state is included into ${\rm (p_{1/2})^{-1}}$ instead,
and the change in the event rate relative to the nominal model is taken as the systematic error.
This results in uncertainties in the 6$-$12\% range for the signal and background modes. 
This extreme case covers the uncertainties of the ${\rm p_{1/2}}$ and ${\rm s_{1/2}}$ spectroscopic strengths.
%%This leads to the dominant error of 6$-$12\%. 
% 
The total error on primary-$\gamma$ production is taken to be 
the sum in quadrature of above two sources.

\begin{figure}[htbp]
  \begin{center}
   \includegraphics[clip,width=7.0cm]{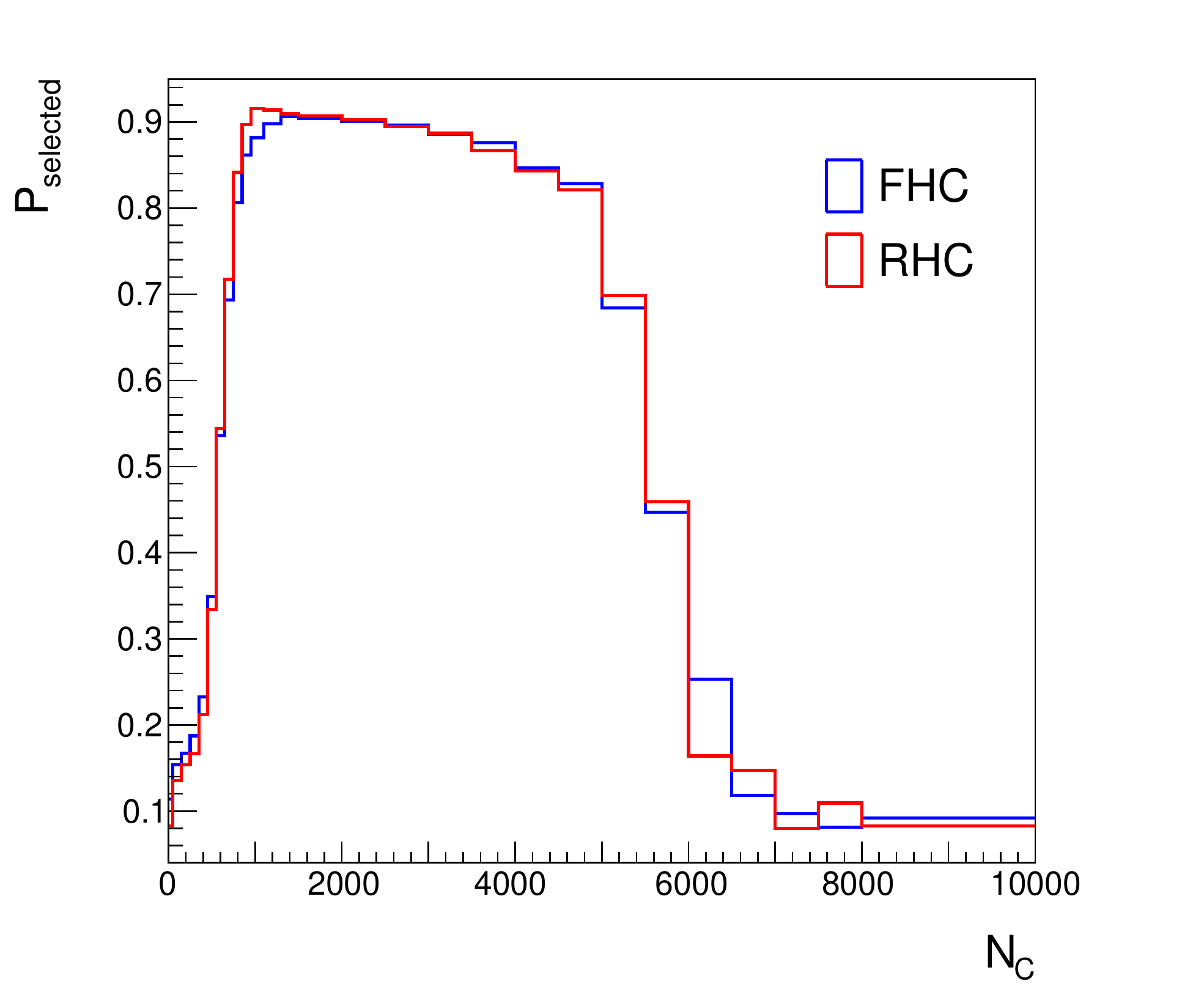}
  \end{center}
  \vspace{-10truept}
  \caption{Probabilities of an event being reconstructed in the energy region of 
           3.49$-$29.49~MeV as a function of the number of Cherenkov photons ($N_{\rm C}$)
	   for FHC and RHC.} 
  \label{fig:psel_nch_dist}
  \end{figure}

%%% Secondary-gamma
The secondary-$\gamma$ emission rate is model-dependent and at present 
there is insufficient data on $\gamma$-ray emission from neutron-oxygen reactions 
at energies above 20~MeV \cite{bib:rcnpe487},
which are most relevant for the present work, making model selection difficult.
Since different models predict different amounts of $\gamma$-ray emission,  
to reduce the impact of such model dependence, 
instead the total number of emitted Cherenkov photons from secondary emission processes is considered. 
First, the probability ($P_{\rm selected}$) of an event being reconstructed 
in the 3.49$-$29.49~MeV energy region of this analysis is estimated 
as a function of the number of emitted Cherenkov photons using MC.
The resulting probabilities for FHC and RHC are shown in Figure~\ref{fig:psel_nch_dist}.
The number of emitted Cherenkov photons ($N_{\rm C}$) can be broken down into 
three parts, 

  %%% Equation : Number of Cherenkov photons 
  \begin{eqnarray}
  \label{eq:ncherenkov}
   N_{\rm C} \simeq N_{\rm C}^{\rm primary\mathchar`-\nu} 
       + N_{\rm C}^{{\rm secondary\mathchar`-}n}  
       + N_{\rm C}^{{\rm secondary\mathchar`-}p}. %\nonumber 
  \end{eqnarray}
  \vspace{2truept}
  %%% 

\noindent 
Here $N_{\rm C}^{\rm primary\mathchar`-\nu}$ denotes the contribution from 
the primary $\gamma$-ray emission and 
$N_{\rm C}^{{\rm secondary\mathchar`-}n}$ ($N_{\rm C}^{{\rm secondary\mathchar`-}p}$) 
is from secondary $\gamma$-rays produced by neutron (proton) interactions in water. 
The systematic uncertainty used in the analysis is estimated by varying the 
contributions from these secondary interactions and calculating the change 
in the selected sample using Figure~\ref{fig:psel_nch_dist}.
%% Second, the number of Cherenkov photons made by the secondary interactions, 
%% $N_{\rm C}^{\rm secondary\mathchar`-n}$ and $N_{\rm C}^{\rm secondary\mathchar`-p}$, 
%% are varied, and then the number of events passing the selection is calculated 
%% using the probability distribution as shown in Figure~\ref{fig:psel_nch_dist}.
%
%% Then by comparing the number of events in the nominal setting to that 
%% in the changed setting, the systematic error is estimated.
% 
%% The size of this variation is determined to be 65\% as explained below. 
% 
% The secondary-$\gamma$ emission consists of two processes: neutron-oxygen reaction 
% and nuclear de-excitation. 
%
The source of uncertainty can be broken down into the initial nucleon-oxygen interaction 
and the subsequent nuclear de-excitation. 
In Ref.~\cite{bib:fitbyma}, proton-carbon data were fit to obtain a constraint
on the nucleon-nucleus scattering cross section. 
Their result showed a 30\% difference between the measured and predicted (GCALOR) cross sections. 
%%% PRD
In the present work, the target nucleus is different but the effect 
is found to be no larger than 5\% in neutrino interaction measurements \cite{bib:t2kccinckoga}, 
so a conservative error of 40\% is adopted.
In order to estimate the impact of $\gamma$-ray emission from fast neutron reactions on oxygen, 
the results of a muon-induced spallation study at SK \cite{bib:skli9} are used.
Since the selected sample contains contributions from such neutron interactions, 
and the measured energy distribution does not differ by more than 50\% from the MC, 
this number is taken as the error estimate. 
For the uncertainty propagation the quadratic sum of these two contributions is used 
and a $\pm 65\%$ variation is applied to both  
$N_{\rm C}^{{\rm secondary\mathchar`-}n}$ and $N_{\rm C}^{{\rm secondary\mathchar`-}p}$. 
The variation producing the largest change in the final sample is used to 
compute the final error and results 
in a $\sim$13\% uncertainty for signal and roughly 20\% for 
the NC-other and CC components. 
In addition, the impact of uncertainties from the final state interaction model 
has been evaluated to be as large as 3\%.
The total uncertainty for each is obtained by summing these two contributions 
in quadrature.

  %%% Figure : Pselected vs. Nc

\subsection{Oscillation parameter and detector response uncertainties}

%%% Oscillation parameter 
Errors on the oscillation parameters, $\theta_{13}$, $\theta_{23}$, and $\Delta m^2_{32}$, 
are taken from Ref.~\cite{bib:t2k2017oa}. 
Varying each of these, the change in the selected number of CC events 
results in $3$$-$$4$\% errors for the FHC and RHC samples. 

%%% Detector response
Errors on each reconstructed parameter used in the event selection, 
$E_{\rm rec}$, {\it dwall}, {\it effwall}, {\it ovaQ}, and $\theta_{\rm C}$ are considered
as detector response uncertainties.
These have been studied using detector calibrations~\cite{bib:sklinac,bib:skdt}, and  
their effect on the final sample is 1\%. 
Similarly, the gain of the SK PMTs was found to vary over the observation period 
and its impact is considered as systematic error in this analysis. 
This gain shift changes the number of PMT hits used to reconstruct energy and produces 
a 3\% error on the final sample.
In total, 3$-$5\% errors are assigned for each interaction mode.

%------------------------------------------------------------------------------
%  Cross Section Results
   \section{Cross Section Results}
   \label{sec:xsec}
%..............................................................................

\begin{figure}[htbp]
  % 1st figure
  \begin{center}
   \includegraphics[clip,width=6.96cm]{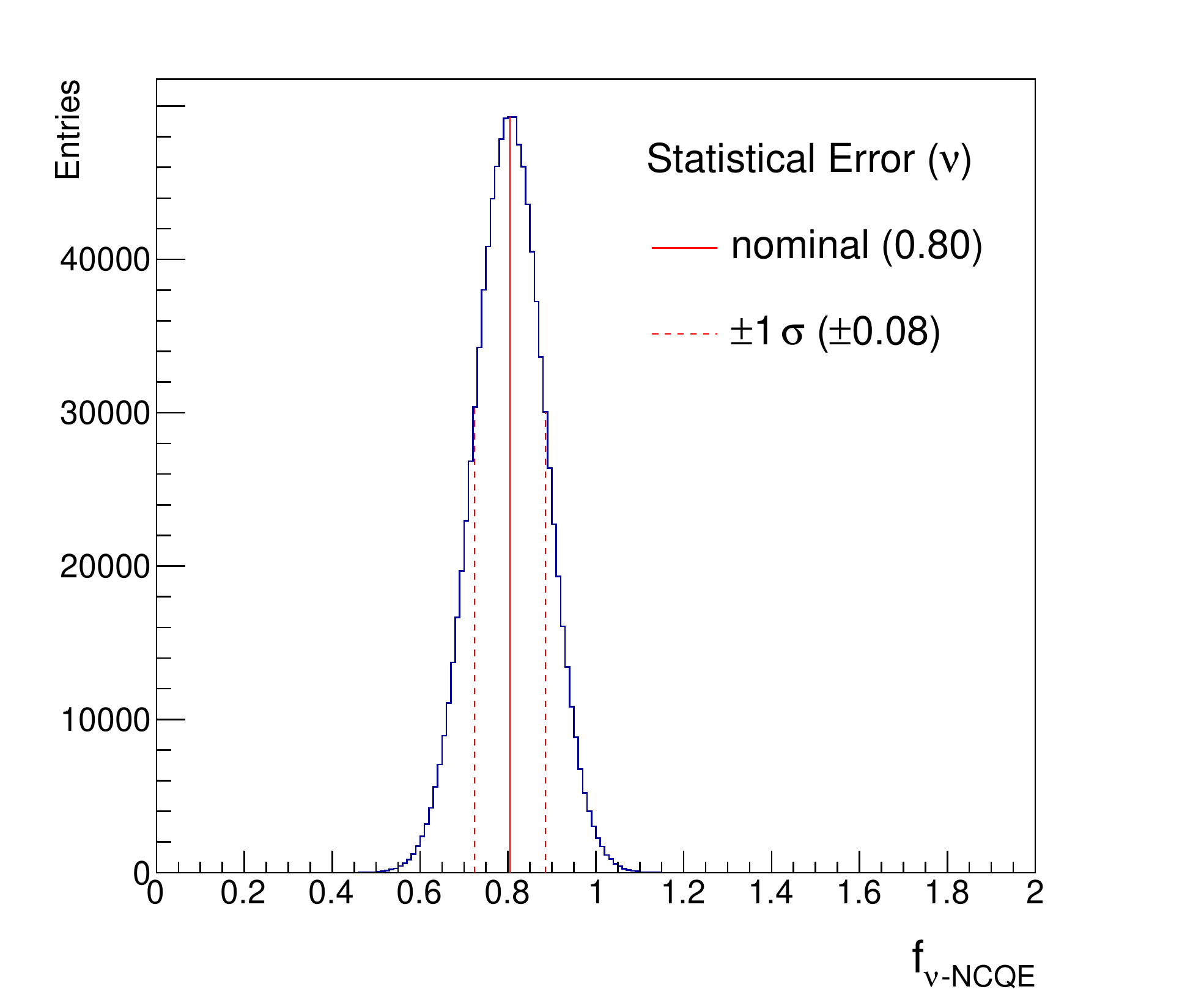}
  \end{center}
  \vspace{-15truept}
  % 2nd figure
  \begin{center}
   \includegraphics[clip,width=6.96cm]{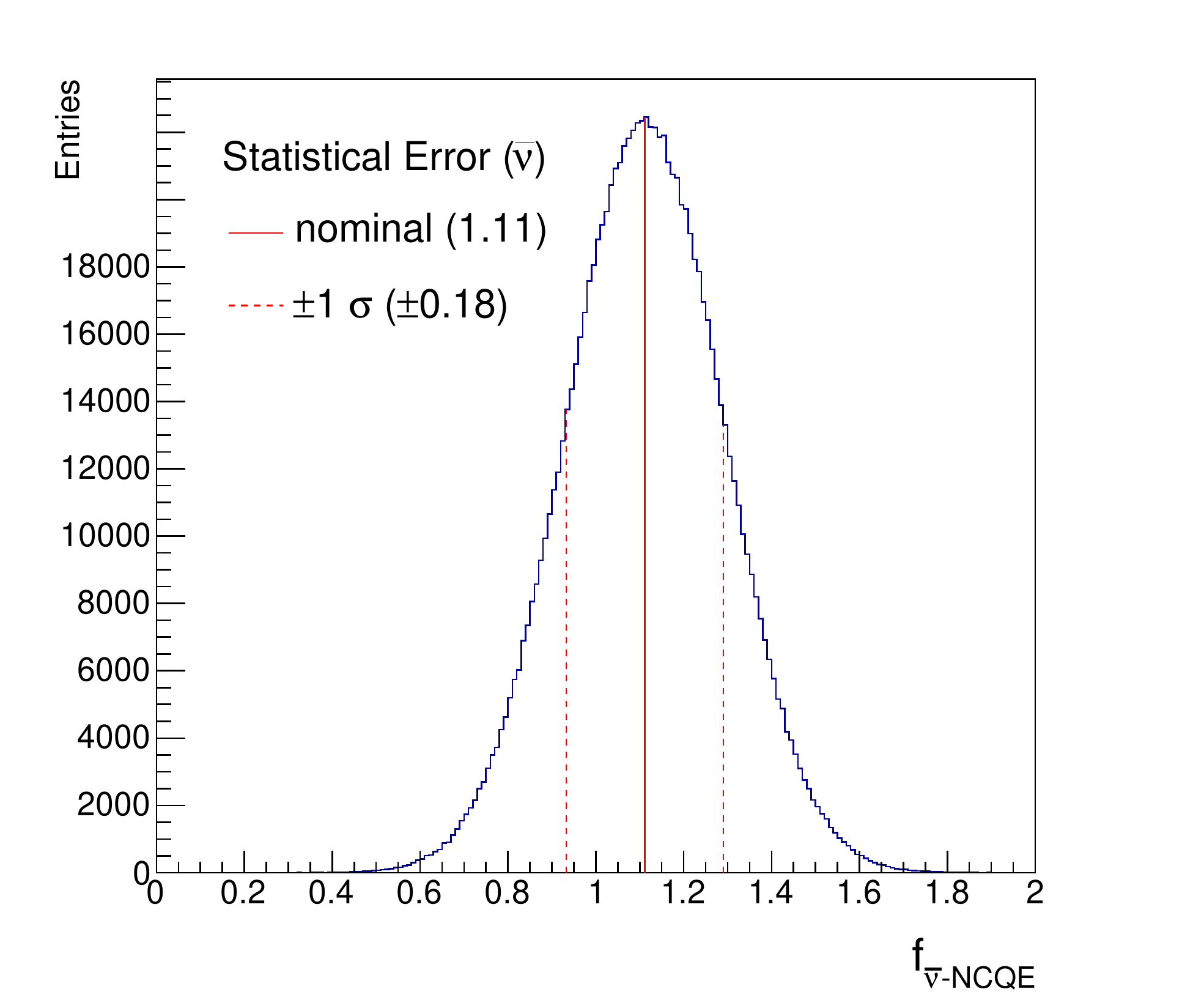}
  \end{center}
  \vspace{-15truept}
  % 3rd figure
  \begin{center}
   \includegraphics[clip,width=6.96cm]{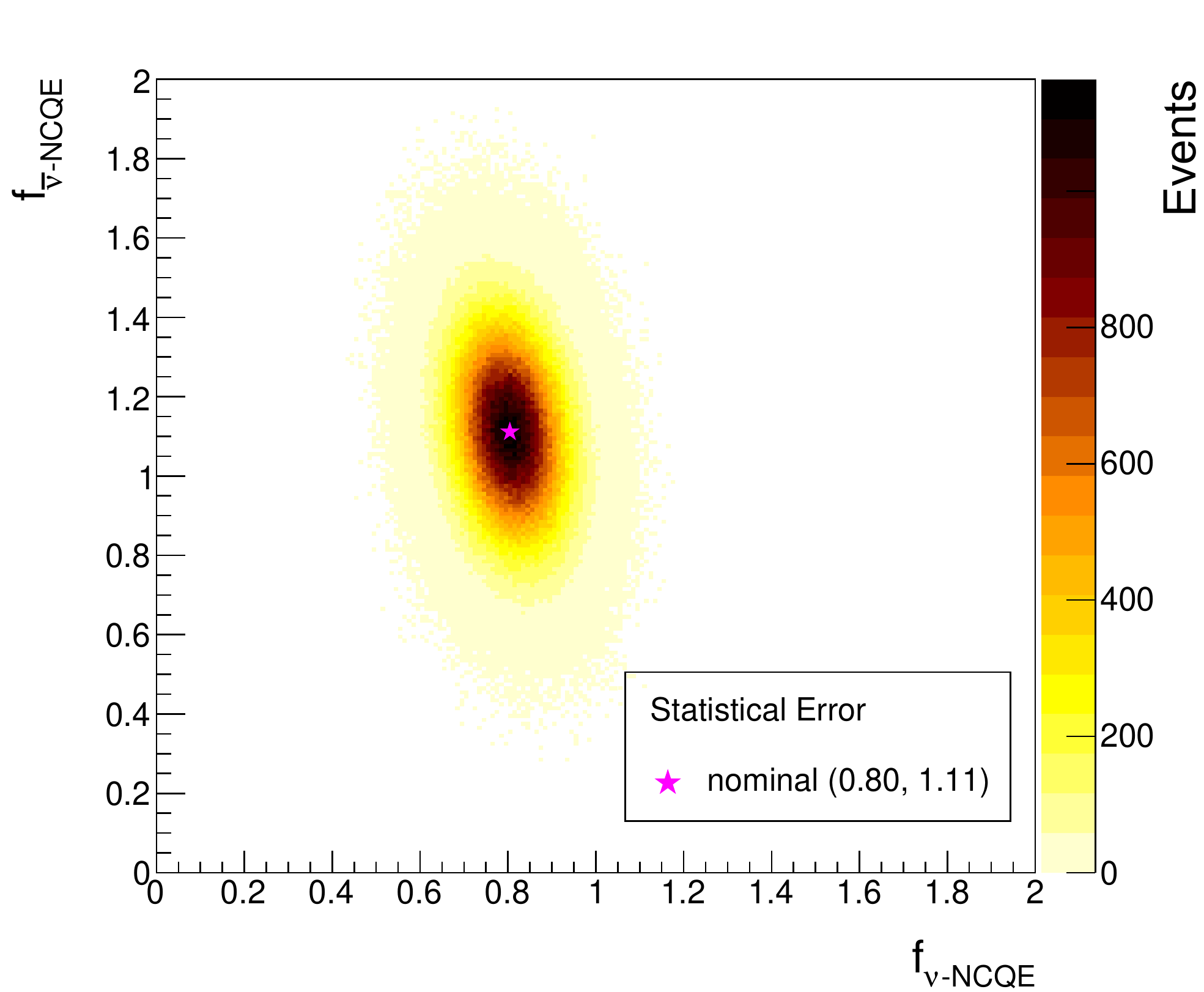}
  \end{center}
  %
  %\vspace{-5truept}
  \caption{Results of the pseudo experiments on the scale factors 
           when the numbers of events are 
           varied based on the statistical uncertainties: 
           $f_{\rm \nu\mathchar`-NCQE} = 0.80 \pm 0.08$ and 
	   $f_{\rm \bar{\nu}\mathchar`-NCQE} = 1.11 \pm 0.18$.}
  \label{fig:toymcstat}
  \end{figure}

\begin{figure}[htbp]
  % 1st figure
  \begin{center}
   \includegraphics[clip,width=6.96cm]{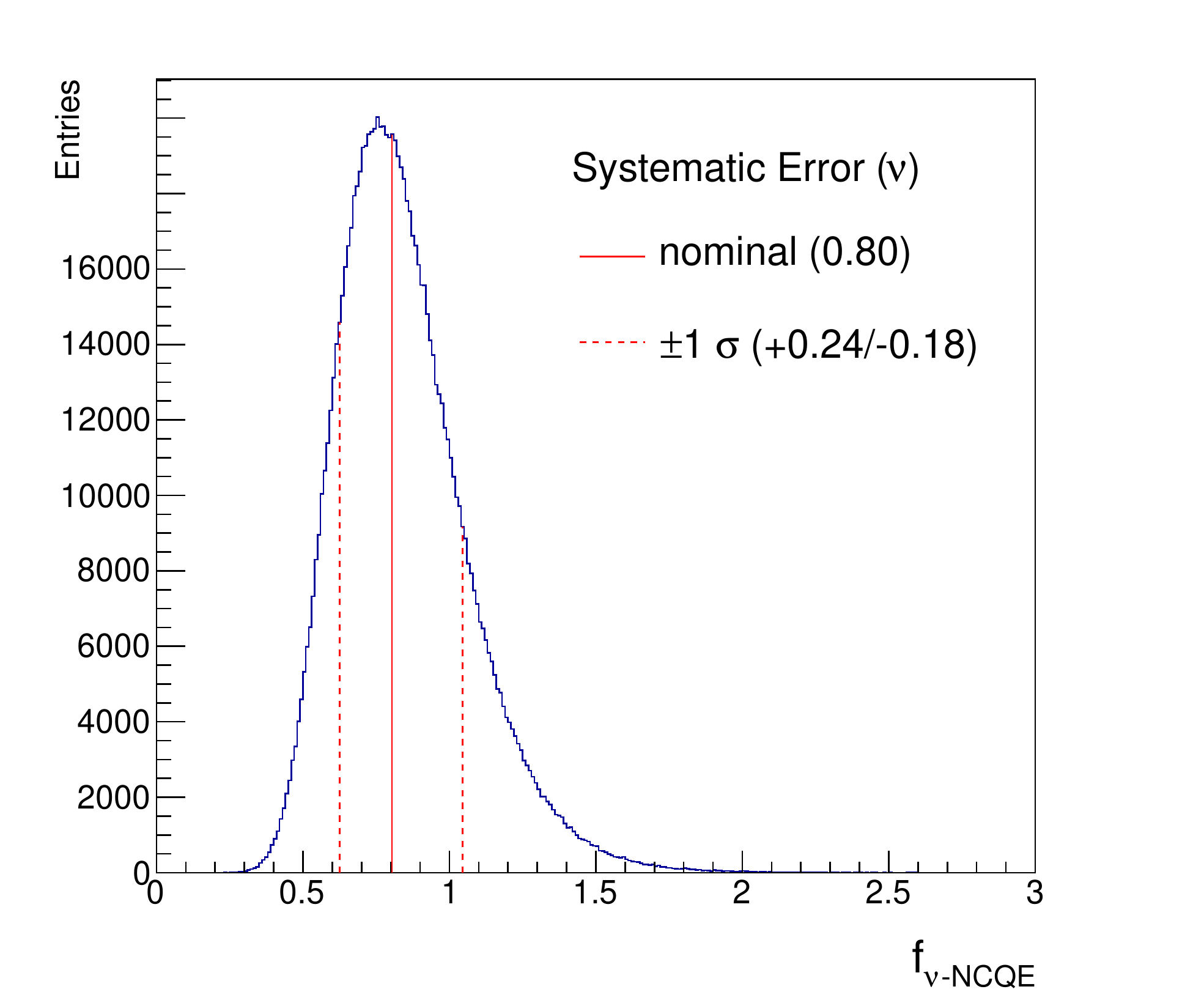}
  \end{center}
  \vspace{-15truept}
  % 2nd figure
  \begin{center}
   \includegraphics[clip,width=6.96cm]{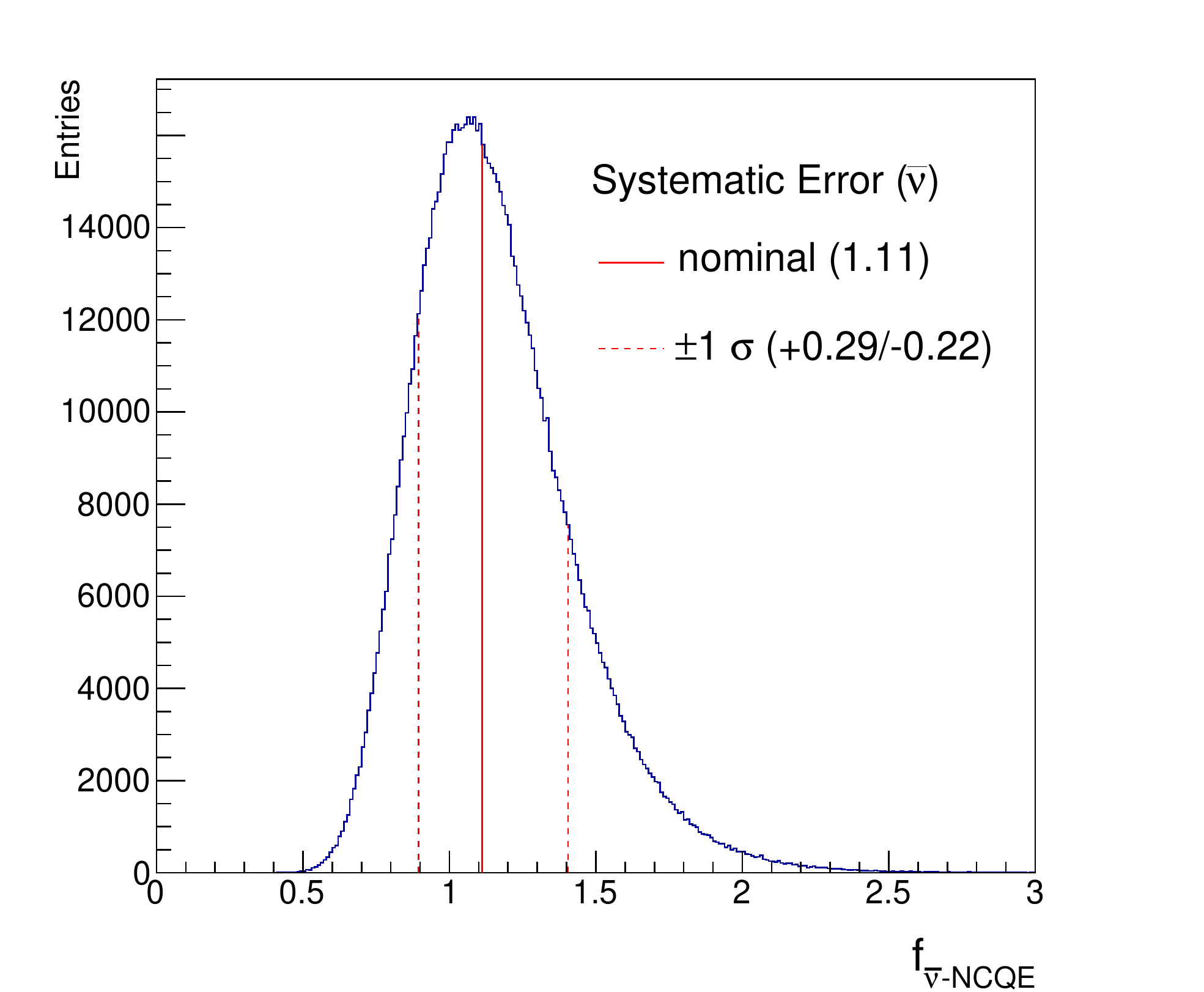}
  \end{center}
  \vspace{-15truept}
  % 3rd figure
  \begin{center}
   \includegraphics[clip,width=6.96cm]{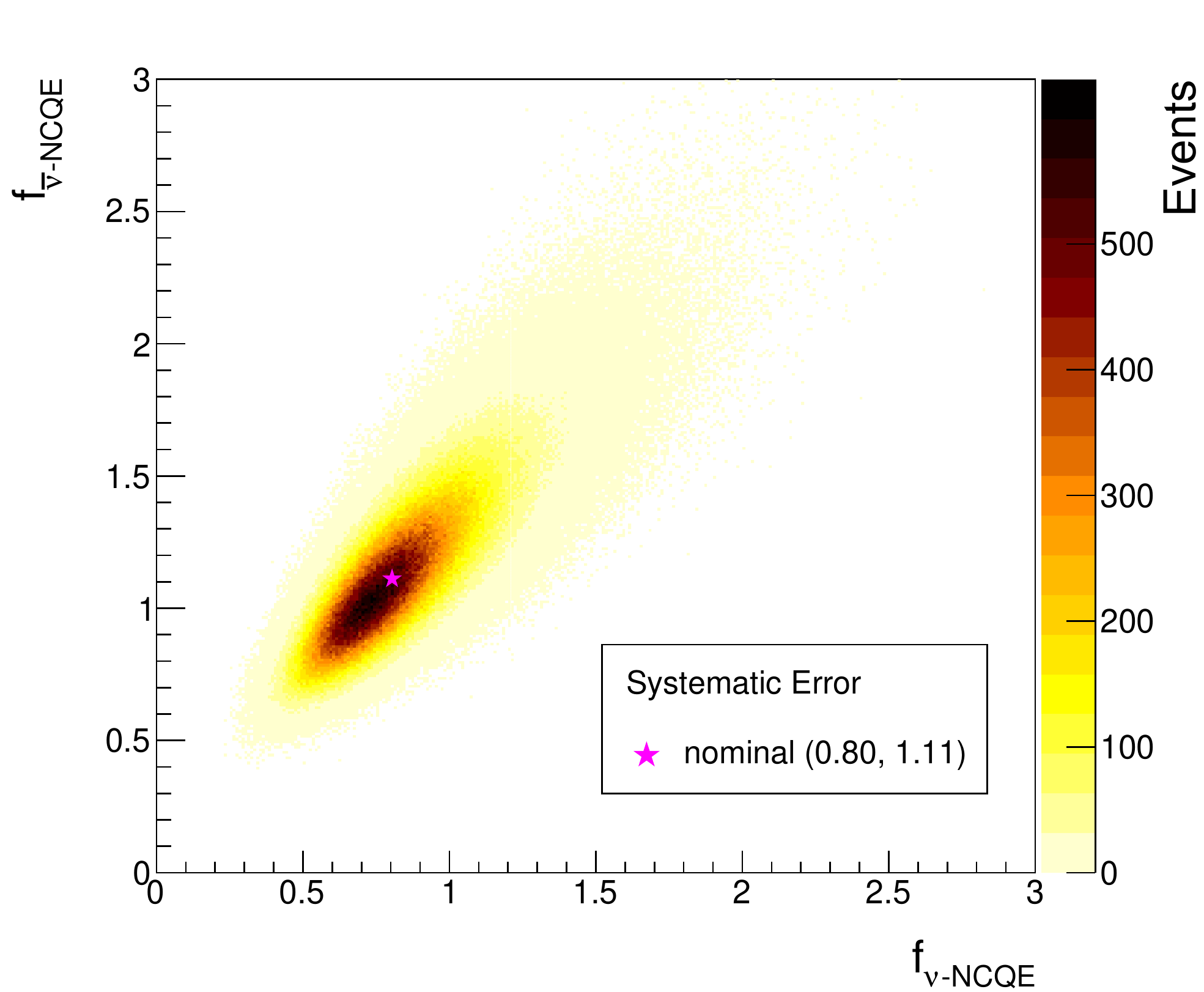}
  \end{center}
  %
  %\vspace{-5truept}
  \caption{Results of the pseudo experiments on the scale factors 
           when the numbers of events are 
           varied based on the systematic uncertainties: 
           $f_{\rm \nu\mathchar`-NCQE} = 0.80^{+0.24}_{-0.18}$ and 
	   $f_{\rm \bar{\nu}\mathchar`-NCQE} = 1.11^{+0.29}_{-0.22}$.
	   The dominant uncertainty source is the secondary-$\gamma$ production model 
	   as described in the text.}
  \label{fig:toymcsyst}
  \end{figure}

The number of observed events in the FHC and RHC data 
($D_{\rm obs}^{\rm FHC}$ and $D_{\rm obs}^{\rm RHC}$, respectively) are expressed as follows:

  %%% Equation : Dobs
  \begin{eqnarray}
  \label{eq:dobs}
   D_{\rm obs}^{\rm mode} = 
     f_{\rm \nu\mathchar`-NCQE} M_{\rm \nu\mathchar`-NCQE}^{\rm mode} 
   &+& f_{\rm \bar{\nu}\mathchar`-NCQE} M_{\rm \bar{\nu}\mathchar`-NCQE}^{\rm mode} 
   \nonumber \\
   &+& M_{\rm NC\mathchar`-other}^{\rm mode} + M_{\rm CC}^{\rm mode} 
   \nonumber \\
   &+& D_{\rm beam\mathchar`-unrelated}^{\rm mode}, %\nonumber 
  \end{eqnarray}
  \vspace{2truept}
  %%% 

\noindent 
where mode $=$ FHC or RHC, 
$M_{\rm \nu\mathchar`-NCQE}^{\rm mode}$, $M_{\rm \bar{\nu}\mathchar`-NCQE}^{\rm mode}$, 
$M_{\rm NC\mathchar`-other}^{\rm mode}$, $M_{\rm CC}^{\rm mode}$, and 
$D_{\rm beam\mathchar`-unrelated}^{\rm mode}$ represent the expected number of $\nu$-NCQE, 
$\bar{\nu}$-NCQE, NC-other, CC, and beam-unrelated events, respectively.
Here, quantities from the data are written with 
a capital $D$ while MC predictions are represented with a capital $M$. 
The factors $f_{\rm \nu\mathchar`-NCQE}$ and $f_{\rm \bar{\nu}\mathchar`-NCQE}$ 
are the measured quantities in the present analysis and serve to scale the NCQE cross section
as predicted in the nominal MC model. 
%
% 
%In the end, the relations below are obtained for each operation mode:
%
%  %%% Equation : Relation
%  \begin{eqnarray}
%  \label{eq:relation}
%   D_{\rm obs}^{\rm FHC} - M_{\rm NCother}^{\rm FHC} - M_{\rm CC}^{\rm FHC} -
%   D_{\rm beam\mathchar`-unrelated}^{\rm FHC} \nonumber \\ [+3truept] =
%   f_{\rm \nu\mathchar`-NCQE} M_{\rm \nu\mathchar`-NCQE}^{\rm FHC} +
%   f_{\rm \bar{\nu}\mathchar`-NCQE} M_{\rm \bar{\nu}\mathchar`-NCQE}^{\rm FHC}, \nonumber \\ [+10truept]
%   %
%   D_{\rm obs}^{\rm RHC} - M_{\rm NCother}^{\rm RHC} - M_{\rm CC}^{\rm RHC} -
%   D_{\rm beam\mathchar`-unrelated}^{\rm RHC} \nonumber \\ [+3truept] =
%   f_{\rm \nu\mathchar`-NCQE} M_{\rm \nu\mathchar`-NCQE}^{\rm RHC} +
%   f_{\rm \bar{\nu}\mathchar`-NCQE} M_{\rm \bar{\nu}\mathchar`-NCQE}^{\rm RHC}. \nonumber
%  \end{eqnarray}
%  \vspace{2truept}
%  %%%
Based on the observed 204 events in FHC mode and the 97 events in RHC mode 
the scale factors are calculated to be
$f_{\rm \nu\mathchar`-NCQE} = 0.80$ and
$f_{\rm \bar{\nu}\mathchar`-NCQE} = 1.11$.
Errors on these factors are evaluated using pseudo experiments generated 
according to random variations of the statistical and systematic uncertainties.
Here, statistical uncertainties are considered for $D_{\rm obs}^{\rm mode}$ 
(the effect of the uncertainty from $D_{\rm beam\mathchar`-unrelated}^{\rm mode}$ is negligible).
Systematic uncertainties are considered for the
$M_{\rm \nu\mathchar`-NCQE}^{\rm mode}$, $M_{\rm \bar{\nu}\mathchar`-NCQE}^{\rm mode}$,
$M_{\rm NC\mathchar`-other}^{\rm mode}$, and $M_{\rm CC}^{\rm mode}$ components. 
%
%%% PRD
The pseudo experiments are generated assuming gaussian distributed error parameters, 
with means and variances as shown in Tables~\ref{tab:reduchist} and \ref{tab:systerr}.
% 
%In the toy experiments, they are varied as Gaussians, with means 1.0 and 
%widths determined by each individual error source, and multiplied by each other.
% 
Correlations among the flux and cross section parameters are not considered 
in this analysis. 
%Correlations among the flux and cross section parameters were found to have a negligible impact 
%on the final result and have therefore been omitted here. 
% 
The systematic uncertainty on primary-$\gamma$ production is considered to be fully correlated 
among the different interaction types and operation modes, and the secondary-$\gamma$ production 
error is treated in the same way, 
since the change of the $\gamma$-ray emission rate should be common 
for the neutrino interaction types and T2K operation modes.
%
% This choice is motivatied by the fact that changes the emission model are expected 
% to have the same impact on each of these sample components in the same way. 
Note that the primary- and secondary-$\gamma$ production uncertainties are uncorrelated. 
Distributions of the calculated scale factors for one million pseudo experiments
are shown in Figures~\ref{fig:toymcstat} and \ref{fig:toymcsyst}.
Here the dominant error is the secondary $\gamma$-ray model uncertainty
as shown in Table~\ref{tab:systerr}.
The factors $f_{\rm \nu\mathchar`-NCQE}$ and $f_{\rm \bar{\nu}\mathchar`-NCQE}$
have a weak negative correlation for variations of the statistical uncertainty 
but a strong positive correlation under the influence of systematic uncertainties. 
In the end, the scale factors are measured as: 

  %%% Equation : a, b
  \begin{eqnarray}
  \label{eq:aandbresult}
   f_{\rm \nu\mathchar`-NCQE} &=&
     0.80 \pm 0.08 ({\rm stat.}) ^{+ {\rm 0.24}}_{- {\rm 0.18}} ({\rm syst.}), \\ [+10truept] %\nonumber \\ [+10truept]
   f_{\rm \bar{\nu}\mathchar`-NCQE} &=&
     1.11 \pm 0.18 ({\rm stat.}) ^{+ {\rm 0.29}}_{- {\rm 0.22}} ({\rm syst.}). %\nonumber
  \end{eqnarray}
  \vspace{2truept}
  %%%

  %%% Figure : Toy MC for Statistical Error
  %%%

  %%% Figure : Toy MC for Systematic Error
  %%%

The predictions of flux-averaged cross sections by NEUT for neutrino and antineutrino 
NCQE interactions on oxygen, $\langle \sigma_{\nu {\rm \mathchar`-NCQE}}^{\rm NEUT} \rangle$ 
and $\langle \sigma_{\bar{\nu} {\rm \mathchar`-NCQE}}^{\rm NEUT} \rangle$, are
calculated as:

  %%% Equation : NEUT Flux-averaged NCQE cross section
  \begin{eqnarray}
  \label{eq:neutfluxavexsec}
   \langle \sigma_{\nu {\rm \mathchar`-NCQE}}^{\rm NEUT} \rangle &=&
     \frac{\sum_{\nu = \nu_\mu, \nu_e}
           \int \sigma_{\nu {\rm \mathchar`-NCQE}}^{\rm NEUT}(E_\nu)
                \phi_\nu (E_\nu) dE_\nu}
          {\sum_{\nu = \nu_\mu, \nu_e} \int \phi_\nu (E_\nu) dE_\nu} 
	  \nonumber \\ [+3truept]
          &=& 2.13 \times 10^{-38} \ {\rm cm^2/oxygen}, \\ [+10truept] %\nonumber \\ [+10truept]
   \langle \sigma_{\bar{\nu} {\rm \mathchar`-NCQE}}^{\rm NEUT} \rangle &=&
     \frac{\sum_{\nu = \bar{\nu}_\mu, \bar{\nu}_e}
           \int \sigma_{\bar{\nu} {\rm \mathchar`-NCQE}}^{\rm NEUT}(E_\nu)
                \phi_\nu (E_\nu) dE_\nu}
          {\sum_{\nu = \bar{\nu}_\mu, \bar{\nu}_e} \int \phi_\nu (E_\nu) dE_\nu} 
	  \nonumber \\ [+3truept]
          &=& 0.88 \times 10^{-38} \ {\rm cm^2/oxygen}. %\nonumber
  \end{eqnarray}
  \vspace{2truept}
  %%%

\noindent
The nominal flux, $\phi_\nu = \phi_\nu^{\rm FHC}$ is used for neutrinos and 
%($\langle E_\nu \rangle = 0.67^{+0.31}_{-0.24}$~GeV), is used for neutrinos and
$\phi_{\bar{\nu}} = \phi_{\bar{\nu}}^{\rm RHC}$ is used for antineutrinos 
%($\langle E_{\bar{\nu}} \rangle = 0.63^{+0.19}_{-0.26}$~GeV) is used for antineutrinos 
in calculations of the flux-averaged NCQE cross sections. 
Note that summation is done over muon and electron (anti)neutrinos in 
Figure~\ref{fig:t2kflux}, though the actual flux at SK contains tau (anti)neutrinos 
due to neutrino oscillations. 
This treatment is justified because the NC cross section is flavor-independent. 
Here the integrations are conducted up to 10~GeV as higher energies have 
a negligible impact on the result. 
The measured flux-averaged NCQE-like cross sections on oxygen nuclei are obtained 
by multiplying the scale factors to each of Eqs.~(\ref{eq:neutfluxavexsec}) and (9), 

  %%% Equation : Observed Flux-averaged NCQE cross section
  \begin{eqnarray}
  \label{eq:obsfluxavexsec}
   \langle \sigma_{\nu {\rm \mathchar`-NCQE}} \rangle &=& 
     f_{\rm \nu\mathchar`-NCQE} \cdot 
     \langle \sigma_{\nu {\rm \mathchar`-NCQE}}^{\rm NEUT} \rangle
     \nonumber \\ [+3truept]
     &=& 1.70 \pm 0.17 ({\rm stat.}) ^{+ {\rm 0.51}}_{- {\rm 0.38}} ({\rm syst.}) 
     \nonumber \\ [+3truept]
     &&\times \ 10^{-38} \ {\rm cm^2/oxygen}, 
     \\ [+10truept] %\nonumber \\ [+10truept]
   \langle \sigma_{\bar{\nu} {\rm \mathchar`-NCQE}} \rangle &=&
     f_{\rm \bar{\nu}\mathchar`-NCQE} \cdot
     \langle \sigma_{\bar{\nu} {\rm \mathchar`-NCQE}}^{\rm NEUT} \rangle
     \nonumber \\ [+3truept]
     &=& 0.98 \pm 0.16 ({\rm stat.}) ^{+ {\rm 0.26}}_{- {\rm 0.19}} ({\rm syst.}) 
     \nonumber \\ [+3truept]
     &&\times \ 10^{-38} \ {\rm cm^2/oxygen}. 
     %\nonumber
  \end{eqnarray}
  \vspace{2truept}
  %%%

\begin{figure}[htbp]
  % 1st figure
   \begin{center}
    \includegraphics[clip,width=8.4cm]{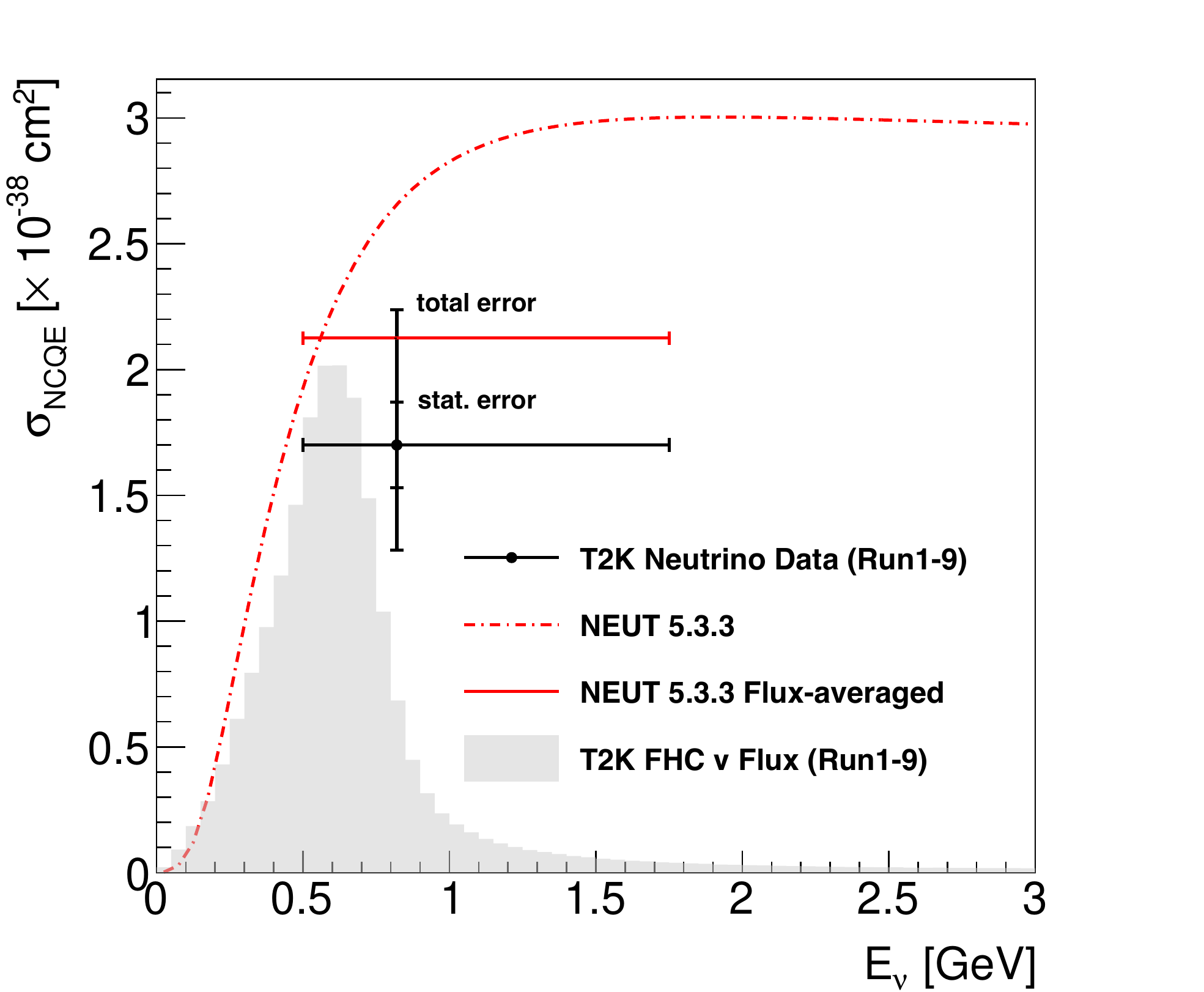}
   \end{center}
  %\vspace{-20truept}
  % 2nd figure
   \begin{center}
    \includegraphics[clip,width=8.4cm]{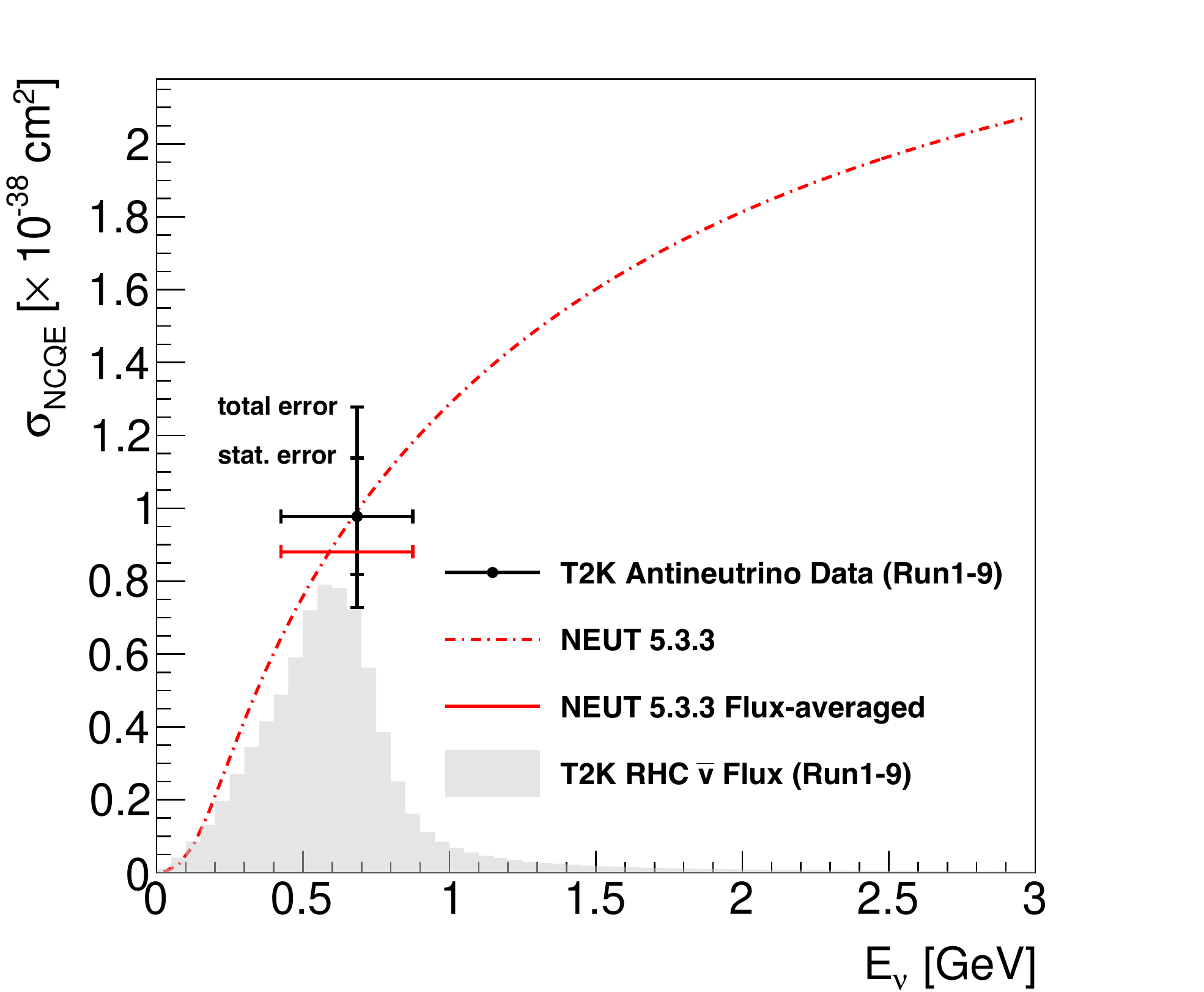}
   \end{center}
  %\vspace{-10truept}
  \caption{The measured $\nu$- (top) and $\bar{\nu}$- (bottom) ${\rm ^{16}O}$ 
           NCQE-like cross sections in comparison with the NCQE cross sections predicted by NEUT.
           The error bars show the statistical error (shorter) and the 
           quadratic sum of statistical and systematic errors (longer). 
           The T2K fluxes for each neutrino beam mode are also shown with 
	   an arbitrary normalization. 
           Data points are placed at the mean flux energies, 0.82~GeV for neutrinos and 
	   0.68~GeV for antineutrions. 
	   Horizontal bars represent the upper and lower range of the mean at 1$\sigma$.}
  \label{fig:ncqexsecresult}
  \end{figure}

\noindent
These measurements are shown together with the predictions from NEUT in Figure~\ref{fig:ncqexsecresult}. 
The neutrino measurement improves over the previous T2K result with FHC data, 
$\langle \sigma_{\nu {\rm \mathchar`-NCQE}} \rangle = 1.55 ^{+ {\rm 0.71}}_{- {\rm 0.35}} ({\rm stat. \oplus syst.})
\times \ 10^{-38} \ {\rm cm^2/oxygen}$ \cite{bib:t2kncqe1to3}.
Covariance matrices of the neutrino and antineutrino flux-averaged NCQE-like cross sections 
are shown for both variations of the statistical and systematic uncertainties in 
Table~\ref{tab:covariance}. 

  %%% Figure : Cross section results

  %%% Table : Covariance of cross sections
  \begin{table}[htbp]
  \begin{center}
  \caption{Covariance of the neutrino and antineutrino cross sections for 
           the statistical (systematic) error case. 
	   The unit of numbers is $(10^{-38}~{\rm cm^2/oxygen})^2$.}
  \label{tab:covariance}
  \vspace{2truept}
   \begin{tabular}{l | r r} \hline \hline
        & $\rm \sigma_{\nu\mathchar`-NCQE}$ & $\rm \sigma_{\bar{\nu}\mathchar`-NCQE}$ \\ \hline 
    $\rm \sigma_{\nu\mathchar`-NCQE}$       & $0.030$ ($0.227$) & $-0.005$ ($0.095$)  \\ 
    $\rm \sigma_{\bar{\nu}\mathchar`-NCQE}$ & $-0.005$ ($0.095$) & $0.025$ ($0.058$) \\ \hline \hline 
   \end{tabular}
  \end{center}
  \end{table}
  %%% 
%  %%% Figure : Covariance of cross sections
%  \begin{figure}[htbp]
%  % 1st figure
%   \begin{center}
%    \includegraphics[clip,width=5.2cm]{./covariance_stat.pdf}
%   \end{center}
%  \vspace{-15truept}
%  % 2nd figure
%   \begin{center}
%    \includegraphics[clip,width=5.2cm]{./covariance_syst_corr.pdf}
%   \end{center}
%  \vspace{-10truept}
%  \caption{Covariance of the neutrino and antineutrino cross sections for 
%           the statistical (top) and systematic (bottom) error cases.}
%  \label{fig:covariance}
%  \end{figure}

%------------------------------------------------------------------------------
%  Discussion
   \section{Discussion}
   \label{sec:discuss}
%..............................................................................

%%%%%
\subsection{NC 2p2h}

Currently, there are no models available in the literature for the NC 2p2h interaction, 
so this channel is not simulated in the present analysis.
Since NC 2p2h interactions involve multi-nucleon knock-out,
not only multiple $\gamma$-rays are expected but additional secondary 
$\gamma$-rays from the recoil nucleons are expected as well. 
It should be noted that if this process exists then the selection in this analysis likely includes 
such events.
However, if the ratio of the NC 2p2h and QE cross sections is similar 
to the corresponding CC ratio, roughly 5$-$10\%~\cite{bib:nieves},
the present measurement will not be sensitive to these events.

%%%%%
\subsection{Comparison with model predictions}

\begin{figure}[htbp]
  % 1st figure
   \begin{center}
    \includegraphics[clip,width=7.8cm]{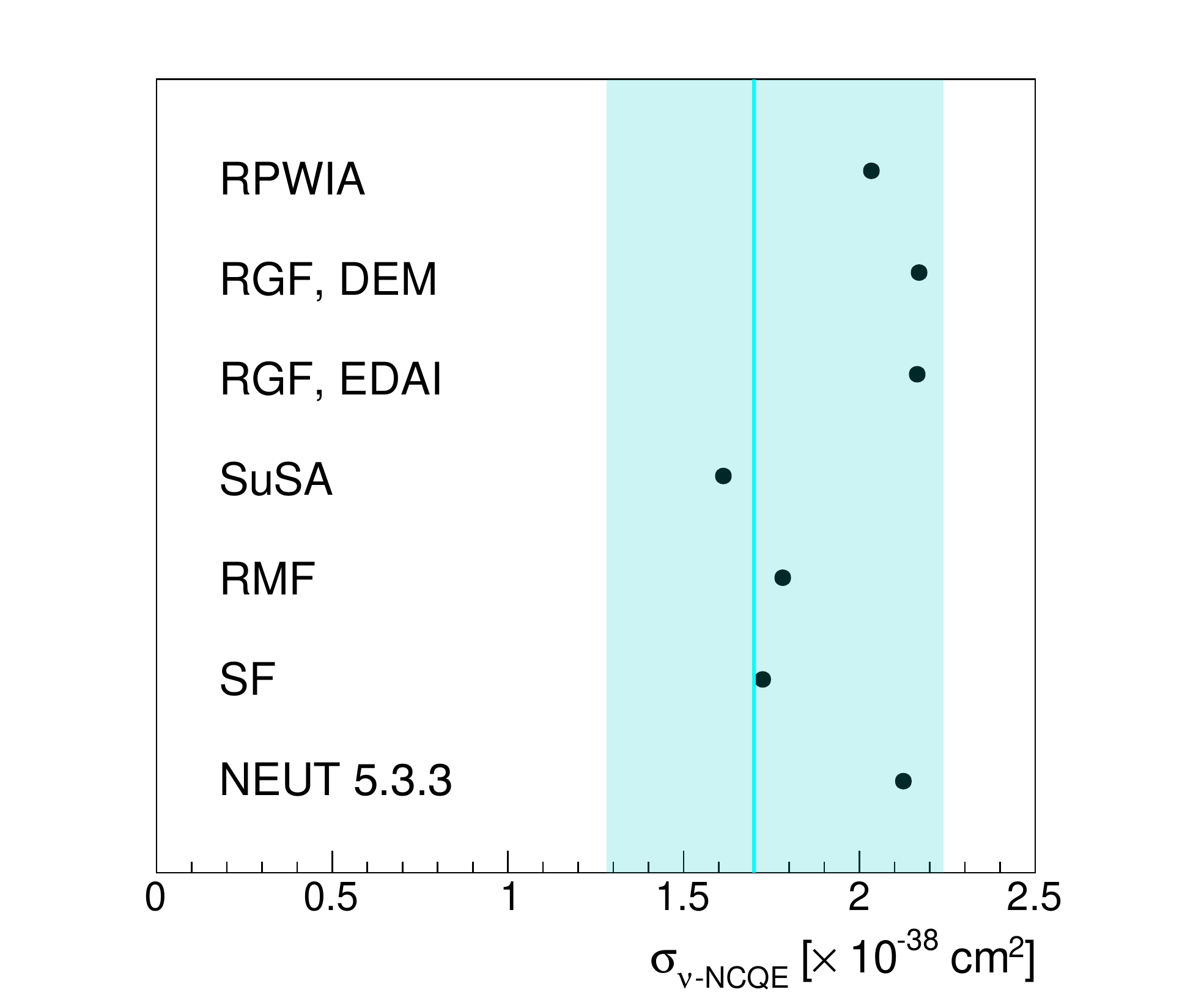}
   \end{center}
  \vspace{-15truept}
  % 2nd figure
   \begin{center}
    \includegraphics[clip,width=7.8cm]{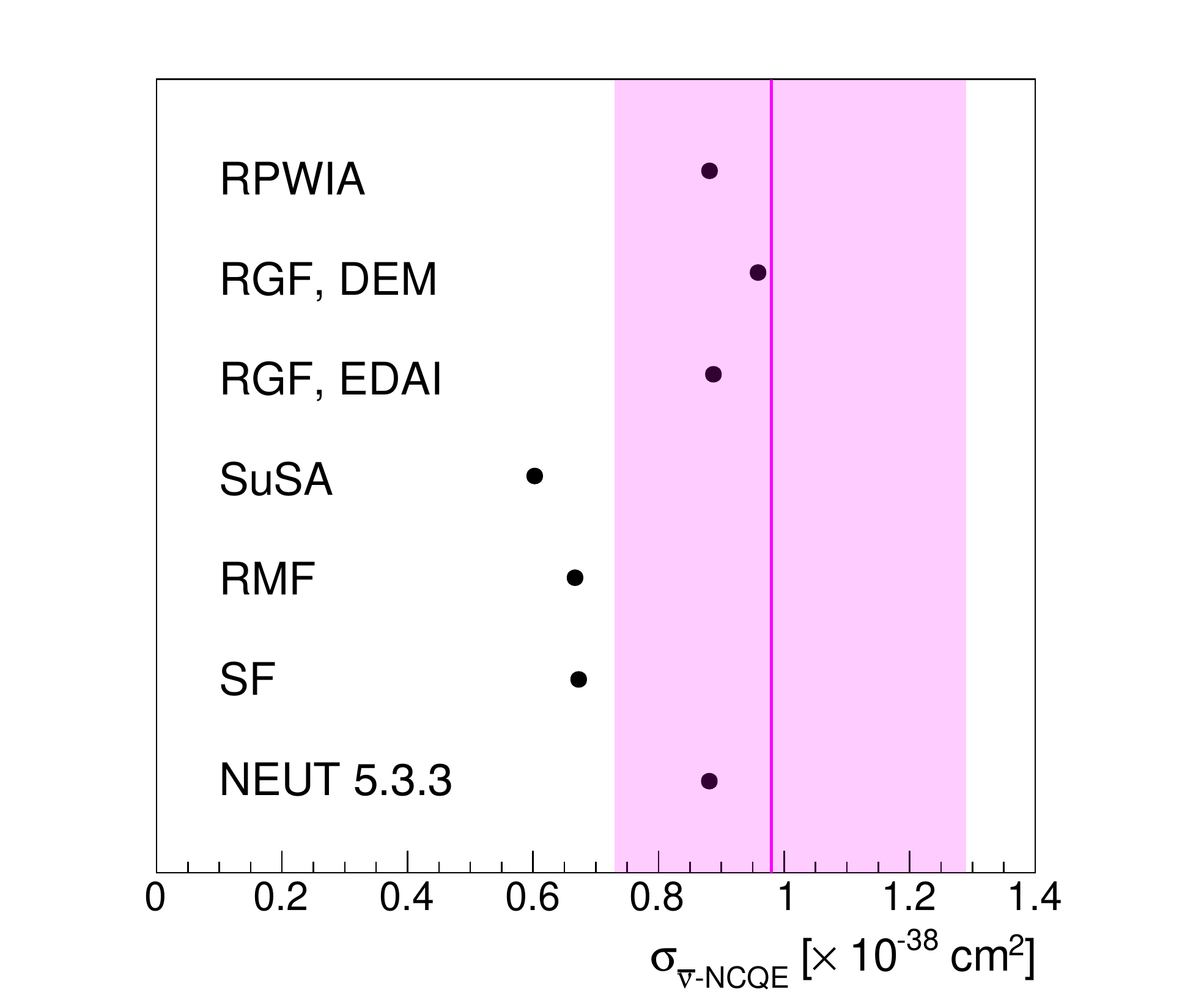}
   \end{center}
  %\vspace{-10truept}
  \caption{Comparison of the measured flux-averaged NCQE-like cross section 
           to the flux-averaged NCQE cross sections by various models for neutrinos (top) 
	   and antineutrinos (bottom).
	   Solid line and shaded area represent the measured mean value and 
	   the 1$\sigma$ uncertainty including both statistical and systematic ones, respectively.}
  \label{fig:model_comparison}
  \end{figure}

%%% PRD
The measured NCQE-like cross sections are tied to NEUT as the underlying model for signal and backgrounds.  
It is interesting to compare the current measurements with various theoretical models.
Six models from Ref.~\cite{bib:ncqetheory} are used in the comparison: the Spectral Function (SF); 
the Relativistic Mean Field (RMF); the Superscaling approach (SuSA); the Relativistic Green's Function 
with two different potentials (RGF EDAI and RGF Democratic); and the Relativistic Plane 
Wave Impulse Approximation (RPWIA) 
\cite{bib:benhar,bib:horowitz,bib:amaro,bib:gonzalez,bib:meucci}. 
The flux-averaged NCQE cross sections for each model are compared in Figure~\ref{fig:model_comparison}.
While the measured result for neutrinos is consistent with all of the models within the 1$\sigma$ error, 
the SF, RMF, and SuSA models lie outside the 1$\sigma$ region for antineutrinos.
%slightly disfavored in the antineutrino case.
% 
However, it is important to note that each model has its uncertainties and 
none of these models contains the NC 2p2h process.  

  %%% Figure : Model comparison

%%%%%
\subsection{Impact on supernova relic neutrino (SRN) searches}

\begin{figure}[htbp]
  % 1st figure
   \begin{center}
    \includegraphics[clip,width=7.8cm]{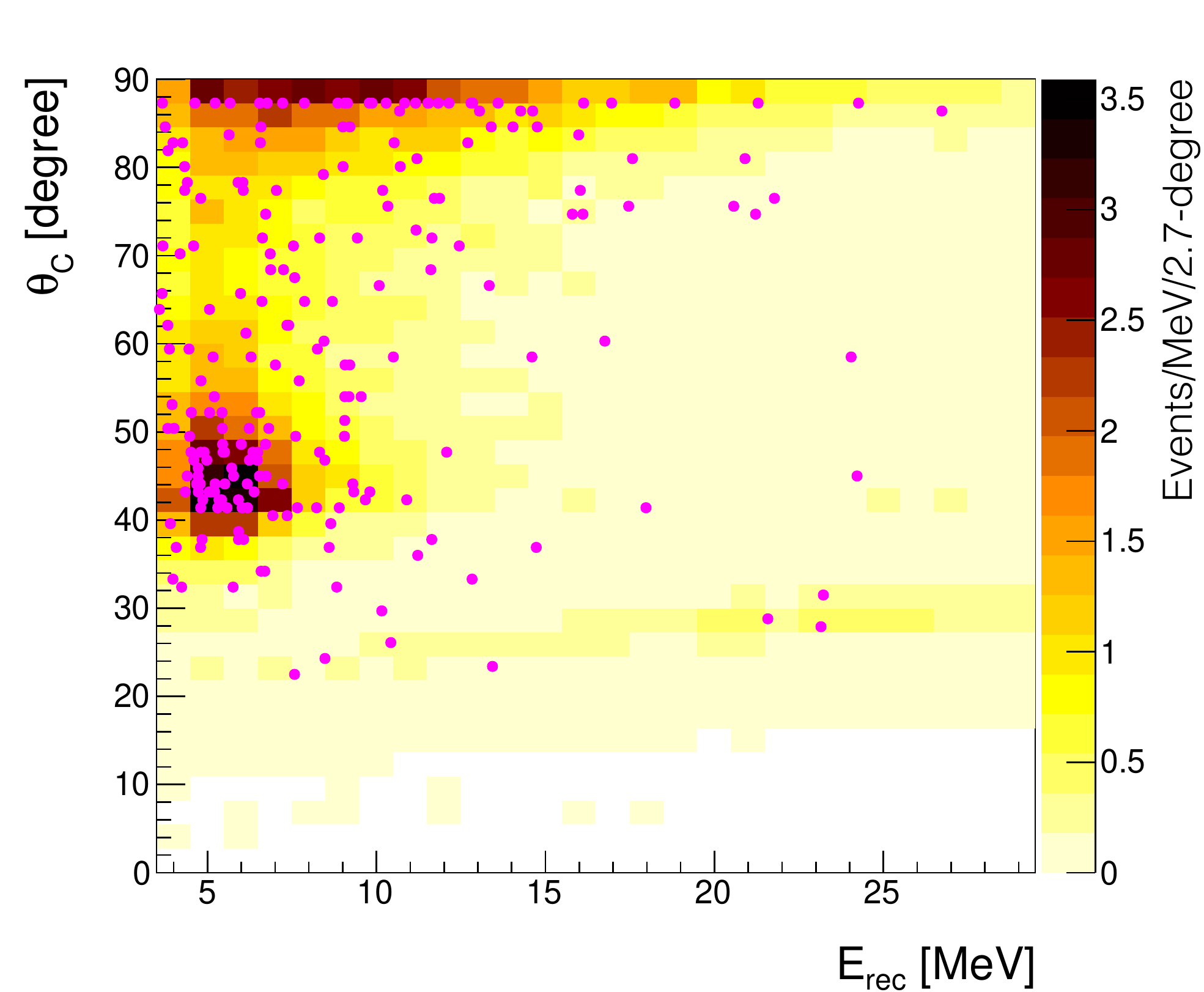}
   \end{center}
  \vspace{-10truept}
  % 2nd figure
   \begin{center}
    \includegraphics[clip,width=7.8cm]{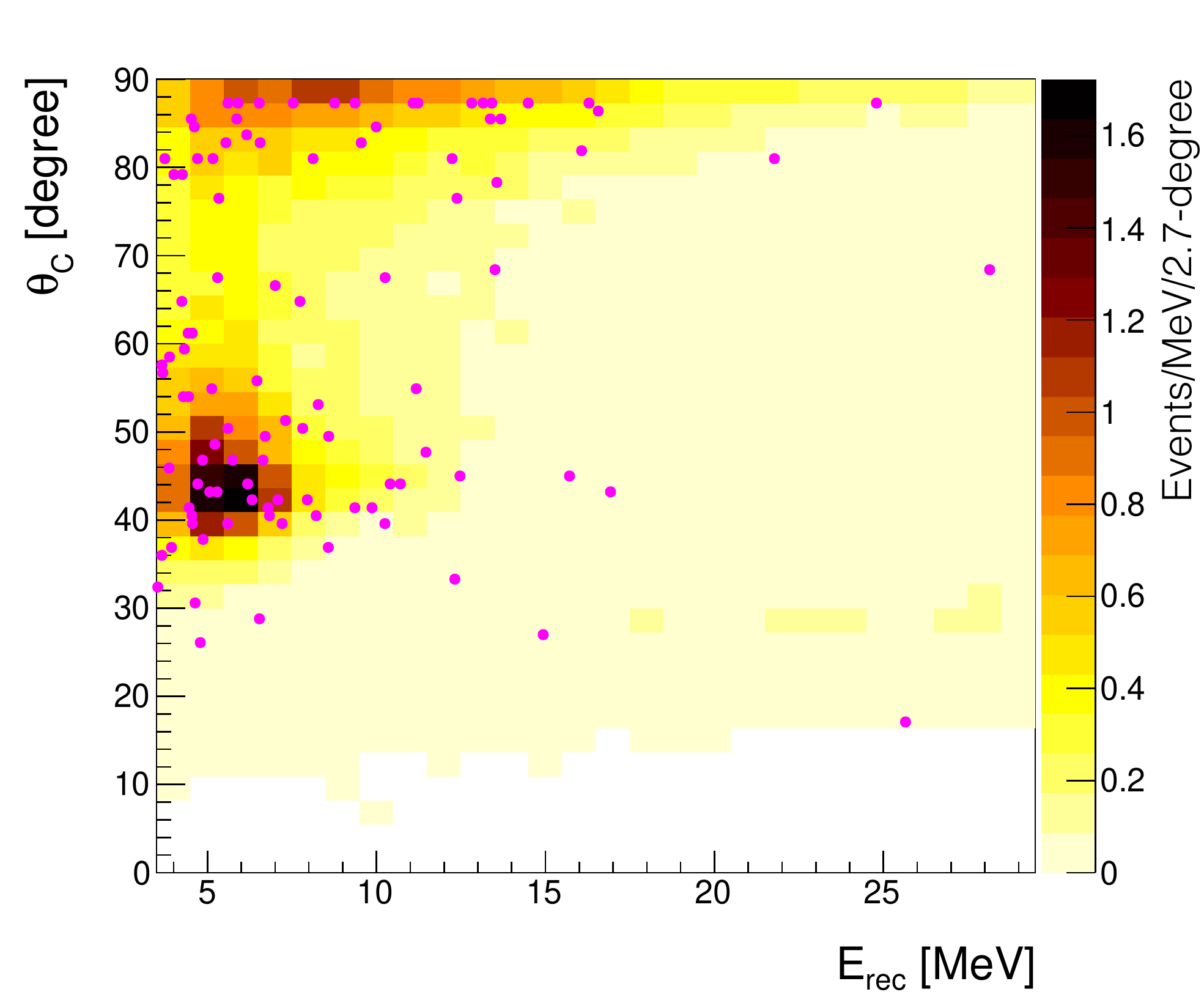}
   \end{center}
  %\vspace{-10truept}
  \caption{Two-dimensional $E_{\rm rec}$$-$$\theta_{\rm C}$ distributions for FHC (top)
           and RHC (bottom) respectively before the CC interaction cut and after 
	   all of the preceding cuts described in Section~\ref{sec:reconsel}.
	   Magenta dots correspond to the observed data.}
  \label{fig:erecangleovaq}
  \end{figure}

\begin{figure}[p]
  % 1st figure
  \begin{center}
   \includegraphics[clip,width=6.6cm]{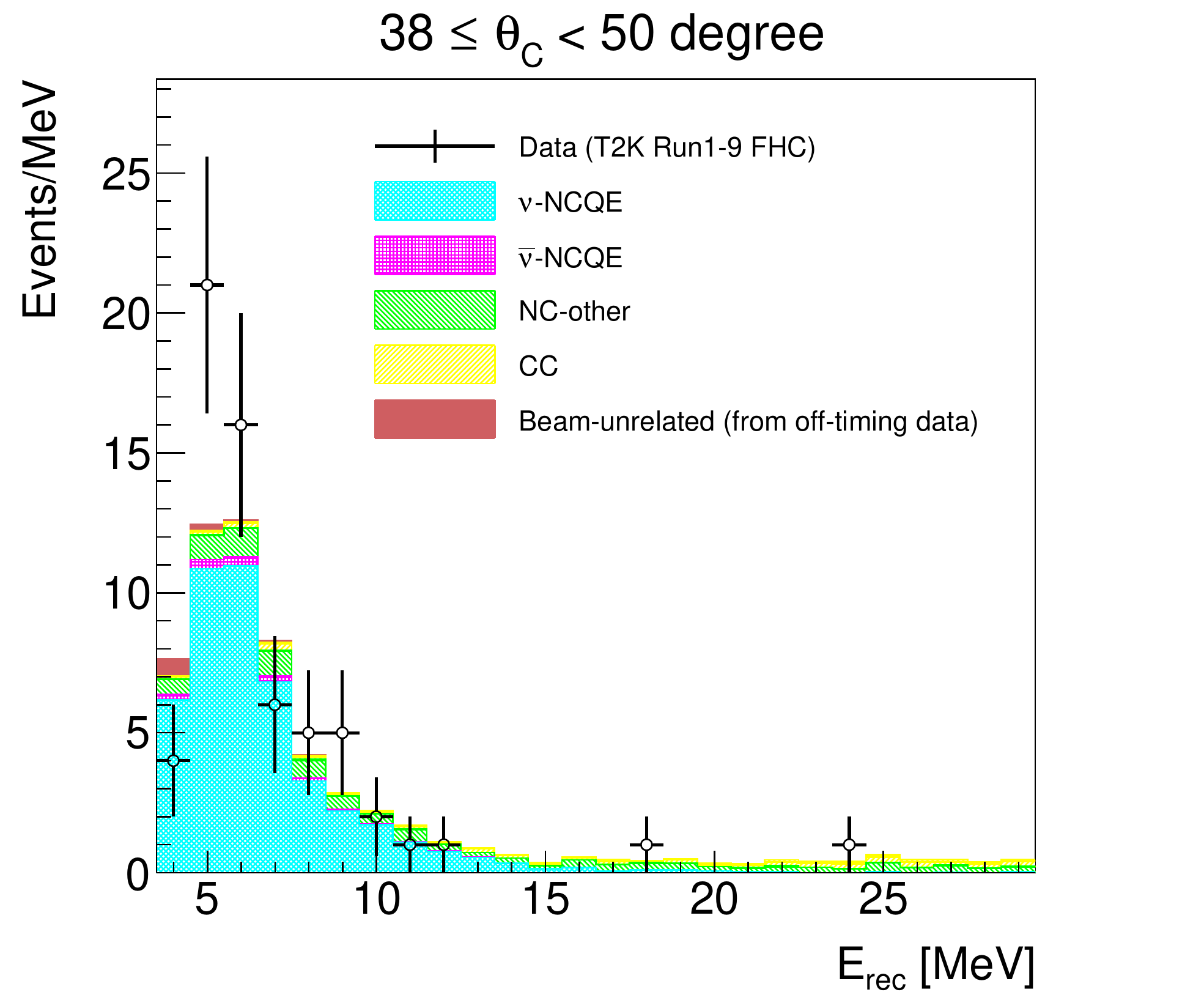}
  \end{center}
  % 2nd figure
  \begin{center}
   \includegraphics[clip,width=6.6cm]{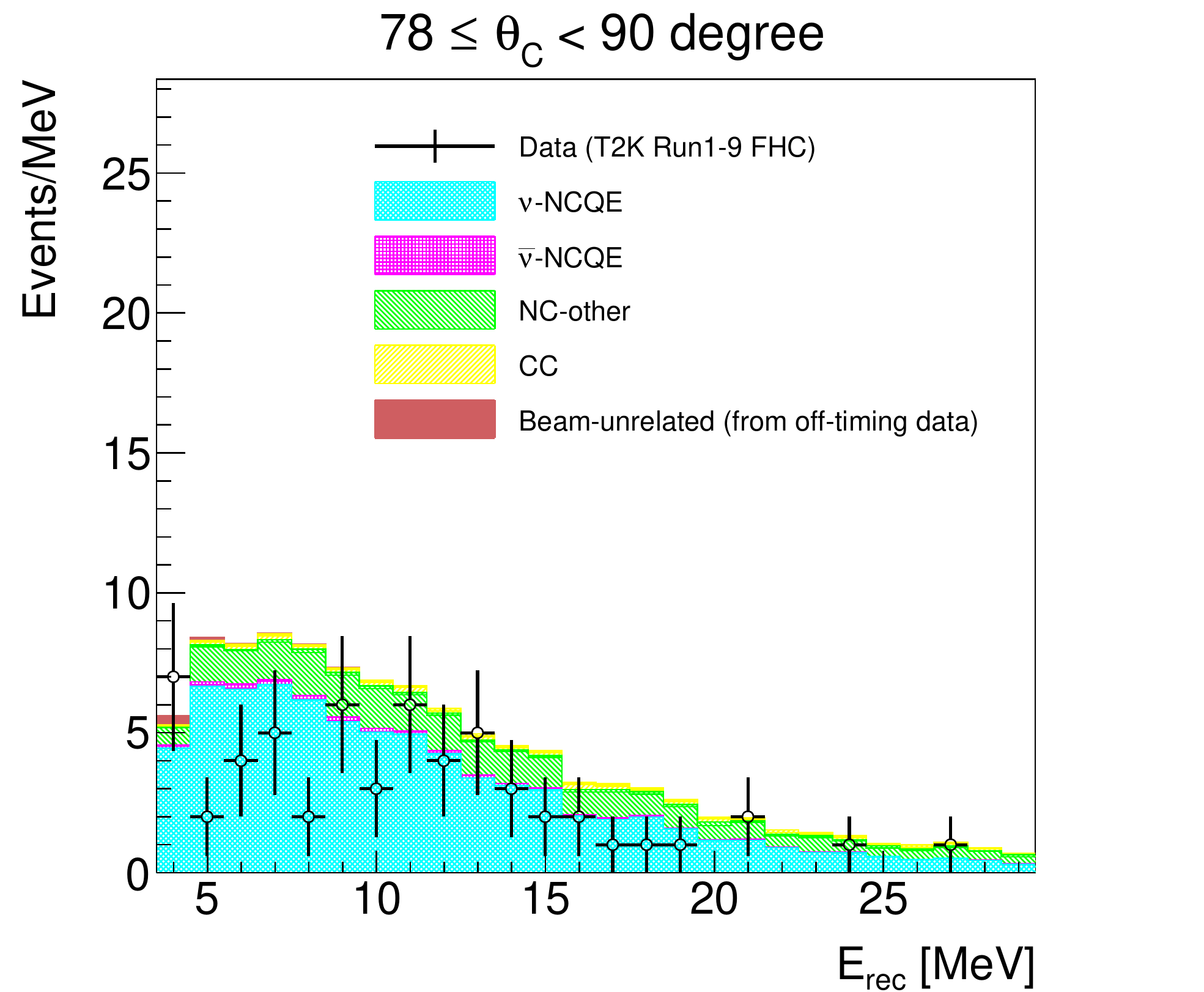}
  \end{center}
  % 3rd figure
  \begin{center}
   \includegraphics[clip,width=6.6cm]{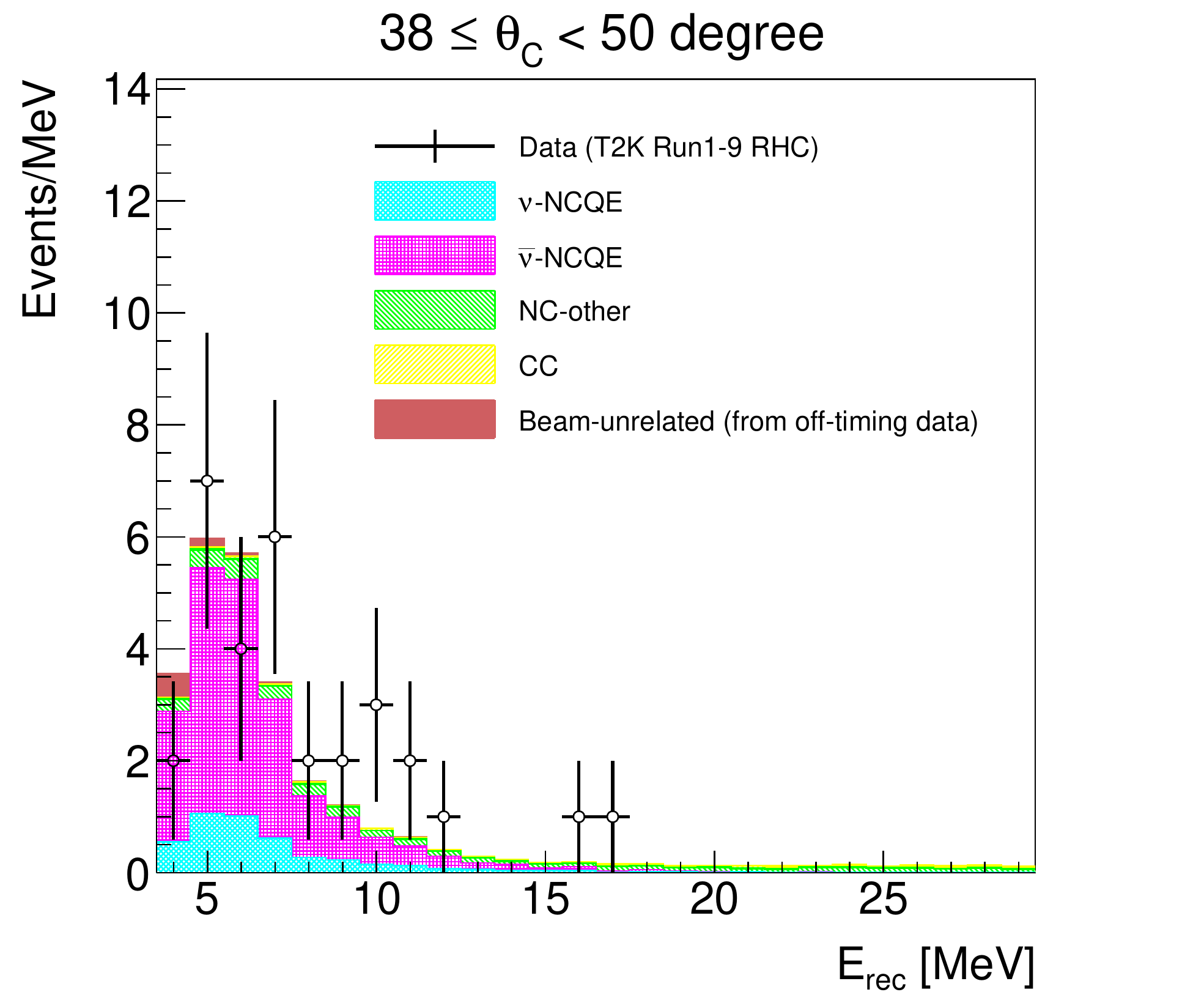}
  \end{center}
  % 4th figure
  \begin{center}
   \includegraphics[clip,width=6.6cm]{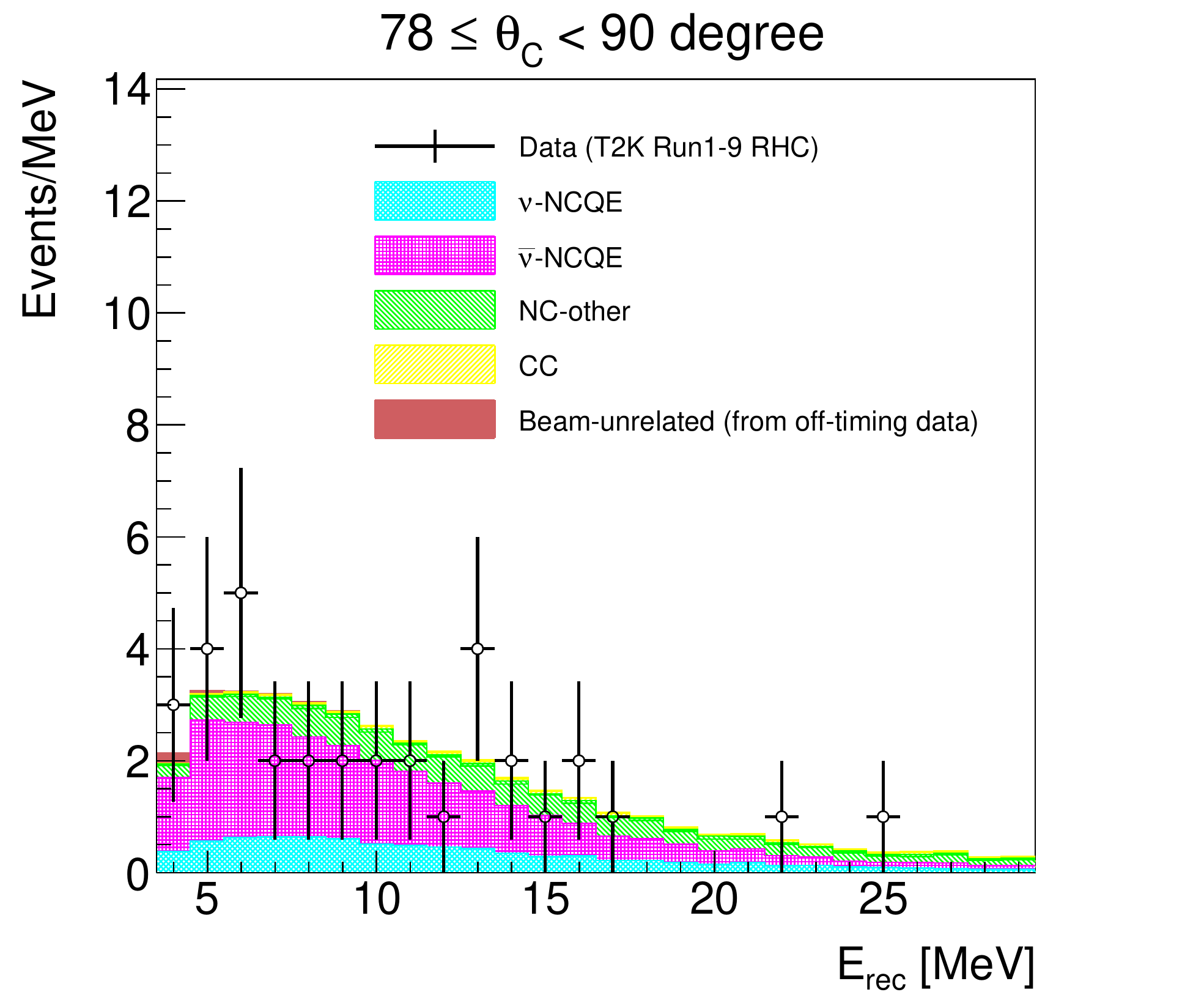}
  \end{center}
  \vspace{-10truept}
  \caption{The $E_{\rm rec}$ distributions for $\theta_{\rm C} \in [38, 50]$~degrees
           and $\theta_{\rm C} \in [78, 90]$~degrees before the CC interaction cut and after 
           all of the preceding cuts described in Section~\ref{sec:reconsel}.
	   Top two figures are the FHC results while the bottom two are the RHC results.}
  \label{fig:erecreg}
  \end{figure}

\begin{figure}[htbp]
  % 1st figure
  \begin{center}
   \includegraphics[clip,width=6.6cm]{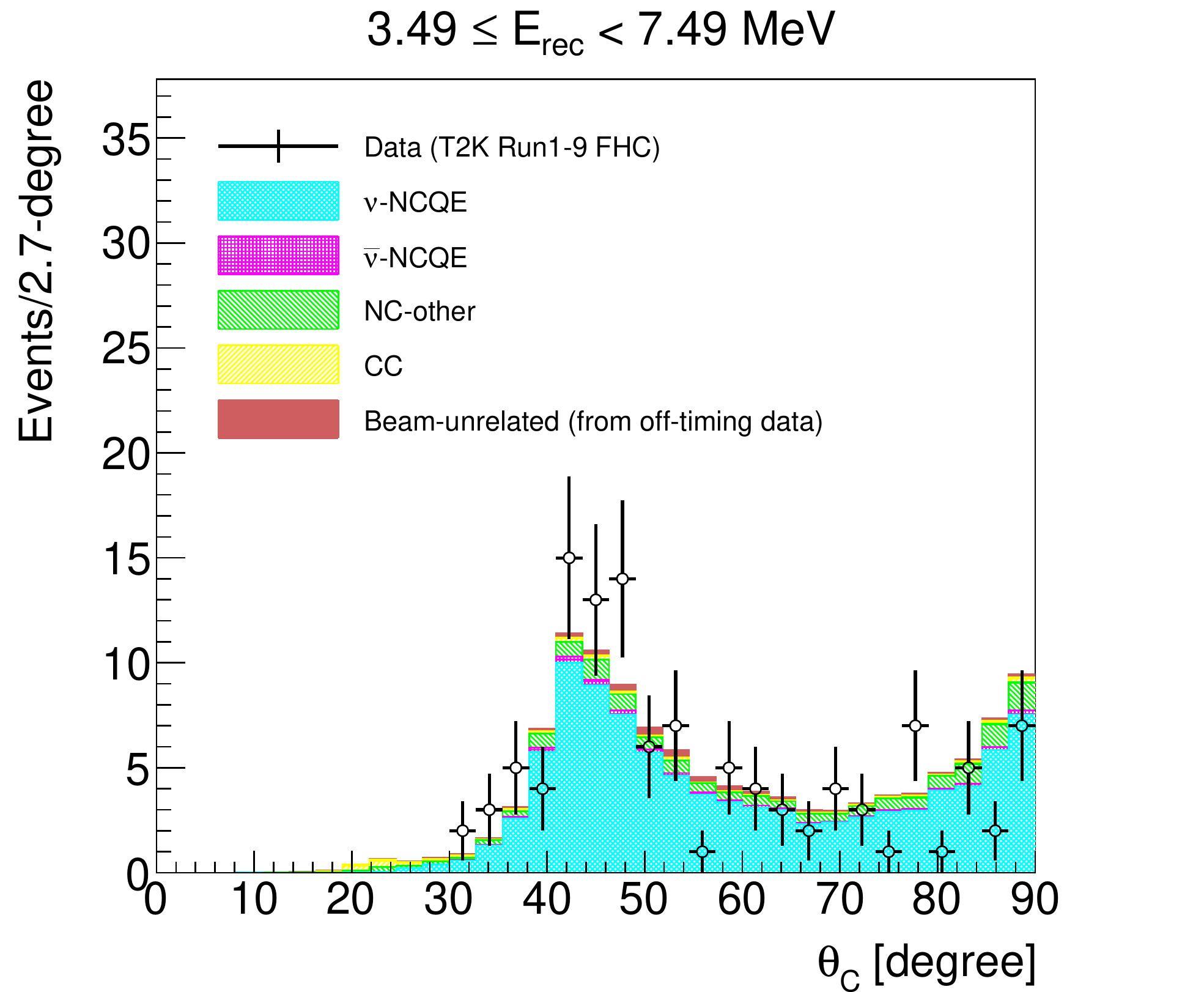}
  \end{center}
  % 2nd figure
  \begin{center}
   \includegraphics[clip,width=6.6cm]{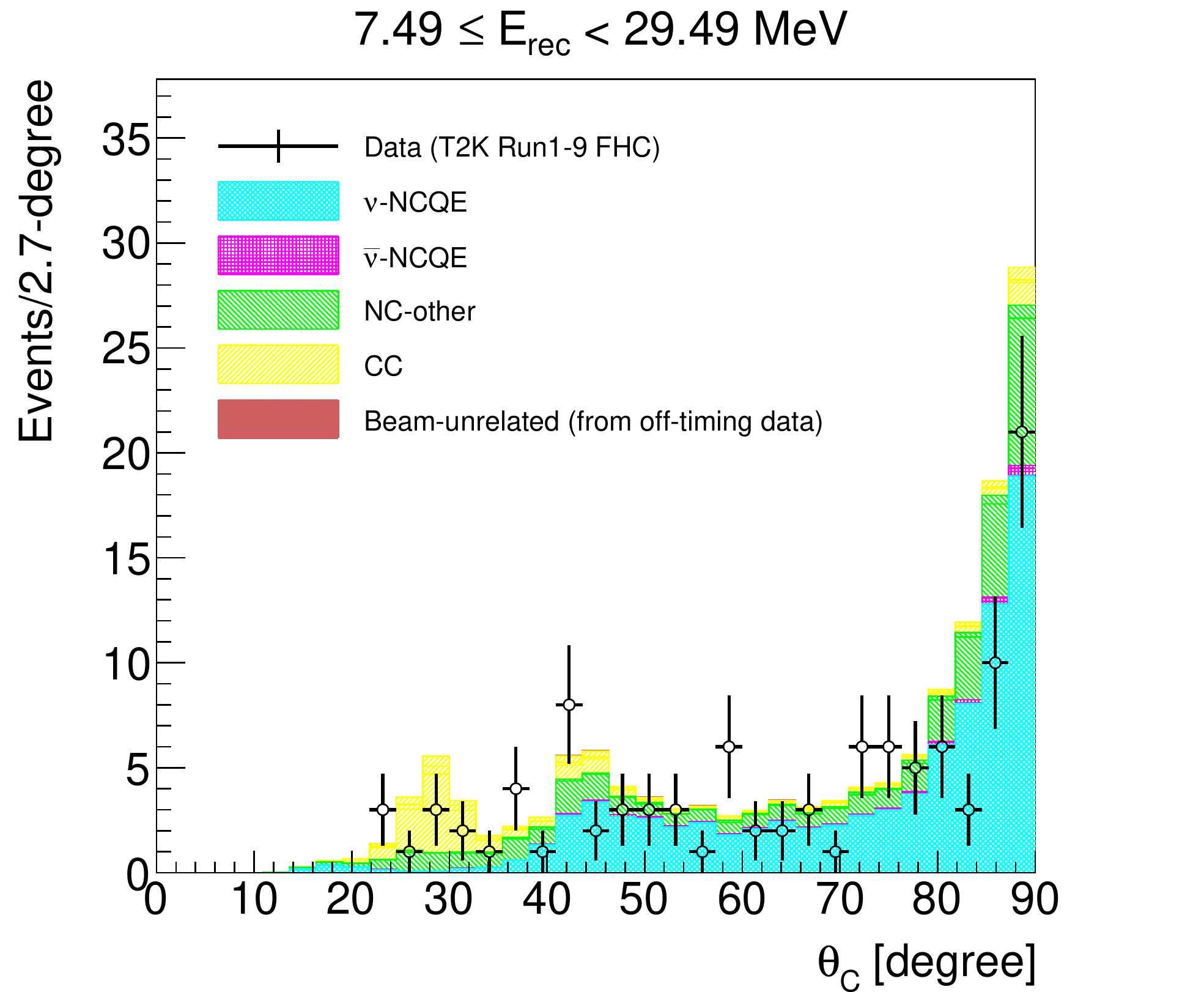}
  \end{center}
  % 3rd figure
  \begin{center}
   \includegraphics[clip,width=6.6cm]{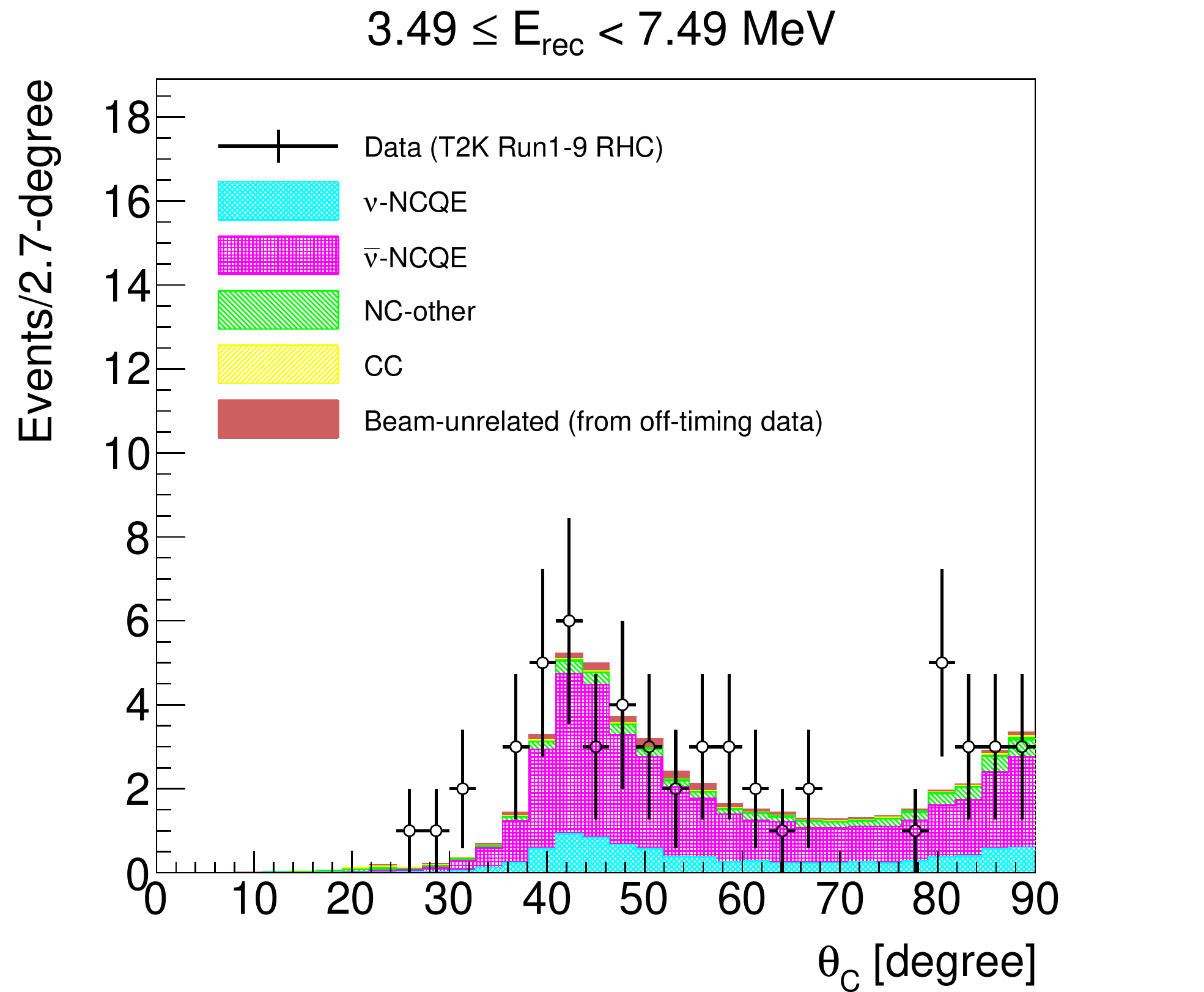}
  \end{center}
  % 4th figure
  \begin{center}
   \includegraphics[clip,width=6.6cm]{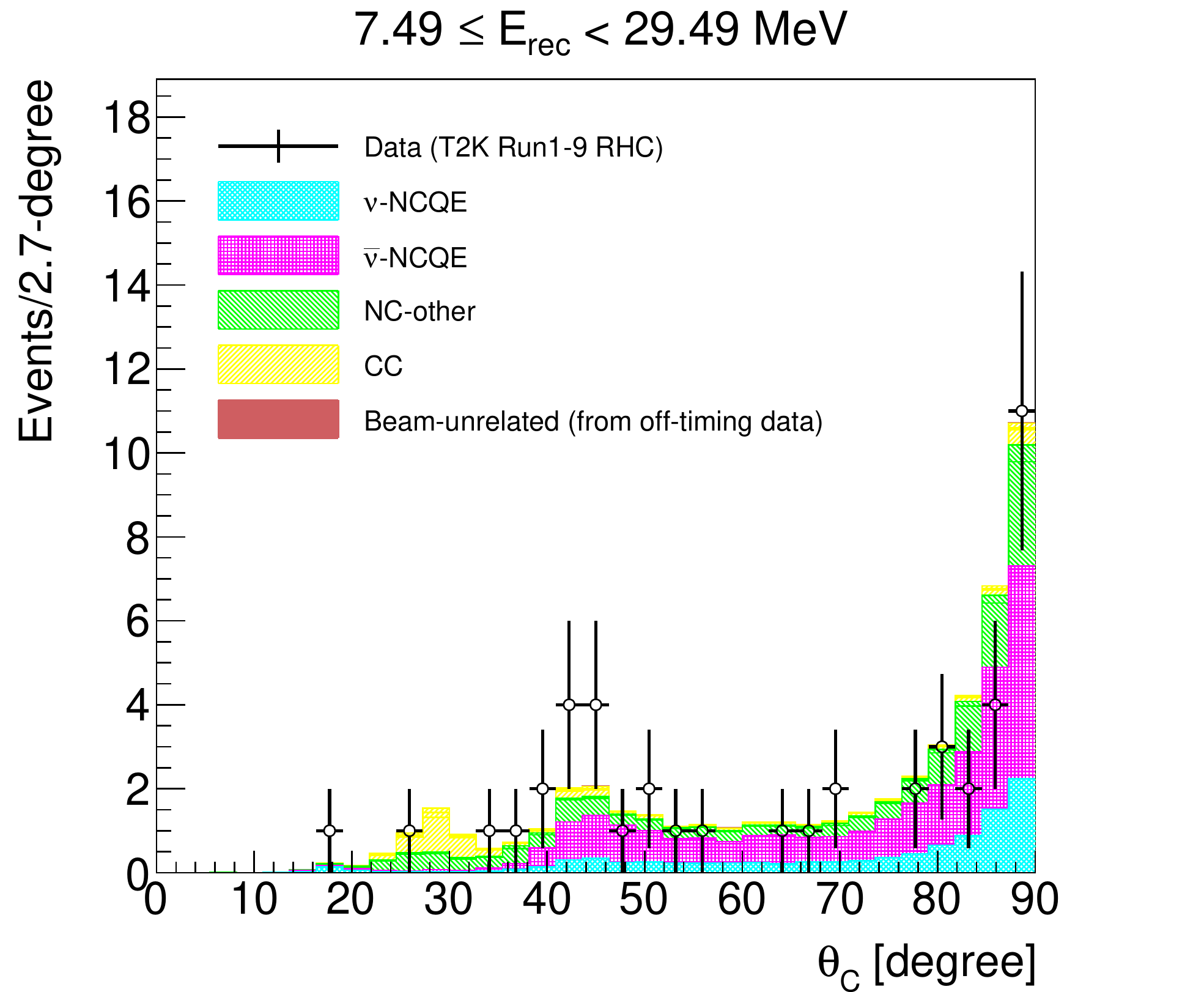}
  \end{center}
  \vspace{-10truept}
  \caption{The $\theta_{\rm C}$ distributions for $E_{\rm rec} \in [3.49, 7.49]$~MeV
           and $E_{\rm rec} \in [7.49, 29.49]$~MeV before the CC interaction cut and 
	   after all of the preceding cuts described in Section~\ref{sec:reconsel},
	   Top two figures are the FHC results while the bottom two are the RHC results.}
  \label{fig:thetacreg}  
  \end{figure}

The present work can be used to estimate NCQE backgrounds
from atmospheric neutrinos to SRN searches. 
Similarly, since $\gamma$-rays from NC 2p2h interactions are also a background to 
such searches the inclusive nature of the current measurement may 
provide useful constraints.
Although the cross section results can be used directly, they suffer from 
large uncertainties from primary- and secondary-$\gamma$ emission models 
as detailed above. 
%% Even with this situation, the current 100\% uncertainty will be reduced.  
% 
If instead one uses the number of events in the expected SRN signal region, 
most uncertainties in Table~\ref{tab:systerr} can be avoided and only errors 
arising from the difference between the T2K beam and atmospheric neutrino fluxes
%%% PRD
($<$10\%)
and detector response error need to be considered.
In the following, the present analysis sample is projected onto the $E_{\rm rec}$$-$$\theta_{\rm C}$ 
phase space used in the SK SRN search and divided into four regions: 
%Although the cross section results can be used, these suffer from 
%large uncertainties of primary- and secondary-$\gamma$ emission models. 
%Even with this situation, the current 100\% uncertainty will be reduced.  
% 
%If instead one uses the number of events in the SRN signal and background regions, 
%all the uncertainties in Table~\ref{tab:systerr} except for the errors due to 
%the flux difference between T2K beam and atmospheric neutrinos will be avoided. 
% 
%In the following, the two-dimensional $E_{\rm rec}$$-$$\theta_{\rm C}$ map 
%is categorized into four regions: 
1) $E_{\rm rec} \in [3.49, 7.49]$~MeV and $\theta_{\rm C} \in [38, 50]$~degrees, 
2) $E_{\rm rec} \in [7.49, 29.49]$~MeV and $\theta_{\rm C} \in [38, 50]$~degrees
3) $E_{\rm rec} \in [3.49, 7.49]$~MeV and $\theta_{\rm C} \in [78, 90]$~degrees, and 
4) $E_{\rm rec} \in [7.49, 29.49]$~MeV and $\theta_{\rm C} \in [78, 90]$~degrees. 
The signal window of the SRN analysis in SK corresponds to region~2 (higher $E_{\rm rec}$ and lower $\theta_{\rm C}$).
Figure~\ref{fig:erecangleovaq} gives the $E_{\rm rec}$$-$$\theta_{\rm C}$ distributions 
from the FHC and RHC data and MC before the CC interaction cut and after all of the preceding cuts
described in Section~\ref{sec:reconsel}. 
Table~\ref{tab:regionevt} summarizes the number of beam events in each region 
calculated from Figure~\ref{fig:erecangleovaq}.
Note that the difference between the observed number of events and predictions in regions~3 and 4 for the FHC sample 
may be attributed to the inaccuracy of the secondary interaction model as explained in Section~\ref{sec:reconsel}.
The $E_{\rm rec}$ distributions for $\theta_{\rm C} \in [38, 50]$~degrees and
$\theta_{\rm C} \in [78, 90]$~degrees for the FHC and RHC samples are given in
Figure~\ref{fig:erecreg}.
Similarly, Figure~\ref{fig:thetacreg} shows the $\theta_{\rm C}$ distributions for 
$E_{\rm rec} \in [3.49, 7.49]$~MeV and $E_{\rm rec} \in [7.49, 29.49]$~MeV.
Here also the FHC distributions for observation and prediction show discrepancies, 
which may be attributed to modeling of the secondary-$\gamma$ emission. 
These distributions can be used to estimate the NCQE background to the SRN search by 
suitable weighting of the MC to data.
% 
%\textcolor{red}{
%Here, it needs to be done to estimate systematic errors on the number of events 
%in each region, in which better understanding secondary-$\gamma$ emission is essential. 
%}
% 
Though beyond the scope of the present work, this is expected to significantly improve the 
current 100\% error on this background used in the SK SRN analysis \cite{bib:sksrn12,bib:sksrn123}. 
%
%To make use of these results in the SRN analysis, it is necessary to 
%consider differences in the neutrino flux (T2K or atmospheric) and 
%the kinematics of the induced nucleons. 
%Then it is possible to estimate the NC background via the unfolding method.

  %%% Figure : Erec-thetaC after ovaQ cut

  %%% Table : Region event number in FHC 
  \begin{table}[htbp]
  \begin{center}
  \caption{Number of observed and predicted events for each region 
           defined in the text.}
  \label{tab:regionevt}
  \vspace{2truept}
   \begin{tabular}{l c c c c} \hline \hline
    FHC                & Region~1 & Region~2 & Region~3 & Region~4 \\ \hline
    Observation        & 47   & 16   & 18   & 40   \\ \hline 
    Prediction (total) & 41.1 & 20.4 & 30.8 & 73.8 \\ 
    $\nu$-NCQE         & 34.8 & 10.7 & 24.4 & 49.6 \\
    $\bar{\nu}$-NCQE   & 1.1  & 0.3  & 0.6  & 1.3  \\
    NC-other           & 3.4  & 5.7  & 4.6  & 19.3 \\
    CC                 & 0.8  & 3.6  & 0.7  & 3.6  \\
    Beam-unrelated     & 1.0  & 0.1  & 0.5  & 0.0  \\ \hline \hline
    RHC                & Region~1 & Region~2 & Region~3 & Region~4 \\ \hline
    Observation        & 19   & 12  & 14   & 21   \\ \hline
    Prediction (total) & 18.6 & 7.3 & 11.9 & 27.0 \\ 
    $\nu$-NCQE         & 3.2  & 1.1 & 2.2  & 5.7  \\
    $\bar{\nu}$-NCQE   & 13.4 & 3.4 & 7.5  & 13.1 \\
    NC-other           & 1.2  & 2.1 & 1.7  & 7.2  \\
    CC                 & 0.1  & 0.7 & 0.2  & 1.0  \\
    Beam-unrelated     & 0.7  & 0.0 & 0.3  & 0.0  \\ \hline \hline
   \end{tabular}
  \end{center}
  \end{table}
  %%%

  %%% Figure : Erec distributions 
  %%%

  %%% Figure : thetaC distributions 
  %%%

%%%%%
\subsection{Future prospects}

At present T2K has collected less than half of its expected POT and 
extensions of the experiment are being considered \cite{bib:t2k2}.
The larger statistics of future data sets motivate several possible 
improvements to the present work.
Systematic errors from the secondary-$\gamma$ production model
can be reduced by incorporating recent measurements of $\gamma$-ray emission 
from neutron-oxygen interactions into MC. 
Measurements using 30, 80, and 250~MeV neutrons have been performed, but only results 
at 80~MeV are available at present~\cite{bib:rcnpe487}.
%There are some possible updates to this analysis in the future.
% 
%First, to reduce the systematic error from secondary-$\gamma$ production model,
%$\gamma$-ray emission from neutron reactions on oxygen should be measured and
%the appropriate model needs to be implemented in MC.
%Such measurements are carried out at Osaka University's Research Center for
%Nucler Physics (RCNP) at neutron energies of 30, 80, and 250~MeV.
%The result of 80~MeV is available in Ref.~\cite{bib:rcnpe487}, but the analyses 
%of other energies are on-going.
%
%% either using hydrogen captures in the present data or gadolinium captures in the upcoming SK-Gd phase, 
Furthermore, neutron tagging at SK, particularly the high-efficiency tagging 
realized in the coming Gd-doped phase of Super-Kamiokande (SK-Gd),
can be used to study the relationship of neutrons, their transport in water, 
and the production of secondary $\gamma$-rays.
%% of investigation of the Cherenkov angle distribution. 
% 
Information on the neutron capture vertex would further constrain the neutron 
kinetic energy in NCQE interactions by measurement of the neutron flight distance 
from the primary interaction vertex.
Neutron information would also allow for differential cross section measurements 
using the reconstructed $Q^{2}$ as well as studies of $\Delta s$ if proton 
and neutron final states can be distinguished.
Finally, using the $\sim$8~MeV $\gamma$ cascade following neutron capture on Gd,
it may be possible to identify the NCQE interactions resulting in the ground state 
nucleus by requiring no activity by the primary-$\gamma$. 

%Going a step further, it would be possible to make a differential cross section 
%measurement using $Q^2$ information.  
%In this case, a more challenging analysis of extracting $Q^2$ information could 
%be conducted and the differential cross section would be measured. 
%
%Another possible analysis is the measurement of $\Delta s$ by separating 
%neutrons from protons in the final state. 
%Here also the neutron tagging and precise knowledge of
%secondary $\gamma$-rays are required.
% 
%In the SK-Gd phase, these analyses are possible and 
%investigation of the nucleon state inside the nucleus could also be possible. 
%Since the sum energy of $\gamma$-rays by Gd capture is $\sim$8~MeV,
%it may be possible to tag the ground state after an NCQE interaction 
%by requiring nothing in the primary sub event. 

%\clearpage
%------------------------------------------------------------------------------
%  Conclusion
   \section{Conclusion}
   \label{sec:conclude}
%..............................................................................

In this paper, neutrino- and antineutrino-oxygen neutral-current quasielastic-like
interactions have been measured using nuclear de-excitation $\gamma$-rays at the T2K far detector,
with data corresponding to $14.94\times10^{20}$~POT in
FHC and $16.35\times10^{20}$~POT in RHC polarities.
Compared to the previous T2K study, the present analysis has improved the event simulation and selection criteria, 
and reduced both systematic and statistical uncertainties. 
In addition, this work presents the first measurement of antineutrino interactions in this channel to date.
The measured flux-averaged NCQE-like cross sections on oxygen nuclei are 
$\langle \sigma_{\nu {\rm \mathchar`-NCQE}} \rangle = 
1.70 \pm 0.17 ({\rm stat.}) ^{+ {\rm 0.51}}_{- {\rm 0.38}} ({\rm syst.}) 
\times 10^{-38} \ {\rm cm^2/oxygen}$ for neutrinos at a flux-averaged energy of 0.82~GeV and 
$\langle \sigma_{\bar{\nu} {\rm \mathchar`-NCQE}} \rangle = 
0.98 \pm 0.16 ({\rm stat.}) ^{+ {\rm 0.26}}_{- {\rm 0.19}} ({\rm syst.})
\times 10^{-38} \ {\rm cm^2/oxygen}$ for antineutrinos at a flux-averaged energy of 0.68~GeV.
Simultaneously treating both FHC and RHC data has resulted in similar sized 
errors for both the neutrino and antineutrino measurements.
These results were found to be consistent with currently available models 
within the measurement precisions. 
In addition, MC and data comparisons in the kinematic regions of interest 
for SRN searches were performed. 
These measurements are expected to improve estimates of backgrounds to those searches 
not only in the present Super-Kamiokande experiment, 
but also in future water Cherenkov detectors such as SK-Gd and Hyper-Kamiokande. 
The data related to the results presented in this paper can be found in \cite{bib:ncqedatarelease}.

%------------------------------------------------------------------------------
%  Acknowledgments
   \section*{Acknowledgments}
%..............................................................................

We thank the J-PARC staff for superb accelerator performance. 
We thank the CERN NA61/SHINE Collaboration for providing valuable particle production data. 
We acknowledge the support of MEXT, Japan; NSERC (Grant No. SAPPJ-2014-00031), 
NRC and CFI, Canada; CEA and CNRS/IN2P3, France; DFG, Germany; INFN, Italy; 
National Science Centre (NCN) and Ministry of Science and Higher Education, Poland; 
RSF (Grant \#19-12-00325) and Ministry of Science and Higher Education, Russia; 
MINECO and ERDF funds, Spain; SNSF and SERI, Switzerland; STFC, UK; and DOE, USA. 
We also thank CERN for the UA1/NOMAD magnet, DESY for the HERA-B magnet mover system, 
NII for SINET4, the WestGrid and SciNet consortia in Compute Canada, and GridPP in the United Kingdom. 
In addition, participation of individual researchers and institutions has been further supported 
by funds from ERC (FP7), ``la Caixa" Foundation (ID 100010434, fellowship code LCF/BQ/IN17/11620050), 
the European Union's Horizon 2020 Research and Innovation programme under 
the Marie Sklodowska-Curie grant agreement no. 713673 and H2020 Grant No. RISE-GA644294-JENNIFER 2020; 
JSPS, Japan; Royal Society, UK; and the DOE Early Career program, USA.

%------------------------------------------------------------------------------
%  References 
%..............................................................................

%\clearpage
\bibliography{reflist}

\end{document}